\documentclass[3p,12pt]{elsarticle}
\pdfoutput=1

\let\counterwithin\relax
\usepackage[linktoc=all,colorlinks=true,citecolor=blue,linkcolor=blue]{hyperref}
\usepackage{amsmath,amssymb}
\usepackage{epsfig}
\usepackage{graphicx}               % Standard graphics package  
\usepackage{url}
\usepackage{titlesec}
\usepackage{cleveref}
\usepackage{color}
\usepackage{listings}
\usepackage{array}
\usepackage{natbib}
\biboptions{numbers,sort&compress}

\RequirePackage[T1]{fontenc}
\RequirePackage[utf8]{inputenc}

\usepackage{titlesec}

\usepackage{tikz}
\usetikzlibrary{trees}
\usetikzlibrary{shapes.geometric}

\tikzstyle{every node}=[draw=black,thick,anchor=west]
\tikzstyle{selected}=[dashed,draw=red,fill=red!30]
\tikzstyle{optional}=[dashed,fill=gray!50]

\usepackage{chngcntr}
\counterwithin{figure}{section}

\usepackage{hyphenat}
\titleclass{\subsubsubsection}{straight}[\subsection]
\newcounter{subsubsubsection}[subsubsection]
\renewcommand\thesubsubsubsection{\thesubsubsection.\arabic{subsubsubsection}}
\titleformat{\subsubsubsection}
  {\normalfont\normalsize\itshape}{\thesubsubsubsection}{1em}{}
\titlespacing*{\subsubsubsection}
{0pt}{3.25ex plus 1ex minus .2ex}{1.5ex plus .2ex}

\newcommand{\bra}[1]{\ensuremath{\left\langle#1\right|}}
\newcommand{\ket}[1]{\ensuremath{\left|#1\right\rangle}}

\newcommand{\bls}{\begin{lstlisting}}
\newcommand{\els}{\end{lstlisting}}

\newcommand{\ttf}{\ttfamily}
\newcommand{\nuSQUIDS}{{\ttfamily nuSQUIDS}}

\newcommand{\pa}[2]{\frac{\partial #1}{\partial #2}}

\newcolumntype{C}[1]{>{\centering\arraybackslash}m{#1}}

\definecolor{mauve}{rgb}{1,0,1}
\definecolor{dkgreen}{rgb}{0,0.6,0}
\newcounter{bla}

\begin{document}

\begin{frontmatter}

\title{\texttt{nuSQuIDS}: A toolbox for neutrino propagation\ifdefined\forjournal\else\tnoteref{t1}\fi}

\author[HU]{Carlos A. Arg\"uelles}
\ead{carguelles@fas.harvard.edu}
\author[UB]{Jordi Salvado}
\ead{jsalvado@icc.ub.edu}
\author[MSU]{Christopher N. Weaver}
\ead{chris.weaver@icecube.wisc.edu}

\address[HU]{Department of Physics \& Laboratory for Particle Physics and Cosmology, Harvard University, Cambridge, MA 02138, USA}
\address[UB]{Departament de F\'isica Qu\`antica i Astrofísica and Institut de Ciencies del Cosmos,
Universitat de Barcelona, Diagonal 647, E-08028 Barcelona, Spain}
\address[MSU]{Dept. of Physics and Astronomy, Michigan State University, East Lansing, MI 48824, USA}

\ifdefined\manualonly
% omit most front matter
\else % not manualonly
\tnotetext[t1]{Code can be found at \url{https://github.com/arguelles/nuSQuIDS}}
\ifdefined\forjournal
\journal{Computer Physics Communications}
\else
\journal{arXiv}
\fi % forjournal

\begin{abstract}
The Neutrino Simple Quantum Integro-Differential Solver (\texttt{nuSQuIDS})
is a C++ code based on SQuIDS that propagates an ensemble of neutrinos
through given media. Neutrino oscillation calculations relevant to
current and next-generation experiments are implemented. 
This includes coherent and non-coherent neutrino interactions in
settings such as the Sun, Earth, or a vacuum.
The code is designed to be accurate and flexible, while at the
same time maintaining good performance. It has a modular design that
allows the user to incorporate new physics in novel scenarios. 
\end{abstract}

\begin{keyword}
Neutrino oscillation, phenomenology, collective neutrino behavior, numerical techniques
\end{keyword}
\fi % manualonly

\end{frontmatter}

\ifdefined\forjournal
{\bf PROGRAM SUMMARY}

\begin{small}
\noindent
{\em Program Title:} Neutrino Simple Quantum Integro-Differential Solver \\
{\em CPC Library link to program files:} (to be added by Technical Editor) \\
{\em Developer's repository link:} \url{https://github.com/arguelles/nuSQuIDS} \\
{\em Code Ocean capsule:} (to be added by Technical Editor)\\
{\em Licensing provisions:} LGPL  \\
{\em Programming language:} C++ with optional Python interface \\
{\em Supplementary material:}                                 \\
{\em Nature of problem:}\\
The evolution of neutrino fluxes includes both coherent flavor changes (oscillations) and non-coherent scattering interactions with matter. Calculating both types of phenomena is necessary for some studies in neutrino physics, and even in cases where it is not required it is convenient to have a full physics description available in a single package. \\
{\em Solution method:}\\
The Simple Quantum Integro-Differential Solver (SQuIDS)~\cite{summary_squids, summary_squids_update} is used to decompose neutrino fields with SU(N) algebras and solve for their evolution with standard ODE solvers (via the GNU Scientific Library~\cite{summary_gsl}). The evolution of the flavor/mass eigenstate relationships are kept updated in parallel with the ODE solution, as are interaction effects with matter. Users may add new material profiles and physics effects. \\
{\em Additional comments including restrictions and unusual features:}\\
Vacuum oscillations are computed analytically, simplifying the ODE problem to be solved. This package supports additions to the neutrino flux over the full course of its evolution, allowing direct treatment of extended sources, and averaging of fast oscillations which cannot be experimentally resolved. The package does not currently implement the complex neutrino interactions relevant at low (O(10 GeV)) neutrino energies. 
   \\

\end{small}
\fi % forjournal

\hypersetup{linkcolor=black}
\tableofcontents
\hypersetup{linkcolor=blue}
\newpage

\ifdefined\manualonly
% omit paper body
\else % not manualonly
\section{Introduction}
\label{sec:intro} 

In recent decades a large body of evidence that neutrinos change
flavor as they propagate macroscopic distances due to the
non-alignment of their mass and flavor eigenstates has accumulated from
solar~\citep{Super-Kamiokande:2010tar, Borexino:2013zhu},
atmospheric~\citep{PhysRevD.91.072004,Super-Kamiokande:2015qek,IceCube:2019dqi,ANTARES:2018rtf},
accelerator~\citep{PhysRevLett.112.181801,
  PhysRevD.93.051104,PhysRevLett.116.151806, PhysRevLett.110.251801}, and
reactor~\citep{An:2013zwz,Abe:2015rcp, Kim:2016yvm} experiments.
Thanks to these remarkable
experimental results and related theoretical calculations, the
  neutrino-mass-induced flavor oscillation paradigm~\citep{Pontecorvo:1967fh,Gribov:1968kq,fukugita2003physics,
  Akhmedov:1999uz,Balantekin:2013kc, GonzalezGarcia:2007ib,Mohapatra:qv, Gouvea:2013fj}
has been firmly established and the three mixing angles, which
parametrize the lepton-mixing matrix, together with the two
squared-mass differences, have been measured to good
precision~\citep{Esteban:2020cvm,deSalas:2017kay,Capozzi:2018ubv,deSalas:2020pgw}. It
is the task of on-going 
and future experiments to determine the neutrino-mass ordering 
and the $CP$-violating phase~\citep{Hewett:2012et,
  Acciarri:2016crz,Aartsen:2014oha, Kouchner:2016pqa,DeRosa:2016ifc}. 
Also, the IceCube Neutrino Observatory has recently made precise measurements of the atmospheric
spectrum above 100~GeV, where the Earth is no longer transparent to
neutrinos~\citep{Donini:2018tsg,Bustamante:2017xuy,IceCube:2020rnc}, allowing new physics
models to be constrained~\citep{Aartsen:2014gkd,TheIceCube:2016oqi}.
The identification of high-energy
extraterrestrial neutrinos~\citep{Aartsen:2014gkd,Aartsen:2015rwa,IceCube:2020wum}
has opened the possibility of exploring new physics at these energies as well~\citep{Arguelles:2015dca, Bustamante:2015waa, Baerwald:2012kc,Arguelles:2019rbn, Esteban:2021tub}. 

Matter effects play a fundamental role in the explanation of solar
neutrinos~\citep{ Davis:1968cp,Bethe:1986ej}, which has motivated the study of new flavor-changing
neutrino interactions~\citep{Barger:1991ae, Roulet:1991sm, GonzalezGarcia:2011my,
  Gonzalez-Garcia:2013usa, Pospelov:2011dp, Kopp:2014nosterile,
  Maltoni:2015kca,Esteban:2018ppq}.
Even though most data can be explained in the standard three
neutrino framework, some puzzling anomalies still remain
~\citep{LSND,Mention:2011rr,MiniBoone:2012dn,Aguilar-Arevalo:2018gpe,Alekseev:2016llm,
  Ko:2016owz, Ashenfelter:2018iov, Abreu:2018pxg,
  Dentler:2018sju}. These may be explained 
by introducing new light neutrino states~\citep{kopp2013sterile,
  Collin:2016rao, Abazajian:2012rf,Blennow:2018hto, Dentler:2018sju, Diaz:2019fwt, Boser:2019rta} and other new physics
~\citep{Palomares-Ruiz:2005zbh,Gninenko:2009ks,Nelson:2010hz,Fan:2012ca,Bai:2015ztj,Bertuzzo:2018itn,Ballett:2019pyw,Arguelles:2018mtc,Dentler:2019dhz,Ahn:2019jbm,deGouvea:2019qre,Abdallah:2020vgg,Hostert:2020oui,Brdar:2020tle,Abdallah:2020biq,Dutta:2021cip,Vergani:2021tgc}.
Also, the interplay between cosmology and neutrino
oscillation has been widely studied in the literature
~\citep{Bergstrom:2014fqa, Giusarma:2016phn,
  Dasgupta:2013la,Hernandez:2016kel,Arguelles:2016uwb,Song:2018zyl,Chu:2018gxk}. 
Generally, neutrinos are good probes to perform 
fundamental physics tests~\citep{Hewett:2012et,Aartsen:2017ibm,Mewes:2018cze,Barenboim:2017vlc}.

Tools are needed to compute neutrino
propagation accurately and reliably.
In classical computations, where the only effect of propagation is neutrino oscillation, libraries such as {\ttf GLoBES}
~\citep{Huber:2007ji}, {\ttf Prob3++}~\citep{prob3pp, Calland:2013vaa},
and {\ttf nuCRAFT}~\citep{Wallraff:2014vl} are
available; for a recent comparison of oscillation calculations comparisons see~\cite{Parke:2019vbs}.
On quantum computers, such as the IBM-Q systems, implementations using two entangled qubits have also been developed~\cite{Arguelles:2019phs}.
Unfortunately, when non-coherent interactions~\cite{Formaggio:2012cpf,Gandhi:1998ri,Zhou:2019frk,Zhou:2019vxt} are
important, these methods are no longer applicable.
In the high-energy regime, when oscillations can be neglected, tools have been developed
to propagate neutrinos taking into account non-coherent interactions; for example, \texttt{TauRunner}~\cite{Safa:2021ghs}, \texttt{NuPropEarth}~\cite{Garcia:2020jwr}, \texttt{nuFATE}~\cite{Vincent:2017svp}, \texttt{JuLiET}~\cite{Yoshida:2003js,shigeruyoshida_2020_4018117}, among others~\citep{Gazizov:2004va,DeYoung:865626}.
Beyond these general-purpose codes, packages have also been developed to propagate tau neutrinos specifically, \textit{e.g.} \texttt{NuTAuSim}~\cite{Alvarez-Muniz:2018owm} and \texttt{nuPyProp}~\cite{NuSpaceSim:2021hgs}.
However, none of the above-mentioned packages are able to take 
coherent and non-coherent neutrino interactions into account simultaneously.
Furthermore, the ability to allow the user to incorporate new physics models or to change the
propagation medium is also limited in some of these packages. {\ttf nuSQuIDS} seeks to
address all of these limitations by providing a highly-customizable
package while at the same time remaining numerically efficient.

In section~\ref{sec:theory}
we review neutrino oscillation and interaction theory, 
in section~\ref{sec:features} we summarize the main features of the code, 
in section~\ref{sec:short_examples} we show the use of the code in a few representative scenarios, 
in section~\ref{sec:comparison} we discuss how this code compares to other software,
in section~\ref{sec:performance} we show the performance of
the library, and finally section~\ref{sec:conclu} presents concluding remarks.

\ifdefined\forjournal
\else
Additionally, an extended appendix that provides
further useful information about the code.
In \ref{sec:tests} we describe the unit tests that
are provided with the library, in \ref{sec:code} we describe
the main classes and structure of the code; and finally, in \ref{sec:python}
we describe the library {\ttf Python} interface. 
\fi %forjournal

\section{Mathematical Framework}
\label{sec:theory}
\subsection{Neutrino Oscillations} 
In this section we briefly review neutrino oscillations
using the density matrix formalism.
We can represent the state of a neutrino ensemble at an energy $E$
and position $x$ using the density matrix. In the weak-interaction flavor-eigenstate basis,
$\{\ket{\nu_\alpha}\}$,  it can be written as
\begin{equation}
\rho(E,x) = \sum_\alpha \phi_\alpha(E,x) \ket{\nu_\alpha}\bra{\nu_\alpha} , 
\label{eq:state}
\end{equation}
where $\phi_\alpha$ specifies the flux of flavor $\alpha$.
These eigenstates can be related with the mass eigenstates,  $\{ \ket{\nu_i}  \}$, by
\begin{equation}
\ket{\nu_\alpha} = \sum_i U^*_{\alpha i} \ket{\nu_i} ,
\label{eq:changebasis}
\end{equation}
where $U$ is a unitary matrix known as the Pontecorvo-Maki-Nakagawa-Sakata (PMNS)
matrix, or neutral lepton mixing matrix. For antineutrinos, the relation
is the same as in 
Eq.~\eqref{eq:changebasis} with $U \to U^*$.
It is customary to parametrize $U$ as a product of complex rotations,
\begin{equation}
\label{eq:mixing}
U(\theta_{ij}\delta_{ij})=R_{N-1\, N} R_{N-2\, N} ... R_{45} R_{35} R_{25} R_{15}
R_{34} R_{24} R_{14} R_{23} R_{13} R_{12}, 
\end{equation}
determined by angles, $\{\theta_{ij}\}$, and phases, $\{ \delta_{ij}
\}$~\cite{SQUIDS, SQUIDSupdate}; see~\cite{Denton:2020igp} for a recent discussion about parameterizations of the PMNS matrix.
When considering the standard three-flavor paradigm,
the following parametrization is often used:
\begin{equation}
U
=
\begin{pmatrix}
c_{12} c_{13} & s_{12} c_{13} & s_{13} e^{-i\delta_{13}} \\ 
- s_{12} c_{23} - c_{12} s_{23} s_{13} e^{i\delta_{13}} & c_{12} c_{23} - s_{12} s_{23} s_{13} e^{i\delta_{13}} & s_{23} c_{13} \\
s_{12}s_{23} -c_{12}c_{23}s_{13}e^{i\delta_{13}} & - c_{12} s_{23} - s_{12} c_{23} s_{13} e^{i\delta_{13}} & c_{23} c_{13}
\end{pmatrix}
\,,
\label{eq:U}
\end{equation}
where $c_{ij} \equiv \cos \theta_{ij}$, $s_{ij} \equiv \sin \theta_{ij}$. In the
three-flavor scenario, we use this parametrization (with
values from \citep{Esteban:2020cvm} by default).
The neutrino ensemble propagation is
described by the following quantum Von Neumann equation\footnote{In this equation and henceforth we set $c = \hbar = 1$.} 
\begin{equation}
\pa{\rho(E,x)}{x} = -i [ H (E,x), \rho(E,x) ].
\label{eq:schrodinger}
\end{equation}

In general, the Hamiltonian, $H$, can always be split into
time-dependent and time-independent parts. For neutrino oscillations,
the following splitting is convenient:
\begin{equation}
H(E,x) = H_0(E)  + H_{1}(E,x) ,
\end{equation}
with
\begin{subequations}
\label{eq:hamiltonian}
\begin{align}
H_0 (E) &= \frac{1}{2E} {\rm diag}( 0 , \Delta m^2_{21},\Delta m^2_{31},\Delta m^2_{41},...,\Delta m^2_{n1}) \label{eq:h0} ,\\
H_1 (E,x) &= \sqrt{2} G_F U^\dagger {\rm diag} ( N_e(x) -
N_{nuc}(x)/2, -N_{nuc}(x)/2, -N_{nuc}(x)/2 , 0,...,0 )U ,\label{eq:hi} 
\end{align}
\end{subequations}
where $n$ is the number of neutrino states, $G_F$ is the Fermi
constant, $N_e(x)$ and $N_{nuc}(x)$ are the electron and nucleon number
densities at position $x$, and $\Delta m^2_{i1}$ are the neutrino mass
square differences.
In writing these equations, we have used the convention that the first
three flavor eigenstates correspond to 
$\nu_e$, $\nu_\mu$, and $\nu_\tau$ respectively, while the rest are assumed to be
sterile neutrinos. $H_0$ arises from the neutrino kinetic
term, whereas $H_1$ incorporates the matter potential induced by coherent
forward scattering~\citep{Mikheev:1986gs,Mikheev:1986wj,Wolfenstein:1977ue}. Note that the matter
potential, $H_1$, given in Eq.~\eqref{eq:hi} for neutrinos changes to
$-H_1^*$ for antineutrinos. 
Since the propagation due to $H_0$ can be solved analytically, it is
convenient to use the  so-called interaction picture. For an operator
$O(x)$ the interaction picture transformed
operator, $O_I(x)$ is defined as
\begin{equation}
O_I(x)=\exp(-iH_0x)O(x)\exp(iH_0x).
\end{equation}
The corresponding evolution equation  for the neutrino state is
\begin{equation}
\pa{\rho_I(E,x)}{x} = -i [ H_{1I} (E,x), \rho_I(E,x) ]~.
\label{eq:schrodinger_int}
\end{equation}

\subsection{Non-coherent Interactions}
So far, we have only incorporated vacuum oscillations and matter
effects through coherent interactions, but we now wish to extend this
formalism to incorporate non-coherent interactions and collective
neutrino behavior. In what follows, we will remove the subindex $I$ and
assume that all operators, unless specified, are in the interaction picture. 
This problem has been extensively discussed in the literature,~\citep{Sigl:1992fn,Duan:2010tk,Strack:qd,Zhang:2013ay,
 Cirelli:mw,Blennow:2007tw,Arguelles:2012cf}, for specificity we
follow the formalism and notation given in~\citep{Gonzalez-Garcia:2005xw}.
The neutrino (antineutrino), $\rho$ $(\bar\rho)$, kinetic equations are
\begin{subequations}
\begin{eqnarray}
\pa{\rho(E,x)}{x} &=& -i [ H_1 (E,x), \rho(E,x) ] - \left\{ \Gamma(E,x),
  \rho(E,x) \right\} + F\left[\rho,\bar\rho;E,x\right] ,\\
\pa{\bar\rho(E,x)}{x} &=& i [ H^*_1 (E,x), \bar\rho(E,x) ] - \left\{ \bar\Gamma(E,x),
  \rho(E,x) \right\} + \bar F\left[\rho,\bar\rho;E,x\right] ,
\end{eqnarray}
\label{eq:kinetic_equations}
\end{subequations}
where $\Gamma$ and $\bar\Gamma$  are functions that incorporate the effect of attenuation due to
non-coherent interactions, and $F$ and $\bar F$  are  functionals on
$\rho$ and $\bar\rho$ that take into account interactions between
different energies for neutrinos and antineutrinos.

In {\ttf nuSQuIDS}, the attenuation terms are
\begin{subequations}
\begin{eqnarray}
\Gamma(E,x) &=& \frac{1}{2} \sum_\alpha  \frac{\Pi_\alpha(E,x)}{
  \lambda^\alpha_{\rm NC}(E,x)+\lambda^\alpha_{\rm CC}(E,x)},\label{eq:gammarhoa} \\
\bar\Gamma(E,x) &=& \frac{1}{2} \sum_\alpha  \frac{\bar\Pi_\alpha(E,x)}{
  \bar\lambda^\alpha_{\rm NC}(E,x)+\bar\lambda^\alpha_{\rm CC}(E,x)
  + \bar\lambda^\alpha_{\rm GR}(E,x)}, \label{eq:gammarhob}
\end{eqnarray}
\end{subequations}
where $\Pi_\alpha(E,x)$ is the neutrino projector onto flavor $\alpha \in
\{e,\mu,\tau\}$, $\lambda^\alpha_{\rm CC}$ ($\lambda^\alpha_{\rm NC}$)
is the charged (neutral) current neutrino interaction length,
given by $(N_{nuc}(x)\sigma^\alpha_{\rm CC(NC)}(E))^{-1}$\citep{Formaggio:2012cpf,Gandhi:1998ri,Zhou:2019frk,Zhou:2019vxt,CooperSarkar:2011pa}, and
$\bar\lambda^e_{\rm GR}$ is the mean free path due to the Glashow
resonance $(N_{e}(x)\sigma^e_{\rm GR}(E))^{-1}$~\citep{Gandhi:1998ri}. Notice that we assume
the matter only contains electrons, protons, and neutrons,
\textit{i.e.} $\bar\lambda^\mu_{\rm GR}=\bar\lambda^\tau_{\rm GR}=0$.
The other interaction terms are as follows:
\begin{subequations}
  \begin{eqnarray}
    F\left[\rho,\bar\rho;E,x\right] &=& \sum_\alpha \Pi_\alpha(E,x)  \int_E^\infty  
    {\rm Tr}\left[\Pi_\alpha(E_{\nu_\alpha},x) \rho(E_{\nu_\alpha},x) \right]
    \frac{1}{\lambda^\alpha_{\rm NC}(E_{\nu_\alpha},x)} \pa{N^\alpha_{\rm
        NC}(E_{\nu_\alpha},E)}{E} dE_{\nu_\alpha}  \label{eq:intronu} \nonumber\\
    &&  + \Pi_\tau (E,x) \int_E^\infty\int_{E_\tau}^\infty  
    {\rm Tr} \left[ \Pi_\tau(E_{\nu_\tau},x)
      \rho(E_{\nu_\tau},x)\right] \nonumber\\
    && \hspace{2cm} \times \frac{1}{\lambda^\tau_{\rm CC}(E_{\nu_\tau},x)}
    \pa{N^{\tau}_{\rm CC} (E_{\nu_\tau},E_\tau)}{E_\tau}
    \pa{N^{\rm all}_{\rm dec}
      (E_\tau,E)}{E}  dE_{\nu_\tau} dE_\tau  \nonumber \\
    &&  + \Big({\rm Br}_e \Pi_e (E,x)+{\rm
      Br}_\mu\Pi_\mu (E,x)\Big) \int_E^\infty\int_{E_\tau}^\infty  
    {\rm Tr} \left[
      \bar\Pi_\tau(E_{\bar\nu_\tau},x)
      \bar\rho(E_{\bar\nu_\tau},x)\right]\nonumber\\
    && \hspace{2cm} \times \frac{1}{\bar\lambda^\tau_{\rm CC} ( E_{\bar\nu_\tau},x)}
    \pa{\bar N^{\tau}_{\rm CC} (E_{\bar\nu_\tau},E_\tau)}{E_\tau}
    \pa{\bar N^{\rm lep}_{\rm dec}
      (E_\tau,E)}{E}  dE_{\bar\nu_\tau}
    dE_\tau 
    \label{eq:Fterm}
  \end{eqnarray}
  \begin{eqnarray}
    \bar F\left[\rho,\bar\rho;E,x\right] &=& \sum_\alpha \bar\Pi_\alpha(E,x)  \int_E^\infty  
    {\rm Tr}\left[\bar\Pi_\alpha(E_{\bar\nu_\alpha},x) \bar\rho(E_{\bar\nu_\alpha},x) \right]
    \frac{1}{\bar\lambda^\alpha_{\rm NC}(E_{\bar\nu_\alpha},x)} \pa{\bar N^\alpha_{\rm
        NC}(E_{\bar\nu_\alpha},E)}{E} dE_{\bar\nu_\alpha}  \label{eq:intronubar} \nonumber\\
    &&  + \bar\Pi_\tau (E,x) \int_E^\infty\int_{E_\tau}^\infty  
    {\rm Tr} \left[ \bar\Pi_\tau(E_{\bar\nu_\tau},x)
      \bar\rho(E_{\bar\nu_\tau},x)\right] \nonumber\\
    && \hspace{2cm} \times \frac{1}{\bar\lambda^\tau_{\rm CC}(E_{\nu_\tau},x)}
    \pa{\bar N^{\tau}_{\rm CC} (E_{\bar\nu_\tau},E_\tau)}{E_\tau}
    \pa{\bar N^{\rm all}_{\rm dec}
      (E_\tau,E)}{E}  dE_{\bar\nu_\tau} dE_\tau  \nonumber \\
    &&  + \Big({\rm Br}_e \bar\Pi_e (E,x)+{\rm
      Br}_\mu\bar\Pi_\mu (E,x)\Big) \int_E^\infty\int_{E_\tau}^\infty  
    {\rm Tr} \left[
      \Pi_\tau(E_{\nu_\tau},x)
      \rho(E_{\nu_\tau},x)\right]\nonumber\\
    && \hspace{2cm} \times \frac{1}{\lambda^\tau_{\rm CC} ( E_{\nu_\tau},x)}
    \pa{N^{\tau}_{\rm CC} (E_{\nu_\tau},E_\tau)}{E_\tau}
    \pa{N^{\rm lep}_{\rm dec}
      (E_\tau,E)}{E}  dE_{\nu_\tau}
    dE_\tau \label{eq:antiFterm}\\ 
    && + \left(\sum_\alpha \bar\Pi_\alpha(E,x)\right) \int_E^\infty {\rm Tr}
    \left[\bar\Pi(E_{\bar\nu_e},x)\bar\rho(E_{\bar\nu_e},x)\right]
    \frac{1}{\bar\lambda_{\rm GR} ( E_{\bar\nu_e},x)}
    \pa{\bar N^e_{\rm GR} (E_{\bar\nu_e},E)}{E}
    d E_{\bar\nu_e}\nonumber
  \end{eqnarray}
\end{subequations}
where ${\rm Br}_\alpha$ is the $\tau$ branching ratio to $\nu_\alpha$,
\begin{subequations}
  \begin{eqnarray}
    \pa{N^{\alpha}_{\rm CC (NC)}
      (E_{\nu_\alpha},E_\alpha)}{E_\alpha}&=&\frac{1}{\sigma^\alpha_{\rm
        CC(NC)} (E_{\nu_\alpha})} \pa{\sigma^\alpha_{\rm
        CC(NC)} (E_{\nu_\alpha},E_\alpha)}{E_\alpha}, \\
    \pa{\bar N^{e}_{\rm GR}
      (E_{\bar\nu_e},E_e)}{E_e}&=&\frac{1}{\bar\sigma^e_{\rm
        GR} (E_{\bar\nu_e})} \pa{\bar\sigma^e_{\rm
        GR} (E_{\bar\nu_e},E_e)}{E_e},
  \end{eqnarray}
\end{subequations}
are the charged-current, neutral-current, and Glashow resonance
interaction. The $\tau$ decay distribution~\citep{Dutta:2000jv} in all modes and leptonic
modes are
\begin{eqnarray}
\pa{N^{\rm lep}_{\rm
    dec}(E_\tau,E)}{E}&=&\frac{1}{{\tilde\Gamma_{\rm lep}}^\tau(E_\tau)}
\pa{\tilde\Gamma_{\rm lep}^\tau(E_{\tau},E)}{E}, \\
\pa{N^{\rm all}_{\rm
    dec}(E_\tau,E)}{E}&=&\frac{1}{{\tilde\Gamma_{\rm all}}^\tau(E_\tau)}
\pa{\tilde\Gamma_{\rm all}^\tau(E_{\tau},E)}{E}, 
\end{eqnarray}
where $\tilde \Gamma^\tau(E_{\tau})=\frac{E_{\tau}}{m_\tau}\tau_\tau$,
in which $m_\tau$ is the $\tau$ mass and $\tau_\tau$ is the $\tau$
lifetime, ``all'' and ``lep'' indicate the all and leptonic $\tau$
decay modes, respectively.

The first term in Eqs.~\eqref{eq:Fterm} and \eqref{eq:antiFterm} accounts for
neutrino re-injection at lower energies coming from
Eqs.~\eqref{eq:gammarhoa} and \eqref{eq:gammarhob} due to neutral-current
interaction.
The second and third term is the injection due to the $\tau$ decay
into $\nu_\tau$ and into other flavors in the leptonic case: this
is known as tau-regeneration. It is important to note that the latter
terms couple the propagation of neutrinos and antineutrinos.
Finally, the last term in Eq.~\eqref{eq:antiFterm} accounts for the
neutrinos produced in the Glashow resonance due to $W^-$ decay.

These equations are implemented by discretizing the problem in bins that the user
specifies. The user should be aware that the discretization used can introduce errors due to
bin spill-over. Special care should be taken for hard spectra,
{\it e.g.} $\phi(E_\nu) = E^{-1}$, where this maybe more relevant and the user is encouraged
to check the effect of discretization by modifying the energy nodes spacing and cross sections' granularity.

\section{Features}
\label{sec:features}

\subsection{Standard oscillations in matter and vacuum}

{\ttf nuSQuIDS} is able to compute the neutrino oscillation probabilities in the presence of matter.
Neutrino oscillations can be computed for up to six neutrino flavors out of the box, though this can be extended to an arbitrary number of
neutrino flavors by producing additional tables using the provided {\ttf Mathematica} files.
The calculation allows setting both the mixing angles and the $CP$-violating phases.
Using the {\ttf Body} and {\ttf Track} objects, {\ttf nuSQuIDS} can compute the oscillation probability in any media provided by the user.
Default media, such as the Earth, the Sun, a constant density slab, and others are provided for immediate use.
In general, no constraints exist in the trajectory of the neutrino oscillation, though in the provided implementations, all neutrinos are assumed
to travel in straight lines, which should cover most applications.

Though some of this functionality is available in other neutrino oscillation
calculators~\citep{Huber:2007ji,prob3pp,Calland:2013vaa,Wallraff:2014vl}, the solution
implemented in {\ttf nuSQuIDS} is unique, and provides advantages over others.
By working in the neutrino interaction basis, the fast oscillations due the the vacuum part of the neutrino Hamiltonian are solved analytically.
This provides significant speed up in propagating neutrinos in dense environments, specially for slowly-changing or small matter densities.
An unusual feature of {\ttf nuSQuIDS} is that it solves the neutrino oscillation problem over a grid of energies;
though solving the problem (when considering only coherent effects) using an individual, fixed neutrino energy is also possible, as in other software.
In the multi-energy operating mode, {\ttf nuSQuIDS} can interpolate the interaction-picture solutions of the neutrino-evolution problem for energy values
that are in between the nodes where the calculation is performed.
This allows accurate and fast evaluation of the neutrino oscillation probability using a relatively coarse grid.
This is particularly important when evaluating the neutrino oscillation probability over a larger number of energies, which is often the case
when re-weighting large Monte Carlo data sets.

\subsection{Extendability to new physics scenarios}

{\ttf nuSQuIDS} implements the standard neutrino oscillation Hamiltonian with matter effects;
however, this can be extended to include additional terms that may appear in the Hamiltonian.
This can be achieved as illustrated in example~\ref{ssec:extphys}, where the user creates an inherited class of the main {\ttf nuSQuIDS} class and implements
the desired new physics operators.
In order to write the operators, the user needs to write in terms of the {\ttf SQuIDS} SU(N) vector type, which represents arbitrary hermitian matrices.
Additionally, {\ttf SQuIDS} implements operations between matrices and scalars that typically arise in writing these terms,
\textit{e.g.} matrix-scalar multiplication, commutator between matrices, \textit{etc}.
Once the new Hamiltonian is defined by the user no further change is needed. The user will be able to use all the functionality
of {\ttf nuSQuIDS} with the new physics.
Finally, {\ttf nuSQuIDS} allows for the serialization of user-defined values which allow for constants used by new physics operators to be serialized as
part of {\ttf nuSQuIDS} serialization.

\subsection{Neutrino transport in dense media}

{\ttf nuSQuIDS} implements non-coherent neutrino interactions that are relevant in very dense media or when neutrinos have high energies.
This functionality is comparable to other neutrino propagators used in neutrino telescopes~\cite{Safa:2021ghs,Garcia:2020jwr,Vincent:2017svp,Yoshida:2003js,shigeruyoshida_2020_4018117,Gazizov:2004va,DeYoung:865626,Alvarez-Muniz:2018owm,NuSpaceSim:2021hgs}.
At high energies, the neutrino interaction length is comparable to the size of the Earth, making the neutrino transport non-trivial.
{\ttf nuSQuIDS} takes into account charged-current and neutral-current interactions with nucleons through deep-inelastic scattering, which is the dominant interaction at the relevant energies~\citep{Formaggio:2012cpf,Gandhi:1998ri,Zhou:2019frk,Zhou:2019vxt,CooperSarkar:2011pa}.
Additionally, it takes into account neutrino-electron scattering, which is only relevant for resonant electron-antineutrino scattering at neutrino energies close to 6.4~PeV~\cite{Glashow:1960zz}.
A subdominant contribution from neutrino-photon scattering is not implemented~\cite{Seckel:1997kk,Alikhanov:2014uja,Alikhanov:2015kla,Zhou:2019frk,Zhou:2019vxt}, but can be easily extended in our framework; see~\cite{Garcia:2020jwr} for a publicly-available implementation of this cross section.
Finally, the production of neutrinos from lepton decay is neglected for all flavors, except for tau neutrinos.
In this case, we account for the contribution of neutrinos from tau decay in the approximation that the tau decays promptly, so it does not undergo significant energy losses.
This approximation is good below approximately 10~PeV and breaks down at higher energies; see~\cite{Safa:2021ghs,Garcia:2020jwr,NuSpaceSim:2021hgs} for recent work on implementing tau neutrino propagation at very high energies.

\subsection{Scenarios with neutrino oscillations and interactions}

{\ttf nuSQuIDS} is able to take into account simultaneously the presence of oscillations and non-coherent interactions in a consistent manner.
For neutrino transport in the Earth, and when standard neutrino oscillations are considered, this scenario is never an issue, since oscillations are important only below $\sim100$~GeV, while interactions appear above $\sim10$~TeV.
However, this is not the case for neutrino transport in larger astrophysical objects, \textit{e.g.} high-energy neutrinos produced in cosmic-ray collisions~\cite{Moskalenko:1991hm,Seckel:1991ffa,Ingelman:1996mj,Masip:2017gvw,Ng:2017aur,Edsjo:2017kjk,Arguelles:2017eao,IceCube:2021koo} or produced from dark matter decay or annihilation~\cite{Srednicki:1986vj,Kamionkowski:1991nj,Cirelli:2005gh,Blennow:2007tw,Barger:2007xf,Liu:2020ckq} in the Sun undergo oscillations and interactions simultaneously.
Additionally, this is not the case for non-standard neutrino oscillations, notably for light sterile neutrinos motivated by short-baseline anomalies, where flavor conversion can happen at high energies where interactions are relevant~\cite{Akhmedov:1988kd,Krastev:1989ix,Chizhov:1998ug,Chizhov:1999az,Akhmedov:1999va,Nunokawa:2003ep,Choubey:2007ji,Barger:2011rc,Esmaili:2012nz,Esmaili:2013vza,Lindner:2015iaa}.

\subsection{Handling of extended sources}

Neutrino emission does not happen as a physical point source, but rather there exist finite emission regions.
However, since the neutrino oscillation and interaction scales are much larger than the production region or detector size, in most scenarios, we can consider the neutrino emission to be point like to good approximation.
Nonetheless, there are scenarios where this approximation is not applicable.
For example, for neutrinos produced from dark matter decay through an intermediate mediator, a scenario known as secluded dark matter~\cite{Pospelov:2007mp}, the neutrinos are produced according to an exponential distribution that can be as large as the radius of the Earth or Sun depending on the model parameters~\cite{Niblaeus:2019gjk,Liu:2020ckq}.
This scenario can be implemented in nuSQuIDS~\cite{Liu:2020ckq} by introducing a source term that follows the exponential yield of neutrinos.
Another relevant scenario is that of production of neutrinos in pion or kaon decay in flight, \textit{e.g.} in accelerator neutrino experiments or in cosmic-ray air-showers.
Though for standard neutrino oscillation parameters the distribution of lengths induced by the varying decay position does not significantly impact the oscillation probability, this is not the case for new physics models that introduce small oscillations scales, \text{e.g.} when heavier neutrinos are introduced.
A similar scenario is that of experiments looking for fast oscillations close to cores of nuclear reactors where the size of the core is relevant.

\subsection{Efficient fast oscillations and fully delocalized neutrino emission}

Neutrino oscillations can vary extremely rapidly with neutrino energy, especially at small energies where the neutrino Hamiltonian in vacuum is large. 
It is numerically expensive to solve the neutrino evolution on a very dense grid or for enough energies to be able to produce a meaningful oscillation pattern.
{\ttf nuSQuIDS} introduces a novel solution to this problem by permitting the evaluation of fast oscillations in a way that the oscillations due to vacuum are tracked analytically.
The evaluation of oscillations on a grid of nodes that has been already computed has been significantly optimized making it possible to evaluate the oscillation probability for large Monte Carlo data sets much faster than computing the individual oscillation probabilities at each of the Monte Carlo event energies.

As noted in the previous point, the issue of fast oscillations and the localization of the neutrino emission is related. 
Neutrinos emitted by an extended source smear out the neutrino oscillation probability statistically.
{\ttf nuSQuIDS} provides an option to analytically perform the smearing of the oscillation probability, in which the user provides the size of the oscillation length that they which to smear.
This can be
understood as the leading order of the Magnus expansion~\cite{Magnus:1954zz}, and can
speed up numerical calculations dramatically in some concrete physics setups,
such as oscillations in the early universe~\cite{Hernandez:2016kel}. 
This calculation has no significant overhead and is much more efficient than performing integrals over the production region.
This feature is most useful in computing the oscillation expected from a nuclear reactor on a nearby detector, where one typically needs to smear over the reactor core size, or in computing the yield of atmospheric neutrino oscillations where one needs to smear over the production region in the atmosphere, which is several kilometers in size.

\subsection{Serializable input and output}

{\ttf nuSQuIDS} allows for the result of a calculation to be serialized into an \texttt{HDF5} file.
The serialization stores all the settings that were used to produce the {\ttf nuSQuIDS} object.
This can be advantageous in several scenarios.
For example, a calculation can be complex and lengthy, and the user may choose to perform it in steps.
In this scenario, the user can stop the calculation at a given point and then recover the {\ttf nuSQuIDS} object where it was left and continue.
Another scenario of interest is that related to not knowing at execution time which energies will need to be computed.
In this case, the user performs a calculation on a grid of energies and then stores the output of {\ttf nuSQuIDS} in a file.
Then, later, when the user wants to evaluate the result at specific energies, they can recover the object and perform the necessary evaluations.
This is also useful when the same calculation will be reused many times. 

\subsection{Techniques for high-performance Monte Carlo evaluation}
{\ttf nuSQuIDS} separates the calculation of a result flux from an input flux at the user's selected energy nodes from evaluating that result flux at arbitrary energies (provided those energies are within the span of energy nodes for which the propagation was performed). 
As discussed previously, this latter step evaluates vacuum oscillations analytically, and interpolates all other effects between the nearest energy nodes. 
As the final flux evaluation is often desired for very large numbers of energies (\textit{e.g.} for every event produced in a Monte-Carlo simulation), it has been optimized for speed, and furthermore to permit concurrent evaluation. 
While the majority of the classes and functions in {\ttf nuSQuIDS} are not safe to access concurrently from multiple threads of execution, such use is specifically supported for all for the flux evaluation functions. 
While not applicable in all cases, concurrent evaluation may provide considerable speed improvements for some users. 

\subsection{{\ttf C++} and {\ttf Python} interfaces}

{\ttf nuSQuIDS} has been written in {\ttf C++} since this compiled language provides the best execution efficiency. 
Extensive use is made internally of templating, inlining, and auto-vectorization to produce efficient machine code. 
In particular, the algebra of SU(N) operators significantly benefits from these types of optimization.
All {\ttf nuSQuIDS} functions can be accessed in {\ttf C++}; however, recently, there has been great interest in designing particle physics software in interpreted languages such as {\ttf Python}.
To facilitate the usage of {\ttf nuSQuIDS} with {\ttf Python}-based physics software, an interface to {\ttf Python} is implemented by means of the Boost library.
Most of the functionality of {\ttf nuSQuIDS} is available through the {\ttf Python} interface, with the notable exception of new body objects and new physics operators. 
These limitations arise because these types of extensions require user-provided functions which must be evaluated very frequently during the propagation calculation, which, if written in {\ttf Python}, would produce a significant amount of overhead.

\section{Usage Examples}
\label{sec:short_examples}
Provided along with the code are a set of working examples demonstrating its use for various purposes. 
Here, we discuss two illustrative use cases: computing neutrino oscillations for a range of energies simultaneously, and extending the base implementation in the library with new physics. 

\subsection{Multiple Energy Propagation}
In this section, we discuss an abbreviated version of the multiple energy propagation example found in the {\ttf examples/Multiple\_energy} subdirectory. 

Unit conversion factors and physical constants are provided by the supporting {\ttf SQuIDS} library's {\ttf Const} class. 
It is typical to allocate an instance of this class to make these conveniently available:

\begin{lstlisting}[frame=leftline, numbers = left,breaklines=true]
squids::Const units;
\end{lstlisting}

Next, we construct the {\ttf nuSQUIDS} object itself. 
For the multiple-energy mode, it requires an array of energies for which calculations will be carried out, which we choose here to be 200 values, spaced logarithmically from 1 GeV to 1,000 GeV. 
We choose the number of neutrino flavors to consider to be four: three active flavors and one sterile. 
These inputs are passed to the class constructor, along with the {\ttf neutrino} constant for the third argument to indicate that we will consider only neutrinos and not anti-neutrinos, and {\ttf false} for the fourth argument to disable the calculation of non-coherent interactions. 
\begin{lstlisting}[frame=leftline, numbers = left,breaklines=true,firstnumber=last]
auto energies = logspace(1.*units.GeV, 1.e4*units.GeV, 200);
unsigned int numneu = 4;
nuSQUIDS nus(energies, numneu, neutrino, false /*no interactions*/);
\end{lstlisting}

Next, we choose the environment (matter profile) through which we will propagate the neutrino flux. 
We choose here the Earth, with path passing through its full diameter, and apply these choices to the {\ttf nuSQUIDS} object:
\begin{lstlisting}[frame=leftline, numbers = left,breaklines=true,firstnumber=last]
double phi = units.pi; // zenith angle for trajectory
auto earth_atm = std::make_shared<EarthAtm>();
auto track_atm = std::make_shared<EarthAtm::Track>(earth_atm->MakeTrack(phi));
nus.Set_Body(earth_atm);
nus.Set_Track(track_atm);
\end{lstlisting}

A variety of other {\ttf Body} classes are provided with the library for media such as vacuum, constant density material, material with user-specified variable density, and the Sun. 
Additionally, users can create their own {\ttf Body} implementations for needs not covered by these. 
Each {\ttf Body} type has an associated {\ttf Track} class, parameterizing paths through it in whatever manner is most appropriate. 

The parameters of the neutrino mixing matrix can be fully customized. 
Here, we take a set of nominal values:

\begin{lstlisting}[frame=leftline, numbers = left,breaklines=true,firstnumber=last]
nus.Set_MixingAngle(0,1,0.563942);
nus.Set_MixingAngle(0,2,0.154085);
nus.Set_MixingAngle(1,2,0.785398);
nus.Set_SquareMassDifference(1,7.65e-05);
nus.Set_SquareMassDifference(2,0.00247);
nus.Set_SquareMassDifference(3,0.1);
nus.Set_MixingAngle(1,3,0.1);
\end{lstlisting}

Mixing angles are expressed in radians, and are labeled by the lower an upper states they connect, with zero-based indexing.
Squared mass differences are specified in eV$^2$ and are always with respect to the lowest-mass eigenstate, so only the upper state index must be specified. 
$CP$-violating phases can also be specified, again in radians and with two state indices. 
If unspecified, they default to zero. 

Next, the initial neutrino flux before propagation must be configured. 
Here, we will use a simple power-law spectrum, purely of the $\nu_\mu$ flavor (determined by the {\ttf k == 1} condition). 
The flux is specified as a two-dimensional array, indexed by neutrino flavor and energy:
\begin{lstlisting}[frame=leftline, numbers = left,breaklines=true,firstnumber=last]
marray<double,2> inistate{energies.size(), numneu};
double N0 = 1.0e18;
for(int i = 0 ; i < inistate.extent(0); i++){
  for(int k = 0; k < inistate.extent(1); k ++){
    inistate[i][k] = (k == 1) ? N0*pow(energies[i],-2) : 0.0;
  }
}
nus.Set_initial_state(inistate,flavor);
\end{lstlisting}
If we choose to work with both neutrinos and anti-neutrinos at the same time, this array would be rank three, with an additional, middle index taking on values zero for neutrinos and one for anti-neutrinos. 

A number of settings are exposed which govern how the ODE solving process operates. 
The most critical are the error tolerances, whose values should be chosen based on the precision needed and the difficulty of the propagation problem; see Sec~\ref{ssec:precision_parameters} for more detailed discussion. 
\begin{lstlisting}[frame=leftline, numbers = left,breaklines=true,firstnumber=last]
nus.Set_rel_error(1.0e-5);
nus.Set_abs_error(1.0e-5);
\end{lstlisting}

After all settings are configured, the propagation itself can be performed:
\begin{lstlisting}[frame=leftline, numbers = left,breaklines=true,firstnumber=last]
nus.EvolveState();
\end{lstlisting}

At this point, the propagated flux can be examined. 
For simplicity, we evaluate it for all three flavors over a grid of energies, and write these results to a file, from which they can be plotted, etc.:
\begin{lstlisting}[frame=leftline, numbers=left, breaklines=true, firstnumber=last]
std::ofstream file("fluxes_flavor.txt");
file << "# E flux-nu_e flux-nu_mu flux-nu_tau\n";
for(double E : logspace(1.*units.GeV, 1.e4*units.GeV, 1000)){
  file << E << " ";
  for(int fl=0; fl<numneu; fl++)
    file << " " <<  nus.EvalFlavor(fl, E);
  file << "\n";
}
\end{lstlisting}
It is worth noting that we can evaluate the final flux on a grid of energies much more dense that was used for the propagation. 
This takes advantage of the interpolation between propagated energies, which automatically includes vacuum oscillations, providing a more precise approximation than a physics-unaware interpolation. 
Fig.~\ref{fig:short_multimode} shows the oscillation probabilities calculated in this example.

\begin{figure}[h]
  \label{fig:short_multimode}
  \centering
  \includegraphics[width=0.7\textwidth]{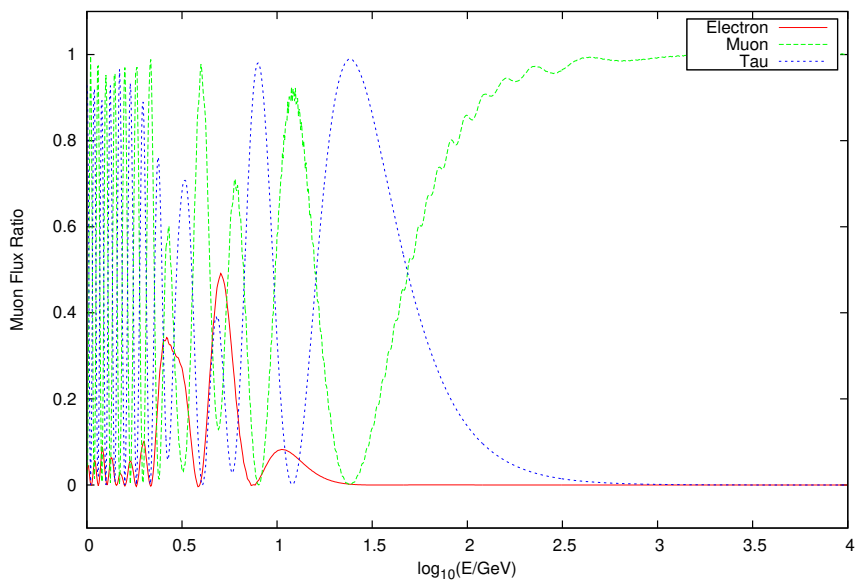} 
  \caption{Output for the multiple-energy mode with oscillations including a sterile neutrino (3+1)} 
\end{figure}

\subsection{Extended Physics}
\label{ssec:extphys}

Besides using the physics phenomena already implemented in {\ttf nuSQuIDS}, users can extend it with various forms of new physics. 
This is accomplished by writing a class derived from the provided {\ttf nuSQUIDS} class, and customizing one or more of the physics member functions used during evolution. 
These functions are {\ttf H0}--the time-independent portion of the Hamiltonian from Eq.~\ref{eq:hamiltonian}, {\ttf HI}--the time-dependent portion of the Hamiltonian from Eq.~\ref{eq:hamiltonian}, {\ttf GammaRho}--the function describing non-coherent attenuation of the neutrino state (combining $\Gamma$ and $\bar\Gamma$ from Eq.~\ref{eq:kinetic_equations}), {\ttf InteractionsRho}--the function describing additions to the neutrino state (combining $F$ and $\bar F$ from Eq.~\ref{eq:kinetic_equations}), {\ttf GammaScalar}--analogous to {\ttf GammaRho} for any scalar fields added to the system, {\ttf InteractionsScalar}--analogous to {\ttf InteractionsRho} for any scalar fields, and the helper {\ttf AddToPreDerive} which can be used for any time/position dependent bookkeeping in the implementation, although only a subset are likely to need customization for any particular physics scenario. 

As an example of this, we demonstrate implementing a non-standard interaction (NSI). 
This requires defining the interaction as an operator using the {\ttf squids::SU\_vector} type, and adding a contribution from it to the time-dependent Hamiltonian. 
First, we define the new class:
\begin{lstlisting}[frame=leftline, numbers = left,breaklines=true]
class nuSQUIDSNSI: public nuSQUIDS {
  private:
    double epsilon_mutau;
    squids::SU_vector NSI;
    std::vector<squids::SU_vector> NSI_evol;
    
    void AddToPreDerive(double x);
    squids::SU_vector HI(unsigned int ei,unsigned int index_rho) const;

  public:
    nuSQUIDSNSI(double epsilon_mutau, marray<double,1> Erange, 
        int numneu, NeutrinoType NT,
	bool iinteraction, double th01=0.563942, 
	double th02=0.154085, double th12=0.785398);
};
\end{lstlisting}
We add three member variables: {\ttf epsilon\_mutau}--the sole NSI parameter we will support, {\ttf NSI}--the operator we construct from it, and {\ttf NSI\_evol} to hold time-evolved copies of {\ttf NSI} as an optimization. 
We will add a new implementation of {\ttf HI} to update the Hamiltonian, as well as using {\ttf AddToPreDerive} to pre-calculate updates to {\ttf NSI\_evol}. 
A new constructor will take the value of the NSI parameter, as well as the standard oscillation angles, which influence the value of the NSI operator. 

\begin{lstlisting}[frame=leftline, numbers=left, breaklines=true, firstnumber=last]
nuSQUIDSNSI(double epsilon_mutau, marray<double,1> Erange, 
  int numneu, NeutrinoType NT, bool iinteraction,
  double th01, double th02, double th12)
  : nuSQUIDS(Erange,numneu,NT,iinteraction),
    epsilon_mutau(epsilon_mutau)
{
  assert(numneu == 3);
  // define the flavor structure as a complex matrix
  gsl_matrix_complex* M = gsl_matrix_complex_calloc(3,3);
  gsl_complex c {{ epsilon_mutau , 0.0 }};
  gsl_matrix_complex_set(M,2,1,c);
  gsl_matrix_complex_set(M,1,2,gsl_complex_conjugate(c));
  NSI = squids::SU_vector(M);
  gsl_matrix_complex_free(M);
  
  Set_MixingAngle(0,1,th01);
  Set_MixingAngle(0,2,th02);
  Set_MixingAngle(1,2,th12);
  // rotate to mass representation
  NSI.RotateToB1(params);

  NSI_evol.resize(ne);
  for(int ei = 0; ei < ne; ei++)
    NSI_evol[ei] = squids::SU_vector(nsun);
}
\end{lstlisting}
The constructor creates the NSI operator in the flavor basis from a matrix representation, then transforms it to the mass basis for later use. 
Note that setting the mixing angles defines the effect of the {\ttf RotateToB1} basis-transformation function. 
The {\ttf NSI\_evol} vector is filled with a set of correctly-sized operator vectors, but their values are not yet computed. 
Due to the simplicity of the example, only three-neutrino scenarios are supported, with other settings being rejected. 

The implementation of {\ttf AddToPreDerive} is responsible for computing the time-evolved operator values, which depend on both the time/position of the system--{\ttf x} and on the time-independent Hamiltonian for each energy, accessed from the base class' {\ttf H0\_array}:
\begin{lstlisting}[frame=leftline, numbers=left, breaklines=true, firstnumber=last]
void AddToPreDerive(double x){
  for(int ei = 0; ei < ne; ei++)
    NSI_evol[ei] = NSI.Evolve(H0_array[ei],(x-Get_t_initial()));
}
\end{lstlisting}
The implementation of {\ttf HI} can then make use of {\ttf NSI\_evol}, knowing that {\ttf AddToPreDerive} will be called to update it for each {\ttf x} before any calls to {\ttf HI} are made. 

The implementation of {\ttf HI} itself is concerned with adding the additional term to the Hamiltonian:
\begin{lstlisting}[frame=leftline, numbers=left, breaklines=true, firstnumber=last]
squids::SU_vector HI(unsigned int ei,unsigned int index_rho) const{
  static const double HI_prefactor=params.sqrt2*params.GF*params.Na*pow(params.cm,-3);
  double CC = HI_prefactor*current_density*current_ye;
  squids::SU_vector potential = (3.0*CC)*NSI_evol[ei];

  if ((index_rho == 0 and NT==both) or NT==neutrino){
      // neutrino potential
      return nuSQUIDS::HI(ei,index_rho) + potential;
  } else if ((index_rho == 1 and NT==both) or NT==antineutrino){
      // antineutrino potential
      return nuSQUIDS::HI(ei,index_rho) + -1*std::move(potential);
  }
  throw std::runtime_error("nuSQUIDS::HI : unknown particle or antiparticle");
}
\end{lstlisting}
A few details to note are that {\ttf current\_density} and {\ttf current\_ye} are member variables of the {\ttf nuSQUIDS} class which are used to obtain the matter properties (density and electron fraction) at the current position, the base {\ttf nuSQUIDS::HI} is invoked to obtain the original, standard model terms of the Hamiltonian, and that when the potential term must have the opposite sign for the anti-neutrino case, we apply the {\ttf std::move} function to it before performing the multiplication by $-1$ to indicate that the object's contents may be overwritten, eliminating the need for a temporary storage buffer. 

The full implementation of the class contains several other useful and convenient features, such as updating the serialization mechanism to save and restore the additional physics parameter; for the details we refer the reader to the file {\ttf examples/NSI/NSI.h} in the code. 
With the {\ttf nuSQUIDSNSI} class fully defined, its usage is analogous to the original {\ttf nuSQUIDS}, with the only major difference being the extra parameters for the constructor; setting the initial state, running the evolution, and reading out the final state are the same. 

While this example makes only a very simple addition to the physics of {\ttf nuSQuIDS}, there are a wide variety possibilities of equal of greater complexity, such as new non-coherent interactions and additional, coupled scalar fields. 

\section{Comparison to Other Software}
\label{sec:comparison}

A number of other software packages are available which perform related calculations. 
In this section, we summarize how a number of these differ from {\ttf nuSQuIDS}. 
Most neutrino propagation software can be divided into categories based on whether it was designed to handle neutrino oscillation physics or neutrino absorption physics. 
{\ttf nuSQuIDS} is unusual in this regard, as it is designed to calculate both types of effect simultaneously. 
Another important categorization is the calculation method used, which tends to have implications for the physics effects which can be readily calculated: 
Three prevalent methods, each with different strengths and weaknesses, are algebraic matrix diagonalization, ordinary differential equation (ODE) integration, and Monte-Carlo (MC) sampling. 
Matrix diagonalization is frequently used for coherent effects (\textit{e.g}. neutrino oscillation) in uniform media, for which the system of differential equations can be written down and solved analytically, as introduced in \cite{barger1980matter}. 
This has the advantage of taking a small, constant time to solve, but loses efficiency when variations in the propagation medium must be approximated by repeating the calculation in many small steps, and is not normally applicable to physics effects which couple fluxes besides those of different flavors at the same energy (such as energy losses due to neural-current interactions). 
ODE integration, the method used by {\ttf nuSQuIDS}, consists of writing down all physics effects in terms of their differential effects on neutrino fluxes and numerically solving the resulting system of equations. 
This provides a quite natural framework for treating variable media and arbitrary coupling between fluxes, but requires care in discretizing the problem (e.g. in neutrino energy), in structuring the problem numerically for handling by a standard ODE integration routine, and in selecting the parameters of the integration routine to get precise results rapidly. 
Monte Carlo sampling is in some sense the more literal method, consisting of simulating individual particles and randomly sampling the interactions they undergo. 
This sampling has both the benefit and disadvantage that it discovers not only the average behavior but also the full distribution of variations from it. 
This is indispensable when the full distribution is required, but has relatively slow convergence to precisely extract the mean behavior if that is all that is sought. 
Monte Carlo sampling can not only naturally treat variable media, but requires no discretization. 

\begin{table}
	\centering
	\begin{tabular}{l | c | c | c | C{5em}}
		\textbf{Package} & \textbf{Method} & \textbf{Oscillation} & \textbf{Absorption} & \textbf{Variable Density}\\
		\hline
    		GLoBES & AD & \checkmark & ~ & ~\\
    		Prob3++ & AD & \checkmark & ~ & ~\\
    		nuCraft & ODE & \checkmark & ~ & \checkmark\\
    		nuFATE & ODE & ~ & \checkmark & \checkmark\\
    		nuPropEarth & MC & ~ & \checkmark & \checkmark\\
    		TauRunner & MC & ~ & \checkmark & \checkmark\\
    		ANIS & MC & ~ & \checkmark & \checkmark\\
    		nuSQuIDS & ODE & \checkmark & \checkmark & \checkmark\\
	\end{tabular}
	\caption{Simple classification of neutrino evolution codes by calculation technique (analytic diagonalization--AD, differential equation integration--ODE, or Monte Carlo sampling--MC), whether they treat neutrino oscillation, absorption, or both, and whether they directly support propagation in variable density media.}
\end{table}

The General Long Baseline Experiment Simulator (GLoBES)~\cite{Huber:2007ji} is a toolkit to enable analyzing the capabilities of neutrino oscillation experiments, and combinations of experiments. 
Calculating neutrino flux evolutions is thus not the package's primary purpose, but is nonetheless one of its key components. 
GLoBES calculates only oscillation effects, via the analytic diagonalization method. 
Via its add-on system, the package can treat new physics effects, including sterile neutrinos, with the {\ttf snu} add-on. 

Prob3++~\cite{prob3pp} is designed specifically to calculate neutrino oscillation probabilities for the Super-Kamiokande collaboration. 
As such, it calculates only oscillation effects, and only for up to three flavors, although Lorentz invariance-violating oscillations are also supported for searches for new physics. 
As the package uses the analytic diagonalization method, only constant-density media are directly supported, although to support atmospheric neutrino measurements, the full Earth may be approximated as a set of spherically concentric shells of material, with a distinct density and electron fraction for each shell. 

nuCraft~\cite{Wallraff:2014vl} is a package created specifically to facilitate searches for sterile neutrino effects in the atmospheric neutrino flux, so it focuses on the computation of oscillation effects with support for four or more neutrino flavors, and has a continuous treatment of the density profile of the Earth. 
Unlike the other pure oscillation calculators, nuCraft is an ODE integration-based system, which makes inclusion of the variable material density convenient. 

Turning to packages designed for higher energies with a focus on neutrino absorption, the All Neutrino Interaction Simulation (ANIS)~\cite{ANIS} is intended for use as an event generator for neutrino observatories. 
For this reason it uses a Monte-Carlo sampling approach, and not only simulates neutrinos passing through the bulk of the Earth, with a continuously-parameterized density profile, but also includes forcing final interactions in a volume defined as the vicinity of a detector. (This volume is typically larger than the instrumented volume of the detector in order to fully sample muons produced by charge-current $\nu_\mu$ interactions entering the detector from outside.)
While ANIS treats all deep-inelastic scattering (DIS) and Glashow resonance (resonant $W^\pm$ production) processes as well as tau propagation and decay, it does not calculate oscillations. 
The code requires DIS cross sections in the somewhat unusual form of Bjorken x, y pairs, pre-sampled at natural frequencies from the desired cross section. 
While this enables the propagation code to run rapidly and without direct dependencies on other software, the user must beware that systematic biases will result if sufficiently large tables are not used. 

nuPropEarth~\cite{nuPropEarth2020} is a newer package which is similar to ANIS, but adds treatment of a number of subdominant interaction processes, such as nuclear effects. 
It also uses a Monte-Carlo sampling method, and does not treat oscillations.
Choosing different tradeoffs than ANIS, nuPropEarth links directly to the GENIE/HEDIS, TAUOLA++, and TAUSIC packages to enable the use of recent and highly detailed cross sections treatments, but with the disadvantage of a considerable set of dependencies required to build the package.

TauRunner~\cite{Safa:2021ghs} is a new {\ttf Python} package that provides similar functionality to nuPropEarth.
It has the advantage that it has less dependence on other software, and allows for arbitrary media and trajectories.
It also uses a Monte-Carlo sampling method, and does not treat oscillations.
It can use several built-in cross-section tables that can be extended by the user.
These tables only account for deep-inelastic scattering.
More importantly, TauRunner properly accounts for the tau energy losses at very high energies, where the on-spot decay approximation used in {\ttf nuSQuIDS} breaks down.

Finally, nuFATE~\cite{nuFATE2017} is a tool for calculation of neutrino flux attenuations at high energies, which uses a particularly efficient differential equation solution method which avoids the need for using a general ODE-integration routine by applying an integrating factor transformation, so that only the material density profile must be integrated, rather than the coupled system of equations. 
While this mechanism does support deep inelastic scattering, the Glashow resonance, and tau-regeneration all in a continuously variable medium, in its current form the package does not support distinct cross sections for different target species, either at the level of individual nucleon types, or different nuclei. 

\section{Performance and Precision}
\label{sec:performance} 

In the tests described in this section, Machine 1 is configured with an Intel Core\texttrademark~i7-6700K CPU, with a `base' frequency of 4.0 GHz. All tests were performed with the `Turbo Boost' feature disabled, preventing frequency scaling. 

The time evolution and basis transformations performed on \lstinline{squids::SU_vector} objects rely heavily on trigonometric functions, specifically sines and cosines, mostly in matched pairs. 
These operations are in turn used by several of the key portions of {\ttf nuSQuIDs}, and for oscillation-only calculations evaluation of trigonometric functions can be the single largest part of time consumed by the library. 
As a result, runtime is significantly dependent on the performance of the underlying math library, which is usually supplied by the operating system, or sometimes the compiler. 
The differences are illustrated in Table~\ref{tab:trig_perf}, which shows measurements performed on Machine 1 with several combinations of operating system and compiler of the cost of computing a single sine function or the combination of sine and cosine for the same argument (which should ideally be performed by an efficient, fused calculation). 
A measurement was also made of the time to run the pseudo-random number generator (PRNG) by itself. 
The time spent for the PRNG varies somewhat, assumedly due to different optimization choices by the various compilers, but amounts to a small fraction of the time for the trigonometric calculation. 
Broadly, it is apparent that the calculations using the math implementations from the older version of the GNU C library (glibc) are significantly slower than the versions using other math libraries. 
The (approximately) factor of two difference in fused sine/cosine performance between glibc 2.17 and 2.28 translates into a 10-25\% difference in the overall runtime of {\ttf nuSQuIDs} oscillation calculations. 
The user is therefore cautioned that the choice of compiler and runtime libraries can be important to maximizing the performance of {\ttf nuSQuIDs}. 
The benchmarks shown in this section will use the GNU compiler, version 8.3.1 (Red Hat 8.3.1-5) on CentOS 8, as this is a combination with good performance which is likely to be relevant to many users. 

\begin{table*}
	\begin{tabular}{ llcccl }
		OS & Compiler & PRNG & Sine & Sine \& Cosine & Math Library \\
		CentOS 7 & gcc 4.8.5 & 6.0 ns & 40.5 ns & 63.5 ns & GNU C Library  2.17 \\
		CentOS 7 & gcc 7.3.1 & 2.7 ns & 37.1 ns & 63.3 ns & GNU C Library  2.17 \\
		CentOS 7 & clang 5.0.1 & 4.8 ns & 96.4 ns & 69.0 ns & GNU C Library  2.17 \\
		CentOS 7 & icc 18.0.3 & 4.0 ns & 20.9 ns & 22.2 ns & libimf 2018.3.222\\
		CentOS 8 & gcc 8.3.1 & 2.7 ns & 26.8 ns & 31.3 ns & GNU C Library  2.28\\
		Darwin 16.7.0 & clang 5.0.0 & 2.9 ns & 19.1 ns & 20.1 ns & libsystem\_m  3121.6.0 \\
		FreeBSD 11.2 & clang 6.0.1 & 2.9 ns & 22.3 ns & 30.8 ns & BSD libc
	\end{tabular}
	\caption{
		Measured time to compute trigonometric functions using different compilers and operating systems on Machine 1. In each case, the same stream of $10^8$ pseudo-random arguments in the domain [-100,100] was used, and the average time per call was computed, including the time necessary for the PRNG to compute the next argument value. 
	}
	\label{tab:trig_perf}
\end{table*}

\subsection{Parameters Affecting Precision}
\label{ssec:precision_parameters} 

The primary quantity defining the size of the problem solved by {\ttf nuSQuIDS} in its multiple energy mode is the number of energy nodes included in the calculation. 
It is important to note that the interpolation provided by {\ttf nuSQuIDS} allows arbitrary energies within the range of the energy nodes to be evaluated after propagating, so it is not necessary to calculate a node for every energy at which the final flux may be needed. 
Rather, the minimum number of nodes should be chosen so that the calculation is suitably precise at each node and when interpolated between nodes. 

It is difficult to give specific guidance for all of the many types of problems to which {\ttf nuSQuIDs} is applied, so users are advised to be prepared to perform their own tests of how many nodes are needed to get good results for their use cases. 
However, it can be noted that for most problems the authors have attempted to compute numbers of nodes in the range 100-300 have typically been found sufficient. 
It is also important that, while it is often convenient to place nodes uniformly in energy or the logarithm of energy, this is by no means required by {\ttf nuSQuIDS}, so it can be advantageous to place nodes more densely in regions of particular interest or where sharp spectral features are expected, such as the vicinity of the Glashow resonance. 

A second key parameter is the requested error tolerance for the ODE integration, which effectively controls the number of steps the integrator must take, while the number of energy nodes causes the amount of work required to compute each step to vary. 
As with the number of nodes, it is impossible to give one-size-fits-all advice on selecting the tolerance. 
In general, problems which include greater variation in the magnitudes of the flux values at different energies (e.g. steep power law fluxes), or for different flavors, require more demanding tolerances in order for the small components to be computed precisely in the presence of the larger components. 
Similar to selecting the number of energy nodes, the necessary tolerance may be found by incrementally tightening it until subsequent calculations produce results which are equivalent to within the desired precision. 
While this process is manual and somewhat annoying, having been performed for one problem, the result can usually be applied to modified versions of the problem without retesting, as long as the changes are not drastic. 

\begin{lstlisting}[float=ht,frame=lines,label=lst:prec_comp,caption={A simple routine for comparing the precision of two different calculations with  {\ttf nuSQuIDS}. Error checking, that numbers of energy ranges, number of flavors, etc. are compatible is advisable, but omitted here for brevity.}]
void compare_fluxes(const nuSQUIDS& nus1, const nuSQUIDS& nus2, 
                    unsigned int oversampling, double tolerance){
  auto er1 = nus1.GetERange(), er2 = nus2.GetERange();
  marray<double,1> samples = er1;
  if(oversampling>1)
    samples=logspace(er1.front(), er1.back(), 
                     er1.size()*oversampling - 1);
  double GeV=squids::Const().GeV;
  double maxErr=0;
  for(double e : samples){
    for(unsigned int fl=0; fl<nus1.GetNumNeu(); fl++){
      for(unsigned int rho=0; rho<nus1.GetNumRho(); rho++){
        double val1 = nus1.EvalFlavor(fl,e,rho);
        double val2 = nus2.EvalFlavor(fl,e,rho);
        double err = std::abs(val2-val1)/val1;
        if(err > tolerance)
          std::cout << "Disagreement: " << err << " for energy " 
            << e/GeV << " GeV flavor " << fl << " rho " << rho
            << " (" << val1 << " vs. " << val2 << ")\n";
        if(err > maxErr)
          maxErr = err;
      }
    }
  }
  if(maxErr > tolerance)
    std::cout << "Maximum disagreement: " << maxErr << std::endl;
}
\end{lstlisting}

A simple comparison function for guiding these selections is shown in Listing~\ref{lst:prec_comp}. 
Two \lstinline{nuSQUIDS} objects are compared over the full energy range for which the first was calculated, with some oversampling factor which may be greater than 1 to check results for energies using interpolation, as well as at the energy nodes of the first object. 
Any evaluation point at which the relative difference of the second \lstinline{nuSQUIDS} object's result from that of the first is greater than the specified tolerance is reported, and the greatest such discrepancy is reported at the end. 
This generic routine may not be suitable in all situations, but gives a basis for users to begin assessing the parameter choices needed for their problems. 
It was used to determine integration error tolerances and in relevant cases node counts for the test shown in the rest of this section. 

\subsection{Propagation Time Scaling with Problem Size}
\label{ssec:propagation_scaling}

\begin{figure}[t]
  \centering
  \includegraphics[width=0.9\textwidth]{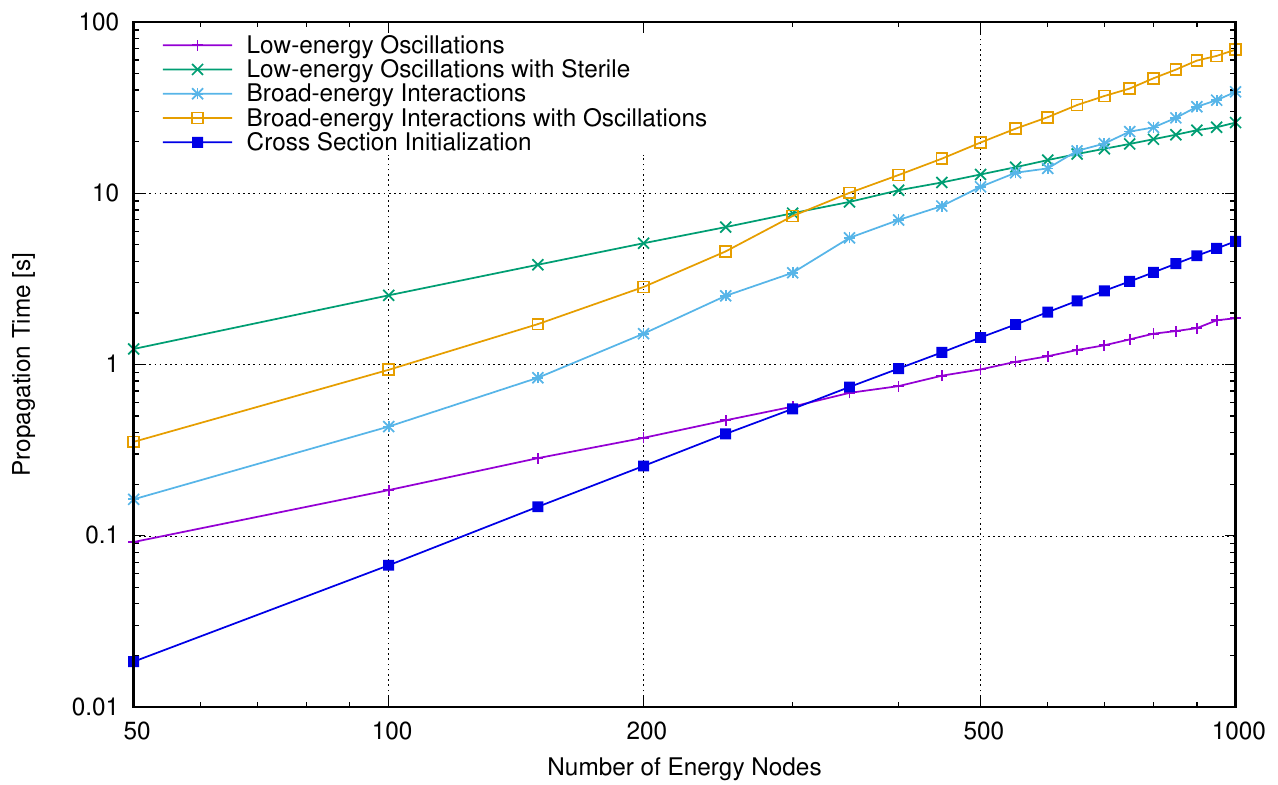}
  \caption{Propagation time scaling with problem size on Machine 1. Only the time to run \lstinline{EvolveState} is included, omitting cross-section preparation time for the cases with interactions. This initialization can be separated by calling \lstinline{InitializeInteractions} manually, and is shown as `Cross Section Initialization'. Propagation with only oscillation effects shows scaling which is very close to linear in the number of energies, while propagation with interactions takes time which is approximately quadratic in the number of energies.}
  \label{fig:propagation_scaling}
\end{figure}

Several tests were performed to demonstrate how the time to perform multi-energy propagations scales with the number of energy nodes included, depending on the use of some of the library's major features. 
The results of these tests are shown in Fig.~\ref{fig:propagation_scaling}. 

The `Low-energy Oscillations' tests used a flat input spectrum  over an energy range of 1~GeV - 10~GeV, initially consisting of $\nu_\mu$ only. 
For these tests, non-coherent interactions were disabled. 
The `Broad-energy Interactions' test used an $E^{-2}$ input spectrum over 10~GeV - $10^7$~GeV with an initially equal mixture of flavors. 
All tests propagated the input flux through the full diameter of the Earth, using {\ttf nuSQuIDS} version of PREM for the density. 
Each propagation was tested for various sets of energy nodes, distributed uniformly in the logarithm of energy, from 50 to 1000 nodes, in increments of 50. 
Each measurement was repeated five times, and the median duration reported, unless the standard deviation of the distribution was greater than 1\% of the mean, which was taken to indicate an unstable set of measurements, in which case the entire set of measurements was repeated. 

The oscillation-only propagations are expected to scale linearly with the number of nodes, as the calculation performed for each integration step consists primarily of evolving the flavor projectors (needed to compute $H_1(E,x)$ and $H^*_1(E,x)$) and commutator forming the first terms on the right hand sides of Eqs.~\ref{eq:kinetic_equations}, which must be repeated for each energy node, neutrino flavor, and particle type (neutrino and anti-neutrino).
This is borne out quite well by the measurements. 

The oscillation calculation including a fourth, sterile neutrino flavor (`Low-energy Oscillations with Sterile') is significantly slower than the standard three-flavor calculation (`Low-energy Oscillations'), as is to be expected. 
The representation of the four-flavor state for each component has sixteen components to the three-flavor state's nine, so a decrease in speed of at least $16/9 = 1.77$ times may be expected, to start with. 
The addition of the sterile flavor also forces the ODE integrator to use a smaller step size for the integration, taking nearly eight times as many steps to complete the propagation. 
The combination of these effects seems to plausibly account for the observed decrease in speed. 

The time for propagations, including non-coherent interactions (`Broad-energy Interactions'), can be expected to scale quadratically in the number of energy nodes, due to the need to calculate the cascading of particles at each node down to all lower energy nodes. 
In addition, calculating the interactions requires an additional initialization step, in which the cross section data governing the cascading processes is gathered. 
Since this extra step is needed only once for a \lstinline{nuSQUIDS} object to be able to perform multiple propagations with the same set of energy nodes, which is often relevant as various hypothesis input fluxes are tested, it is accounted separately in Fig.~\ref{fig:propagation_scaling} as `Cross Section Initialization'. 
Both the cross section initialization and the propagations with interactions show approximately the expected quadratic scaling, and the calculation of both interactions and oscillations together (`Broad-energy Interactions with Oscillations') exhibits the additional cost of including both types of effects. 
A brief departure from the expected power law form of the propagation time scaling is observed at intermediate numbers of energy nodes is also observed for the tests with interactions; in Fig.~\ref{fig:propagation_scaling}, this appears approximately between 200 and 300 energy nodes. 
This is believed to be a result of the calculation's memory use growing comparable in size to the CPU's last-level cache (8 MB, for Machine 1). 
It appears reliably, but moves to higher node counts for CPUs with larger cache sizes. 
See Sec.~\ref{ssec:memory} for a discussion of memory use and some representative numbers which support this hypothesis. 

\subsection{Flux Evaluation Performance}

\begin{figure}[t]
  \centering
  \includegraphics[width=0.9\textwidth]{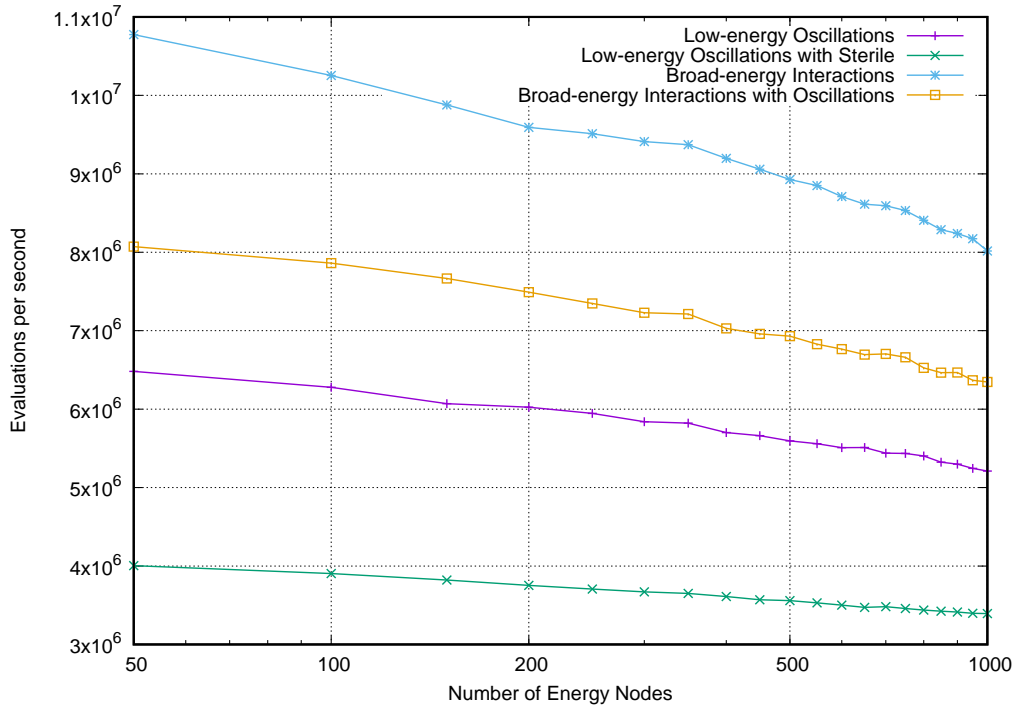}
  \caption{Flux evaluation speed scaling with problem size on Machine 1. As expected, only a weak dependence on the problem size is observed. }
  \label{fig:evaluation_scaling}
\end{figure}

The purpose of propagating a flux with {\ttf nuSQuIDS} is to determine the spectrum or composition after propagation, so the efficiency with which this information can be extracted is the second critical half of the library's performance. 

Evaluation of the propagated flux for a given energy, flavor, particle type combination consists of three parts: Mapping the physical energy to a logical position within the energy nodes of the propagated state, interpolating the state components from the nearest nodes to form the requested state, and application of the suitably evolved $H_0$ operator to that state. 
Locating the nearest energy nodes for interpolation depends on the total number of such nodes, due to the allowance for arbitrary node distributions, and is performed using a binary search. 
The interpolation of the state, computation and evolution of the $H_0$ operator, and application of the operator to the interpolated state do not depend on the number of nodes, but do depend on the size of each state component, determined by the number of neutrino flavors considered. 

Fig.~\ref{fig:evaluation_scaling} shows how the rate of flux evaluations scales with the number of energy nodes treated for the four example problems introduced in Sec.~\ref{ssec:propagation_scaling}. 
As expected, due to the binary search for adjacent nodes, a weak dependence on number of nodes is observed. 
A marked drop in performance also occurs between the three-flavor and four-flavor oscillation calculations, as expected from the corresponding increase in the complexity of both the states and operators. 

\subsubsection{Concurrent Evaluation}

\begin{figure}[t]
  \centering
  \includegraphics[width=0.9\textwidth]{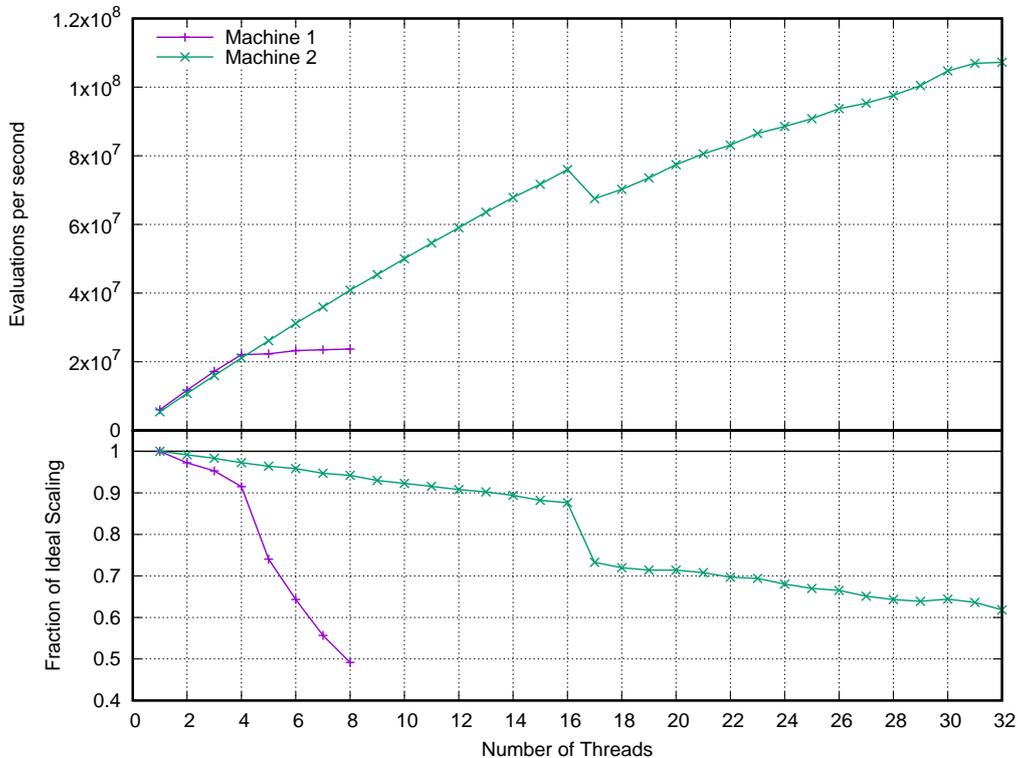}
  \caption{Flux evaluation speed scaling with number of threads. Fairly good linearity is observed on both machines until the number of physical CPU cores is exhausted, at 4 and 16 threads, respectively. Both processors tested support 2-way simultaneous multithreading, but using it for this task is not productive on Machine 1 as shown by the lack of significant increase in total throughput when the number of threads is increased from 4 to 8, while Machine 2 initially suffers a penalty (the drop in throughput from 16 to 17 threads), but recovers and shows increased throughput as all logical CPU cores are used. }
  \label{fig:evaluation_threads}
\end{figure}

The \lstinline{nuSQUIDS} object is not generally concurrency-safe for modification, but concurrent use of non-modifying member functions, including \lstinline{EvalFlavor}, is safe and supported. 
This is potentially useful when computing large numbers of flux evaluations for tasks such as weighting large numbers of simulated neutrino interactions in a detector to form an expected observation. 
Fig.~\ref{fig:evaluation_threads} shows an example of the application of this technique to evaluating the flux computed by the `Low Energy Oscillations' calculation introduced previously. 
A problem size of 200 energy nodes was tested with a variable number of worker threads each being assigned to perform $10^7$ flux evaluations. 
Due to the limited number of processor cores available to Machine 1, a second hardware configuration was introduced for this test. 
Machine 2 used an AMD Ryzen\texttrademark~9 3950X CPU, having a `base' operating frequency of 3.5 GHz, with frequency scaling disabled. 

Scaling to many flux evaluation threads appears to work well until the number of physical CPU cores is saturated, in spite of memory allocations which are required by the internals of the calculation. 
An important optimization to enable this result was thread-local reuse of memory buffers, avoiding the need to serialize execution between threads even when the system memory allocator might use globally shared state. 
For numbers of evaluation threads beyond the available number of physical processor cores, simultaneous multithreading (SMT) is used, and produces mixed results which depend significantly on the CPU architecture. 
SMT does not enable a useful increase in throughput for Machine 1, while on Machine 2, additional SMT threads provide less marginal benefit than full physical cores, but aside from a small loss in total throughput when a small number of physical cores are running multiple SMT threads, throughput generally benefits. 
As this is not a thorough survey of available CPU models, users are advised to perform their own checks to determine whether use of SMT will be productive on their hardware. 

\subsection{Memory Requirements}
\label{ssec:memory}

For a {\ttf nuSQuIDS} object configured with $M$ energy nodes, $N$ neutrino flavors, $P$ particle types, and $T$ target species for interactions (if non-coherent interactions are enabled), neglecting constant terms, the amount of memory used may be broken down as follows. 
Essentially all data allocated by the {\ttf nuSQuIDS} object is either in the form of machine pointers or double-precision floating-point numbers, assuming that the non-pointer components of \lstinline{squids::SU_vector} are small enough that padding and alignment requirements fit them into space equivalent to two additional pointers, making the direct size of an \lstinline{SU_vector} three pointers. 
The underlying {\ttf squids::SQuIDS} object then requires:

\begin{subequations}
\label{eq:squids_memory}
\begin{align}
(PN^2 + 1)M &~\textrm{doubles}\\
9(P+1)M &~\textrm{pointers}
\end{align}
\end{subequations}

This omits some amount of memory which the GSL ODE integration library will allocate for its own internal state. 
The size of this state will be proportional to $PN^2M$ doubles, with the constant of proportionality typically being around 10 (e.g. it is 12 when using the {\ttf nuSQuIDS} default Runge-Kutta-Fehlberg 4-5 stepper, RKF45). 

Without structures needed for non-coherent interactions, the {\ttf nuSQuIDS} object additionally allocates:

\begin{subequations}
\label{eq:nusquids_memory}
\begin{align}
(2PN^3 + N^2 + 6PN + 5)M + (P+1)N^3 + 3(P+1)N &~\textrm{doubles}\\
3(2PN + 1)M + 3(P+1)N &~\textrm{pointers}
\end{align}
\end{subequations}

The data structures for computing non-coherent interactions (neglecting any memory allocated by the cross section and $\tau$ decay objects themselves) require:

\begin{subequations}
\label{eq:intereaction_memory}
\begin{align}
(2(NT+1)P + 1)M^2 + (PN^2 + (2T + 3)PN + 12)M &~\textrm{doubles}\\
3PM &~\textrm{pointers}
\end{align}
\end{subequations}

If the external neutrino source feature is used, it requires a buffer of another $PNM$ doubles. 

It should be noted that $P$ can only take on the values 1 and 2, $T$ is typically 2, and $N$ is generally 3 or 4. 

As an example, a modestly-sized three-flavor neutrino oscillation calculation might use 100 energy nodes and not consider non-coherent interactions, so $M=100$, $N=3$, $P=2$ (if both neutrinos and anti-neutrinos are considered), and $T$ is irrelevant. 
Assuming typical data sizes of 8 bytes for both pointers and double-precision floating-point numbers, such a calculation will then require $24435 * 8$ bytes or approximately 191 kilobytes of memory, plus approximately another 170 kilobytes allocated by GSL (for the RKF45 stepper). 

Due to the quadratic scaling with number of energy nodes (of the differential charged- and neutral-current interaction cross sections and $\tau$ decay spectra), memory requirements are much higher for calculations with non-coherent interactions. 
An interaction calculation with 100 energy nodes and all interaction effects enabled would have $M=100$, $N=3$, $P=2$, and $T=2$, leading to a memory requirement of $322235 * 8$ bytes, or approximately 2.5 megabytes, and increasing to 300 nodes ($M=300$) would require $2706435 * 8$ bytes, or approximately 20.6 megabytes (with the size of the GSL state also increasing to 506 kilobytes). 
Use of the external neutrino source feature, even with this larger problem size, would require only an additional $1800 * 8$ bytes or 14 kilobytes. 

\subsection{Efficient Use}
\label{ssec:efficient_use}

Besides configuring the number of energy nodes and the error tolerance to the lowest and loosest values which will provide the desired precision, there are several other considerations for using {\ttf nuSQuIDS} with the best possible efficiency. 

One simple trick is to turn off features which are not needed for a given calculation. 
If neutrino oscillations are not relevant in the regime covered by a given propagation problem, they can be disabled, and the same is true for non-coherent interactions. 
With oscillations disabled, {\ttf nuSQuIDS} is able to skip computing the evolution of the flavor projection operators, and with interactions disabled it can omit obtaining interaction cross sections, updating interaction lengths for the matter density at each step, and computing the contribution to each energy node from the neutrinos cascading down from higher energy nodes. 
Additionally, some particular types of interactions can be enabled or disabled as needed, in particular, tau regeneration and Glashow resonance effects. 
For example, to propagate a high-energy atmospheric neutrino flux, not only are oscillations potentially unnecessary, but the flavor composition of the flux may mean the tau regeneration is irrelevant as well. 
If a flux is being calculated which involves no anti-neutrinos, the effects of the Glashow resonance will not appear, and they may be turned off without changing the result. 
In general, {\ttf nuSQuIDS} attempts to detect possibilities at runtime to minimize work performed, and this is fairly successful in the case of tau regeneration, so turning this feature off may not produce much change in computation time. 
Handling of the Glashow resonance, though, requires additional interaction length updates which the existing structure of the code is unable to optimize away even when it turns out to be unused, so turning this capability off when it is not needed can have a useful impact. 

The cross-section data required to compute non-coherent interactions is dependent only on the placement of the energy nodes, not on the flux being propagated. 
As a result, if many fluxes are to be calculated over the same set of nodes, it is most efficient to reuse the same \lstinline{nuSQUIDS} object, merely resetting the initial flux with \lstinline{Set_initial_state} before calling \lstinline{EvolveState} again, so that this initialization is not repeated. 
As shown in \ref{fig:propagation_scaling} the initialization can take 5-15\% of the time for the propagation, so this can represent an important savings over many propagations. 

\section{Conclusions}
\label{sec:conclu}

\nuSQUIDS~  provides a complete solution to neutrino propagation while accounting for neutrino oscillations and scattering. 
We find that it is competitive with dedicated neutrino oscillation calculators, when the number of evaluations of the oscillation probability
is large, which is the typical case of interest in neutrino experiments.
The library also has a {\ttf Python} extension. Finally, the \nuSQUIDS~ library has been designed to be easily extendable 
to both new media and new physics scenarios. User contributions to the project are encouraged and a number of additional
models have already been implemented and are linked in {\ttf resources/user\_contributions}.

\section*{Acknowledgements}

We thank: Alexander Tretting for improving the fast oscillations treatment;
Jakob Van Santen for implementing the Glashow resonance cross sections;
Benjamin Jones for early adoption of this code onto the sterile analysis;
Melanie Day for early adoption of this code onto her analysis, carefully checking the non-standard interaction extension, and providing feedback;
Tianlu Yuan for comments on neutrino cross section utilities and testing the performance of the code in realistic analysis scenarios;
Subir Sarkar for providing the code to generate neutrino-nucleon cross section tables;
Kotoyo Hoshima and Aaron Vincent for dedicated comparisons and checks of this code;
Zander Moss for performing an independent check of several core features of the code;
Felix Kallenborn for carefully reading through the code, writing a GPU version of the algorithm, and pointing out an improve handling of the GSL ODE solver;
Jos\'e Carpio and Zander Moss for pointing out the need to improve the Earth density interpolation;
Mauricio Bustamante for testing the Python V3 support;
Austin Schneider for reading through this draft and testing the Ubuntu support;
Ibrahim Safa for reading through this draft;
Alejandro Diaz for reading through this draft;
Tom Studdard for providing feedback and contribution of neutrino decoherence;
Jos\'e Carpio, Alberto Gago, and Eduardo Massoni for providing feedback and implementing nuSQuIDS in GLoBES;
Zander Moss, Marjon Moulai, and Janet Conrad for providing feedback and contribution of neutrino decay extension;
Shivesh Mandalia and Teppei Katori for checking, providing feedback, and contribution on the Lorentz Violation model implementation;
Joshua Highnight and Jessie Micallef for providing a modified version of Prob3++ with NSI for detailed comparisons; and
Thomas Ehrhardt for performing detailed comparison between this code speed and accuracy versus Prob3++ CPU and GPU version.
Pedro Machado, Aaron Vincent, and Logan Wille for providing the Solar model files and useful discussion in the construction of the SolarTransition extension.
CAA is supported by the Faculty of Arts and Sciences of Harvard University, the Alfred P. Sloan Foundation, and was supported by U.S. National Science Foundation (NSF) grant PHY-1912764 through this work.
JS is supported by  European ITN project H2020-MSCA-ITN-2019/860881-HIDDeN, the Spanish grants FPA2016-76005-C2-1-P, 
PID2019-108122GB-C32, PID2019-105614GB-C21.

\bibliographystyle{elsarticle-num}
\ifdefined\forjournal
\bibliography{nusquids_journal}
\else
\bibliography{nusquids}
\fi % forjournal
\fi % manualonly

\ifdefined\forjournal
% no appendices
\else
\ifdefined\manualonly
% let appendices be main sections
\else
\appendix
\fi % manualonly

\section{Examples}
\label{sec:examples}
To illustrate use of the library in concrete scenarios we provide
a set of examples. These examples are located in different sub-folders inside a folder
called {\ttf examples}. A specific example can be compiled using the
makefile by running {\ttf make examples example\_name}, and omitting the
name compiles all the examples. 
Some of the examples contain a {\ttf Gnuplot} script to plot the output file.

\subsection{Single energy \textnormal{({\ttf examples/Single\_energy})}}
\label{sec:single}
This example illustrates the use of the simplified mode to compute the
propagation of the neutrinos for a single energy. 

In the following we will go though the main file of the example.
First, we construct the nuSQuIDS object using the signature that
requires only the number of flavors and specification of the use of
neutrinos or antineutrinos.

\begin{lstlisting}[frame=leftline, numbers = left,breaklines=true, label = ex:sin1]
  nuSQUIDS nus(3,neutrino);
\end{lstlisting}

We next set the value of the oscillation parameters, giving the values for the mixing angles in radians, the value for the mass
square difference in electron-volt squared, and a value for the $CP$
phase in radians.
In order to do so we use the {\ttf Set\_MixingAngle} member function
whose first two arguments are the indices of the rotation, starting
from zero. The $CP$ phase function is similar.
In the case of the square mass difference, the first argument of the function is
the mass eigenvalue index and the second argument the value. These
square mass differences are always with respect to the lightest mass state (whose index is zero).
 
\begin{lstlisting}[frame=leftline, numbers = left,breaklines=true, label = ex:sin1,firstnumber=last]
  nus.Set_MixingAngle(0,1,0.563942);
  nus.Set_MixingAngle(0,2,0.154085);
  nus.Set_MixingAngle(1,2,0.785398);

  nus.Set_SquareMassDifference(1,7.65e-05);
  nus.Set_SquareMassDifference(2,0.00247);

  nus.Set_CPPhase(0,2,0.0);
\end{lstlisting}

We declare a structure that contains all the physical units and
constants we will need. 
\begin{lstlisting}[frame=leftline, numbers = left,breaklines=true, label = ex:sin1,firstnumber=last]
  squids::Const units;
\end{lstlisting}

This line sets the energy of the neutrino to be propagated.
\begin{lstlisting}[frame=leftline, numbers = left,breaklines=true, label = ex:sin1,firstnumber=last]
  nus.Set_E(10.0*units.GeV);
\end{lstlisting}

Next, we define the medium in which we are going to propagate the neutrinos. 
The atmospheric Earth object used here, {\ttf earth\_atm}, contains the
geometry and density of the Earth, including a simple treatment of the atmosphere; see~\ref{sec:body} for more details. We also construct a trajectory in the
earth for a zenith angle {\ttf phi}, {\ttf earth\_atm\_track}. Finally,
we apply these objects to the nuSQuIDS object.

\begin{lstlisting}[frame=leftline, numbers = left,breaklines=true, label = ex:sin1,firstnumber=last]
  double phi = acos(-1.0);
  auto earth_atm = std::make_shared<EarthAtm>();
  auto earth_atm_track = std::make_shared<EarthAtm::Track>(
                           earth_atm->MakeTrackWithCosine(phi));
  nus.Set_Body(earth_atm);
  nus.Set_Track(earth_atm_track);
\end{lstlisting}

Now we set the initial neutrino flavor composition. In this example we start with
a pure muon neutrino state, represented by an {\ttf marray} (multi-dimensional array) of rank 1 and length {\ttf
  \{3\}}, corresponding to the number of flavors, and with value given
by {\ttf\{0,1,0\}}.
We indicate that this state is described in the flavor basis when applying it to the {\ttf nuSQuIDS} object:

\begin{lstlisting}[frame=leftline, numbers = left,breaklines=true, label = ex:sin1,firstnumber=last]
  marray<double,1> ini_state({3},{0,1,0});
  nus.Set_initial_state(ini_state,flavor);
\end{lstlisting}

We set the numerical error for the GSL differential equation solver.
The parameters and errors are defined as in the standard GSL libraries.

\begin{lstlisting}[frame=leftline, numbers = left,breaklines=true, label = ex:sin1,firstnumber=last]
  nus.Set_rel_error(1.0e-20);
  nus.Set_abs_error(1.0e-20);
\end{lstlisting}

Finally, we show the result before and after propagation. The
propagation is done by calling the function {\ttf nus.EvolveState()}

\begin{lstlisting}[frame=leftline, numbers = left,breaklines=true, label = ex:sin1,firstnumber=last]
  std::cout << "In state" << std::endl;
  for (double EE : nus.GetERange()){
    std::cout << EE/units.GeV << " ";
    for(int i = 0; i < 3; i++){
      std::cout << nus.EvalFlavor(i) << " ";
    }
    std::cout << std::endl;
  }
  // We do the calculation                                                                                  
  nus.EvolveState();
  
  // Output the result                                                                                 
  std::cout << "Out state" << std::endl;
  for (double EE : nus.GetERange()){
    std::cout << EE/units.GeV << " ";
    for(int i = 0; i < 3; i++){
      std::cout << nus.EvalFlavor(i) << " ";
    }
    std::cout << std::endl;
  }
\end{lstlisting}

\subsection{Multiple Energy \textnormal{({\ttf examples/Multiple\_energy})}}
This is a more realistic example where we consider an ensemble of
neutrinos in an energy range. With this setup,
we can model the neutrino energy and flavor changes, 
including the effects of charged and neutral current interactions.
In the following we describe the code for the example.

As before, we construct the {\ttf Const} object to have a set of useful physical parameters.
\begin{lstlisting}[frame=leftline, numbers = left,breaklines=true, label = ex:sin1]
  squids::Const units;
\end{lstlisting}

In the following, we allow the choice to compute
the propagation for three active neutrinos or three active and one sterile.

\begin{lstlisting}[frame=leftline, numbers = left,breaklines=true, label = ex:sin1,firstnumber=last]
  std::cout << "(3) Three Active Neutrinos, " << "(4) 3+1 Three Active and One Sterile Neutrino" << std::endl;
  unsigned int numneu;
  std::cin >>numneu;
  if( not(numneu==3 || numneu==4)){
    throw std::runtime_error("Only (3) or (4) are valid options");
  }
\end{lstlisting}

In the next line, we construct the {\ttf nuSQUIDS} object. For the
multiple energy constructor, we need to provide the following arguments:
a list of neutrino energy nodes
(\lstinline[columns=fixed,breaklines=true]{logspace(1.*units.GeV,1.e4*units.GeV,200)}),
number of neutrino flavors ({\ttf numneu}= 3 or 4), neutrino or
anti-neutrino type ({\ttf neutrino}), and whether to compute non-coherent scattering
interactions ({\ttf false}). 
\begin{lstlisting}[frame=leftline, numbers = left,breaklines=true,
  label = ex:sin1,firstnumber=8]
  nuSQUIDS nus(logspace(1.*units.GeV,1.e4*units.GeV,200),numneu,neutrino,false);
\end{lstlisting}

As in the single energy mode, we must define the body and the path
where neutrinos will propagate.

\begin{lstlisting}[frame=leftline, numbers = left,breaklines=true,
  label = ex:sin1,firstnumber=last]
  double phi = acos(-1.);
  auto earth_atm = std::make_shared<EarthAtm>();
  auto track_atm = std::make_shared<EarthAtm::Track>(
                     earth_atm->MakeTrackWithCosine(phi));
  nus.Set_Body(earth_atm);
  nus.Set_Track(track_atm);
\end{lstlisting}

We set the neutrino oscillation parameters, and if a sterile
neutrino is considered, we set its parameters to $\theta_{24}=0.1$ and
$\Delta m_{41}^2=0.1{\rm eV}^2$.
\begin{lstlisting}[frame=leftline, numbers = left,breaklines=true,label = ex:sin1,firstnumber=last]
  nus.Set_MixingAngle(0,1,0.563942);
  nus.Set_MixingAngle(0,2,0.154085);
  nus.Set_MixingAngle(1,2,0.785398);
  nus.Set_SquareMassDifference(1,7.65e-05);
  nus.Set_SquareMassDifference(2,0.00247);

  if(numneu==4){ // sterile neutrino parameters
    nus.Set_SquareMassDifference(3,0.1);
    nus.Set_MixingAngle(1,3,0.1);
  }
\end{lstlisting}

Next we set some of the integration parameters: the maximum
step size for the evolution, the GSL stepper algorithm (all the steppers in
the GSL libraries can be used), and maximum relative and absolute error tolerances.

\begin{lstlisting}[frame=leftline, numbers = left,breaklines=true,label = ex:sin1,firstnumber=last]
  nus.Set_h_max( 500.0*units.km );
  nus.Set_GSL_step(gsl_odeiv2_step_rk4);
  nus.Set_rel_error(1.0e-5);
  nus.Set_abs_error(1.0e-5);
\end{lstlisting}

We set to {\ttf true} the progress bar options in order to display the
progress of the propagation.

\begin{lstlisting}[frame=leftline, numbers = left,breaklines=true,label = ex:sin1,firstnumber=last]
  nus.Set_ProgressBar(true);
\end{lstlisting}

We can ask {\ttf nuSQuIDS} to give back the array containing the values of
the energy nodes. We use this here to fill a multi-dimensional array with the initial
state of the system; in this case a $E^{-2}$ power-law for the muon
flavor and zero for the other flavors.
Unlike the single-energy example, the array now has rank 2, with dimensions for both energy and flavor. 
 
\begin{lstlisting}[frame=leftline, numbers = left,breaklines=true,label = ex:sin1,firstnumber=last]
  marray<double,1> E_range = nus.GetERange();
  marray<double,2> inistate{E_range.size(),numneu};
  double N0 = 1.0e18;
  for ( int i = 0 ; i < inistate.extent(0); i++){
      for ( int k = 0; k < inistate.extent(1); k ++){
        inistate[i][k] = (k == 1) ? N0*pow(E_range[i],-2) : 0.0;
      }
  }
  nus.Set_initial_state(inistate,flavor);
\end{lstlisting}

We can then evolve the state:

\begin{lstlisting}[frame=leftline, numbers = left,breaklines=true,label = ex:sin1,firstnumber=last]
  nus.EvolveState();
\end{lstlisting}

Finally, we write the propagated fluxes to a file and ask whether to run the plotting script:

\begin{lstlisting}[frame=leftline, numbers = left,breaklines=true,label = ex:sin1,firstnumber=last]
  std::ofstream file("fluxes_flavor.txt");
  
  int Nen =1000;
  double lEmin=0;
  double lEmax=4;
  
  file << "# log10(E) E flux_i fluxRatio_i . . . ." << std::endl;
  for(double lE=lEmin; lE<lEmax; lE+=(lEmax-lEmin)/(double)Nen){
    double E=pow(10.0,lE)*units.GeV;
    file << lE << " " << E << " ";
    for(int fl=0; fl<numneu; fl++){
      file << " " <<  nus.EvalFlavor(fl, E) << " " <<  nus.EvalFlavor(fl, E)/(N0*pow(E,-2));
    }
    file << std::endl;
  }
  file.close();
  std::string plt;
  std::cout << std::endl <<  "Done! " << std::endl <<
  "  Do you want to run the gnuplot script? yes/no" << std::endl;
  std::cin >> plt;
  if(plt=="yes" || plt=="y")
  return system("./plot.plt");
\end{lstlisting}

Note that while we explicitly propagated 200 energies, we evaluate the resulting flux at 1000 energies, taking advantage of the automatic interpolation. 

\begin{figure}[h!]
  \label{fig:multimode}
  \centering
  \includegraphics[width=0.7\textwidth]{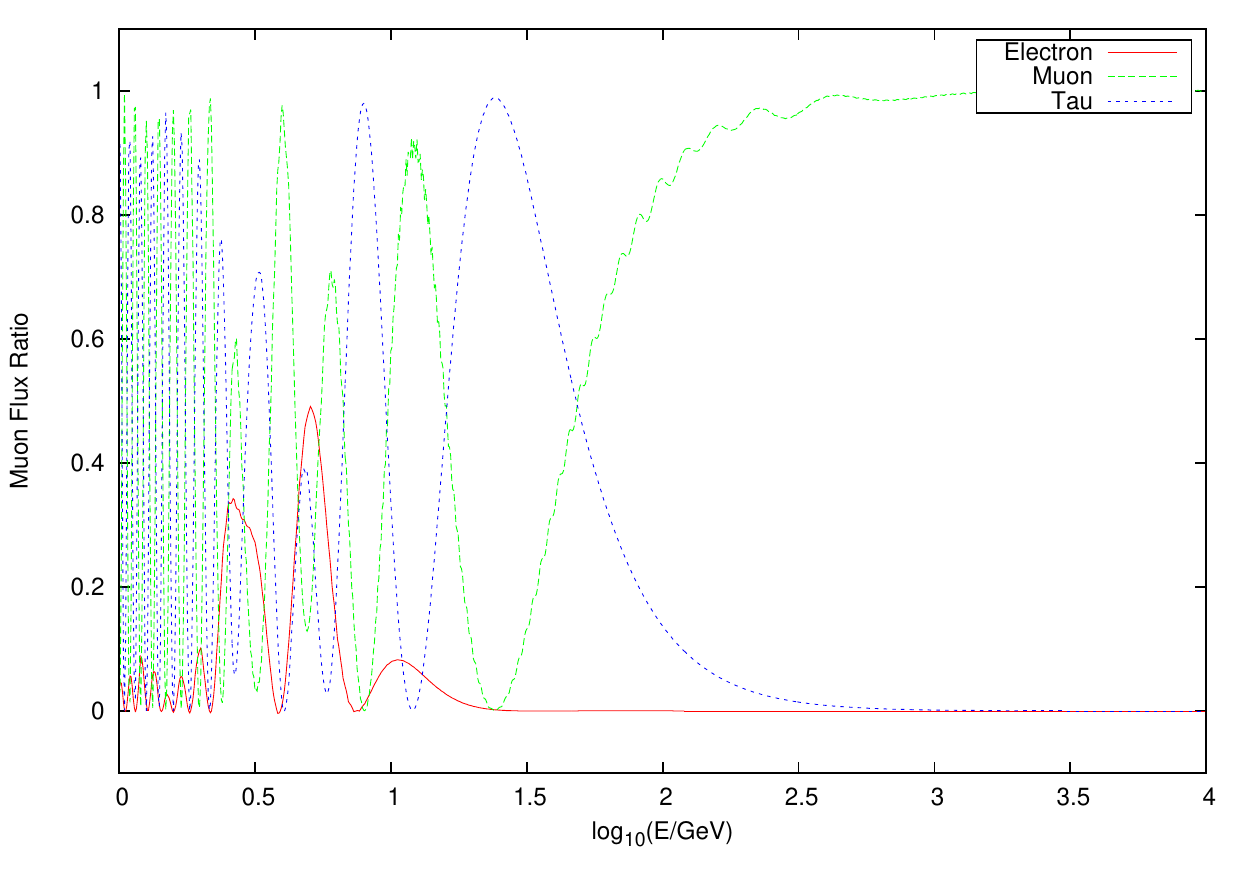} 
  \caption{Output for the multiple-energy mode with sterile neutrino (3+1)} 
\end{figure}
 
In Fig.~\ref{fig:multimode}, we show the results of running this example
for the case with sterile neutrino.

\subsection{Read and Write \textnormal{({\ttf examples/HDF5\_Write\_Read})}}
\label{sec:readwrite}
In this example, we illustrate the use of the functions to
read and write {\ttf HDF5} files. The file contains the full
information of the system: all settings, body, track (including the
current position along the track), and the density matrix at each node.
This serialization allows saving the system
in the middle of a complex propagation, and reading it later to continue
the calculation. Saving the system and then loading it from
the file restores the system to the saved state, allowing the user to
compute any quantum-mechanical observable.

This example is based upon the multiple energy example, but
split into two parts. The first part computes the evolution and saves the
state of the system ({\ttf write.cpp}). The second reads the
state, extracts the flavor fluxes, and prints them to a file ({\ttf read.cpp}).

In the following, we describe lines that are relevant for reading and writing.

In the file {\ttf write.cpp}, we save the system before and after the
evolution:
\begin{lstlisting}[frame=leftline, numbers =
  left,breaklines=true,label = ex:sin1]
  nus.WriteStateHDF5("./initial_state.hdf5");
  nus.EvolveState();
  nus.WriteStateHDF5("./final_state.hdf5");
\end{lstlisting}

In order to recover that state in {\ttf read.cpp}, we use the
constructor that takes the name of the file as an argument:

\begin{lstlisting}[frame=leftline, numbers =
  left,breaklines=true,label = ex:sin1]
  nuSQUIDS inus("./initial_state.hdf5");
  nuSQUIDS fnus("./final_state.hdf5"); 
\end{lstlisting}

\subsection{Construction and use of bodies \textnormal{({\ttf
      examples/Bodies})}}
\label{sec:body}
Two of the main classes in the {\ttf nuSQuIDS} library are the body and the
track. 
These properties of the {\ttf nuSQuIDS} object itself do not have defaults, and always need to
be specified.
In this example, we show how to use the objects already implemented in
the library, and also how to create derived types from the classes
already defined.

The folder {\ttf examples/Bodies/} contains new body and track definitions
in the files {\ttf exBody.h} and {\ttf exBody.cpp}.
It also contains the file {\ttf main.cpp}, whose main function 
uses all the pre-existing objects in nuSQUIDS and the new objects
of these examples.

\subsubsection{Construct a derived body}

Here is an example of how to define a derived class from the EarthAtm
body. In this simple case the object is just an Earth model with some
parameters that weight the relative densities in the different layers
of the Earth's inner core, outer core, and mantle.

% TODO: Is this actually a good idea? Seems like a recipe for user code to be unnecessarily broken if we add things
The new classes may be placed inside the {\ttf nusquids}
namespace. We define a new class called {\ttf EarthMod}, which is a derived class of {\ttf
  EarthAtm}. Because of this, the new class will have all the properties,
values, and functions of the parent class. 
\begin{lstlisting}[frame=leftline, numbers = left,breaklines=true,label = ex:sin1]
namespace nusquids{

class EarthMod: public EarthAtm{
public:
\end{lstlisting}

Our object's constructors require the following inputs:
base Earth model file name ({\ttf earthmodel}), inner core weight
({\ttf frho1}), outer core weight ({\ttf frho2}), and mantle weight ({\ttf
  frho3}).

\begin{lstlisting}[frame=leftline, numbers = left,breaklines=true,label = ex:sin1,firstnumber=last]
  EarthMod(std::string earthmodel, double frho1, double frho2, double frho3);
\end{lstlisting}
The {\ttf Mod} function allows changing the parameters once the
object is already initialized.
\begin{lstlisting}[frame=leftline, numbers =
  left,breaklines=true,label = ex:sin1,firstnumber=last]
  void  Mod(double frho1, double frho2, double frho3);
};

\end{lstlisting}

The implementations of the functions that set the density arrays with
the modified values are given in the file {\ttf exBody.cpp}.

\subsubsection{Use of the bodies}

The {\ttf main.cpp} file defines a nuSQUIDS object and sets different bodies and
tracks. For each of these, it evolves the system and displays the resulting oscillation
probabilities.
For simplicity we use the single energy mode Sec.~\ref{sec:single}, but the use of the body
and track would be the same for the multiple energy case.

First, we construct the nuSQUIDS object for three neutrinos, then we set the
oscillation parameters and the neutrino energy.

\begin{lstlisting}[frame=leftline, numbers =
  left,breaklines=true,label = ex:sin1]
  nuSQUIDS nus(3,neutrino);
  nus.Set_MixingAngle(0,1,0.563942);
  nus.Set_MixingAngle(0,2,0.154085);
  nus.Set_MixingAngle(1,2,0.785398);
  nus.Set_SquareMassDifference(1,7.65e-05);
  nus.Set_SquareMassDifference(2,0.00247);
  nus.Set_CPPhase(0,2,0.0);
  squids::Const units;
  nus.Set_E(10.0*units.GeV);
\end{lstlisting}

\begin{enumerate}
\item {\ttf Earth}

The first example is the {\ttf Earth} body. In this case, the track is
parametrized by the chord length through the Earth, which is useful for treating the baseline of an experiment with a fixed source and detector. 
Here we define the body and track. For the track we need to specify the initial and final
position as well as the baseline: 
\begin{lstlisting}[frame=leftline, numbers =
  left,breaklines=true,label = ex:sin1,firstnumber=last]
  double baseline = 500.0*units.km;
  std::shared_ptr<Earth> earth = std::make_shared<Earth>();
  std::shared_ptr<Earth::Track> earth_track = std::make_shared<Earth::Track>(0.0,baseline,baseline);
\end{lstlisting}
We apply the Body and Track to the nuSQuIDS object:
\begin{lstlisting}[frame=leftline, numbers =
  left,breaklines=true,label = ex:sin1,firstnumber=last]
  nus.Set_Body(earth);
  nus.Set_Track(earth_track);
\end{lstlisting}

We then set the initial state of the system and print it:
\begin{lstlisting}[frame=leftline, numbers =
  left,breaklines=true,label = ex:sin1,firstnumber=last]
  marray<double,1> ini_state({3},{0,1,0});
  nus.Set_initial_state(ini_state,flavor);
  
  std::cout << "In state" << std::endl;
  for (double EE : nus.GetERange()){
    std::cout << EE/units.GeV << " ";
    for(int i = 0; i < 3; i++){
      std::cout << nus.EvalFlavor(i) << " ";
    }
    std::cout << std::endl;
  }
\end{lstlisting}
We set the numerical error and maximum step for the GSL integrator:
\begin{lstlisting}[frame=leftline, numbers =
  left,breaklines=true,label = ex:sin1,firstnumber=last]
  nus.Set_h_max( 200.0*units.km );
  nus.Set_rel_error(1.0e-12);
  nus.Set_abs_error(1.0e-12);
\end{lstlisting}

Finally, we evolve the state and print it:
\begin{lstlisting}[frame=leftline, numbers =
  left,breaklines=true,label = ex:sin1,firstnumber=last]
  nus.EvolveState();
  std::cout << "Out state" << std::endl;
  for (double EE : nus.GetERange()){
    std::cout << EE/units.GeV << " ";
    for(int i = 0; i < 3; i++){
      std::cout << nus.EvalFlavor(i) << " ";
    }
    std::cout << std::endl;
  }
\end{lstlisting}
These last steps in the code are the same in all the bodies examples and so are omitted for the following cases.

\item {\ttf EarthAtm}

In this example we use the {\ttf EarthAtm} body. 
This body includes both the Earth and its atmosphere. 
The track is defined by the zenith angle of the trajectory, and begins at the top of the atmosphere. 
Because the track depends on the height of the atmosphere being treated, it is most conveniently constructed from the body, using the {\ttf MakeTrack} function, which takes a zenith angle and creates a track with the necessary details of the body automatically included. 
\begin{lstlisting}[frame=leftline, numbers =
  left,breaklines=true,label = ex:sin1,firstnumber=last]
  double phi = acos(-1.0);
  auto earth_atm = std::make_shared<EarthAtm>();
  auto earth_atm_track = std::make_shared<EarthAtm::Track>(earth_atm->MakeTrack(phi));

  nus.Set_Body(earth_atm);
  nus.Set_Track(earth_atm_track);
\end{lstlisting}

\item {\ttf earth\_mod}

  In this case, we use the modified Earth object.
  As for the {\ttf EarthAtm} object, the track is defined by the zenith angle of
  the trajectory. In the constructor we set all the weights to $0.1$.
  Finally, we set the body and track for the {\ttf nuSQUIDS} object.
  
\begin{lstlisting}[frame=leftline, numbers =
  left,breaklines=true,label = ex:sin1,firstnumber=last]
  double phi = acos(-1.0);
  std::shared_ptr<EarthMod> earth_mod = std::make_shared<EarthMod>(0.1,0.1,0.1);
  std::shared_ptr<EarthMod::Track> earth_mod_track = std::make_shared<EarthMod::Track>(phi);  

  nus.Set_Body(earth_mod);
  nus.Set_Track(earth_mod_track);
\end{lstlisting}

\item {\ttf VariableDensity}

In this case, we use the variable density body and a track of $200{\rm km}$.
First, we define the density, position, and electron fraction arrays
with the corresponding values.
\begin{lstlisting}[frame=leftline, numbers =
  left,breaklines=true,label = ex:sin1,firstnumber=last]
  int N=40;

  std::vector<double> x_arr(N);
  std::vector<double> density_arr(N);
  std::vector<double> ye_arr(N);

  double size = 1000.0*units.km;
  for(int i = 0; i < N; i++){
    x_arr[i] = size*(i/(double)N);
    density_arr[i] = fabs(cos((double)i));
    ye_arr[i] = fabs(sin((double)i));
  }
\end{lstlisting}

Now we construct the body and the track. The constructor for the
variable density takes as an input the position, density, and electron
fraction vectors. Finally, like before, we set the body and the track on the {\ttf nuSQUIDS} object.
\begin{lstlisting}[frame=leftline, numbers =
  left,breaklines=true,label = ex:sin1,firstnumber=last]

  std::shared_ptr<VariableDensity> vardens = std::make_shared<VariableDensity>(x_arr,density_arr,ye_arr);
  std::shared_ptr<VariableDensity::Track> track_vardens = std::make_shared<VariableDensity::Track>(0.0,200.0*units.km);

  nus.Set_Body(vardens);
  nus.Set_Track(track_vardens);
\end{lstlisting}

\item {\ttf Vacuum}

This is a trivial case where the density and electron fraction are
zero. We only need to give the baseline as an argument to construct
the track. In this example, we set the baseline to $500{\rm km}$.

\begin{lstlisting}[frame=leftline, numbers =
  left,breaklines=true,label = ex:sin1,firstnumber=last]
  double baseline_2 = 500.0*units.km;
  std::shared_ptr<Vacuum> vacuum = std::make_shared<Vacuum>();
  std::shared_ptr<Vacuum::Track> track_vac = std::make_shared<Vacuum::Track>(baseline_2);
  
  nus.Set_Body(vacuum);
  nus.Set_Track(track_vac);
\end{lstlisting}

\item {\ttf ConstantDensity}

In the case of constant density an analytic approximation can be
used to propagate the neutrinos if non-coherent interactions are
disabled in the construction of the {\ttf nuSQUIDS} object. The full
Hamiltonian of the system is diagonalized and exponentiated. 

We set the density to $100{\rm g/cm}^3$, the electron fraction to $0.3$, and the
baseline to $500{\rm km}$.
\begin{lstlisting}[frame=leftline, numbers =
  left,breaklines=true,label = ex:sin1,firstnumber=last]

  double density = 100.0;
  double ye = 0.3;
  std::shared_ptr<ConstantDensity> constdens = std::make_shared<ConstantDensity>(density,ye);
  double baseline_3 = 500.0*units.km;
  std::shared_ptr<ConstantDensity::Track> track_constdens =   std::make_shared<ConstantDensity::Track>(0.0,baseline_3);

  nus.Set_Body(constdens);
  nus.Set_Track(track_constdens);
\end{lstlisting}
\end{enumerate}

\subsection{Cross Sections \textnormal{({\ttf
      examples/Xsections})}}

One of the important features of {\ttf nuSQuIDS} is the possibility of consistently handling non-coherent interactions and oscillation
behavior. The physical quantity that encodes how often these scattering
interactions happen between the neutrinos and the media is the cross
section.
{\ttf nuSQuIDS} has implemented two kinds of deep inelastic scattering (DIS) interactions: charged- and
neutral-current, as well as resonant $W^-$ production on electrons (the Glashow resonance). It takes into account
the neutrinos produced by the decay of short lived charged particles
such as $\tau^\pm$ and $W^\pm$. For the neutral-current, we always include
the outgoing neutrino.
Other particles produced in the interactions, such as hadrons or long-lived
charged leptons, are ignored along the evolution.
This information is organized and stored in the cross-section class.
This class requires the user to provide the total cross-section for
each flavor and current in units of cm$^2$. It also requires
the singly-differential neutrino cross-sections with respect to
the outgoing neutrino energy in units of cm$^2$/GeV.
See ~\ref{sec:neutrino_cross_section_from_tables} for more details of how this data must be specified. 

{\ttf nuSQuIDS} includes by default deep inelastic neutrino-nucleon
cross-sections as well as neutrino-electron
cross-sections~\citep{Gandhi:1998ri, CooperSarkar:2011pa}.

In this example, we construct a new cross section object to be used by
{\ttf nuSQuIDS} instead of the default one.
Every cross section must be a class of {\ttf NeutrinoCrossSections}
and implement at least two member functions:
{\ttf SingleDifferentialCrossSection} and {\ttf TotalCrossSection}.

\begin{lstlisting}[frame=leftline, numbers =
  left,breaklines=true,label = ex:sin1]
  class LinearCrossSections : public NeutrinoCrossSections {
    private:
    const squids::Const units;
    const double GF = 1.16639e-23; // eV^-2
    const double mp = 938.272e6; // proton mass eV
    const double CC_to_NC; // proportion of CC to NC which goes from 0 to 1.
    public :
    LinearCrossSections(double CC_to_NC):CC_to_NC(CC_to_NC){assert( CC_to_NC <= 1.0  && CC_to_NC >= 0.0 );}
    LinearCrossSections():LinearCrossSections(0.5){}
    double TotalCrossSection(double Enu, NeutrinoFlavor flavor, NeutrinoType neutype, Current current) const override;
    double SingleDifferentialCrossSection(double E1, double E2, NeutrinoFlavor flavor, NeutrinoType neutype, Current current) const override;
  };  
 
\end{lstlisting}

We use a toy total cross section that scales linearly with
the neutrino energy and the corresponding differential cross sections
linear in the outgoing neutrino energy; see implementation in {\ttf
  exCross.cpp}. We also allow changing the proportion of charge-to-neutral
  current at the construction of the object.

We construct two cross section libraries with the new cross section
associated with an isoscalar nucleon target: one with only charged-current interactions
{\ttf ncs\_cc} and another with only neutral-current {\ttf
  ncs\_nc}. Then we construct the corresponding {\ttf nuSQUIDS} objects {\ttf nus\_cc} and {\ttf nus\_nc}.
We also disable neutrino oscillations in this example by means of the
option {\ttf Set\_IncludeOscillations(false)}.

\begin{lstlisting}[frame=leftline, numbers =  
    left,breaklines=true,label = ex:sin1]
  auto ncs_cc=std::make_shared<CrossSectionLibrary>(CrossSectionLibrary::MapType{
    {isoscalar_nucleon, std::make_shared<LinearCrossSections>(0.0)}});
  auto ncs_nc=std::make_shared<CrossSectionLibrary>(CrossSectionLibrary::MapType{
    {isoscalar_nucleon, std::make_shared<LinearCrossSections>(1.0)}});

  nuSQUIDS nus_cc(logspace(Emin,Emax,200),numneu,neutrino,true,ncs_cc);
  nus_cc.Set_IncludeOscillations(false);
  nuSQUIDS nus_nc(logspace(Emin,Emax,200),numneu,neutrino,true,ncs_nc);
  nus_nc.Set_IncludeOscillations(false);
\end{lstlisting}

We can see the final to initial flux ratios for both cases in fig~\ref{fig:crossext}.

\begin{figure}[h!]
  \label{fig:crossext}
  \centering
  \includegraphics[width=0.7\textwidth]{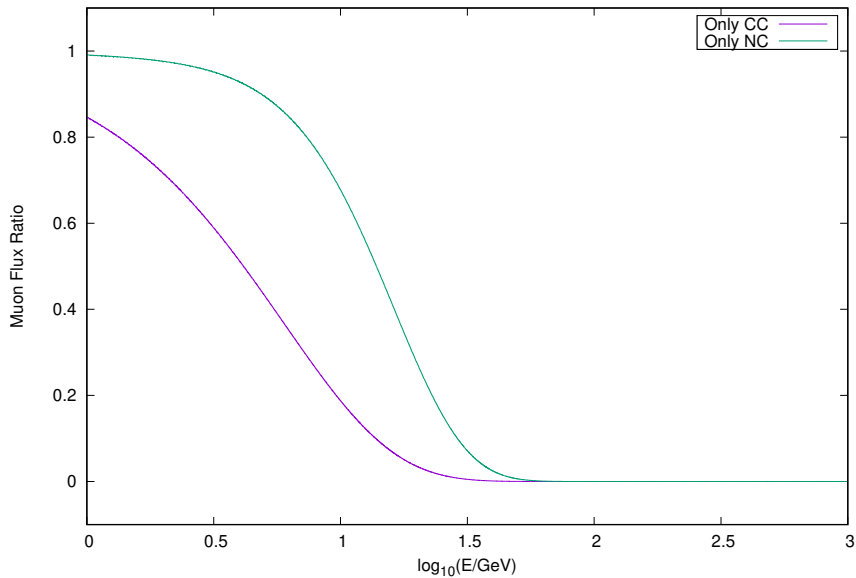} 
  \caption{Final to initial flux ratio as a function of the logarithm
    of the neutrino energy for the example toy cross-sections with
    only charged-current (CC) and only neutral-current (NC) interactions.} 
\end{figure}

\subsection{Constant density multiple layers \textnormal{({\ttf
      examples/Constant\_density\_layers})}}
Since the constant density allows us to do very fast computation using
the diagonalization of the full Hamiltonian here we show an example
where we concatenate the evolution of the neutrinos through different
layers of constant density.

We construct three different layers: first $100{\rm km}$ in
vacuum, second $50{\rm km}$ in matter with density $3.5{\rm g/cm}^3$ and
 electron fraction $0.5$, and finally $200{\rm km}$ in matter with
 density $10{\rm g/cm}^3$ and electron fraction $0.1$.
 
\begin{lstlisting}[frame=leftline, numbers =
  left,breaklines=true,label = ex:sin1]
  const double layer_1 = 100.*units.km;
  std::shared_ptr<Vacuum> vacuum = std::make_shared<Vacuum>();
  std::shared_ptr<Vacuum::Track> track_env0 = std::make_shared<Vacuum::Track>(layer_1);

  const double layer_2 = 50.*units.km;
  std::shared_ptr<ConstantDensity> constdens_env1 = std::make_shared<ConstantDensity>(3.5,0.5); 
  std::shared_ptr<ConstantDensity::Track> track_env1 = std::make_shared<ConstantDensity::Track>(layer_2);

  const double layer_3 = 200.*units.km;
  std::shared_ptr<ConstantDensity> constdens_env2 = std::make_shared<ConstantDensity>(10.,0.1);
  std::shared_ptr<ConstantDensity::Track> track_env2 = std::make_shared<ConstantDensity::Track>(layer_3);
\end{lstlisting}

Now in order to evolve the system we set the corresponding body
and track and evolve every layer.

\begin{lstlisting}[frame=leftline, numbers =
  left,breaklines=true,label = ex:sin1]
  nus.Set_Body(vacuum);
  nus.Set_Track(track_env0);
  nus.EvolveState();

  nus.Set_Body(constdens_env1);
  nus.Set_Track(track_env1);
  nus.EvolveState();

  nus.Set_Body(constdens_env2);
  nus.Set_Track(track_env2);
  nus.EvolveState();
\end{lstlisting}

Finally, we write the output for the different flavor fluxes in a text file.

\subsection{BSM extension: Non-Standard  Interaction (NSI) \textnormal{({\ttf
      examples/NSI})}}
\label{sec:NSI}
In this example we illustrate how to construct a derived class of
{\ttf nuSQUIDS} including a new physics term. This procedure is similar to
other new physics setups~\ref{sec:LV}.

The implementation of {\ttf nuSQUIDSNSI} class is in {\ttf NSI.h}.
We will go through the implementation to see what needs to be
added to the base class.

First, we declare the NSI matter potential as a {\ttf SU\_vector}
called {\ttf NSI}. By default {\ttf nuSQuIDS}  works in the interaction
picture, thus every term we add to the Hamiltonian has to be properly
evolved with $H_0$. Since $H_0$ depends on the energy we do this in
every node. For optimization we will compute and store the evolved NSI
term in a vector of {\ttf SU\_vector} objects called {\ttf NSI\_evol}.

The {\ttf HI\_prefactor} contains all the numerical factors multiplying the
operator and {\ttf epsilon\_mutau} is the strength of the NSI $\mu$-$\tau$ 
non-diagonal component. Notice that we set all other NSI
contributions to zero.

\begin{lstlisting}[frame=leftline, numbers =
  left,breaklines=true,label = ex:sin1]
class nuSQUIDSNSI: public nuSQUIDS {
  private:
    squids::SU_vector NSI;
    std::vector<squids::SU_vector> NSI_evol;
    std::unique_ptr<double[]> hiBuffer;
    double HI_prefactor;
    // nsi parameters
    double epsilon_mutau;

\end{lstlisting}

As previously noted, we need to compute the NSI operator in the
interaction picture prior to adding it to the r.h.s. of the
differential equation. We compute this for every energy node in the
{\ttf AddToPreDerive} function. This function is called inside the
{\ttf PreDerive} function before evaluating the derivatives to allow
the user to pre-compute terms used in the derivative.
The evolution of the NSI term by $H_0$ is done by the {\ttf
  SU\_vector::Evolve} member. This function is optimized and performs
the evolution of the operator analytically.

\begin{lstlisting}[frame=leftline, numbers =
  left,breaklines=true,label = ex:sin1,firstnumber=last]
    void AddToPreDerive(double x){
      for(int ei = 0; ei < ne; ei++){
        NSI_evol[ei] = NSI.Evolve(H0_array[ei],(x-Get_t_initial()));
      }
    }
\end{lstlisting}

The following auxiliary member functions allow {\ttf nuSQUIDSNSI} to
save the new physics parameters into the hdf5 files when the
{\ttf WriteStateHDF5} is called (See Sec.~\ref{sec:readwrite} for an overview of reading and writing).

\begin{lstlisting}[frame=leftline, numbers =
  left,breaklines=true,label = ex:sin1,firstnumber=last]
    void AddToReadHDF5(hid_t hdf5_loc_id){
      // here we read the new parameters now saved in the HDF5 file
      hid_t nsi = H5Gopen(hdf5_loc_id, "nsi", H5P_DEFAULT);
      H5LTget_attribute_double(hdf5_loc_id,"nsi","mu_tau" ,&epsilon_mutau);
      H5Gclose(nsi);
    }

    void AddToWriteHDF5(hid_t hdf5_loc_id) const {
      // here we write the new parameters to be saved in the HDF5 file
      H5Gcreate(hdf5_loc_id, "nsi", H5P_DEFAULT, H5P_DEFAULT, H5P_DEFAULT);
      H5LTset_attribute_double(hdf5_loc_id, "nsi","mu_tau",&epsilon_mutau, 1);
    }
\end{lstlisting}

Here we overload the {\ttf HI} function where we return the original {\ttf nuSQUIDS}
matter potential plus the new NSI contribution.

\begin{lstlisting}[frame=leftline, numbers =
  left,breaklines=true,label = ex:sin1,firstnumber=last]
    squids::SU_vector HI(unsigned int ei,unsigned int index_rho) const{
      double CC=HI_prefactor*body->density(*track)*body->ye(*track);

      squids::SU_vector potential(nsun,hiBuffer.get());

      potential = (3.0*CC)*NSI_evol[ei];

      if ((index_rho == 0 and NT==both) or NT==neutrino){
          return nuSQUIDS::HI(ei,index_rho) + potential;
      } else if ((index_rho == 1 and NT==both) or NT==antineutrino){
          return nuSQUIDS::HI(ei,index_rho) + (-1.0)*std::move(potential);
      } else{
          throw std::runtime_error("nuSQUIDS::HI : unknown particle or antiparticle");
      }
    }
\end{lstlisting}

The constructor first calls the
{\ttf nuSQUIDS} constructor and then sets the oscillation
parameters and the NSI operator.

\begin{lstlisting}[frame=leftline, numbers =
  left,breaklines=true,label = ex:sin1,firstnumber=last]
  public:
  nuSQUIDSNSI(double epsilon_mutau, marray<double,1> Erange,
              int numneu, NeutrinoType NT, bool iinteraction,
              double th01=0.563, double th02=0.154, double th12=0.785):
              nuSQUIDS(Erange,numneu,NT,iinteraction),
	      hiBuffer(new double[nsun*nsun]),
              epsilon_mutau(epsilon_mutau)
  {
    assert(numneu == 3);
    // defining a complex matrix M which will contain our flavor
    // violating flavor structure.
    gsl_matrix_complex * M = gsl_matrix_complex_calloc(3,3);
    gsl_complex c {{ epsilon_mutau , 0.0 }};
    gsl_matrix_complex_set(M,2,1,c);
    gsl_matrix_complex_set(M,1,2,gsl_complex_conjugate(c));
    
    NSI = squids::SU_vector(M);
    
    Set_MixingAngle(0,1,th01);
    Set_MixingAngle(0,2,th02);
    Set_MixingAngle(1,2,th12);
    
    // rotate to mass reprentation
    NSI.RotateToB1(params);
    NSI_evol.resize(ne);
    for(int ei = 0; ei < ne; ei++){
      NSI_evol[ei] = squids::SU_vector(nsun);
    }
    gsl_matrix_complex_free(M);
    
    HI_prefactor = params.sqrt2*params.GF*params.Na*pow(params.cm,-3);
  }
\end{lstlisting}

The next function sets the value of {\ttf epsilon\_mutau} changing the
 {\ttf SU\_vector NSI} object accordingly. We construct the {\ttf
   SU\_vector} operator as a complex GSL matrix that represents the
 operator in the flavor basis. Since {\ttf nuSQuIDS} solves the propagation
 in the basis where $H_0$ is diagonal, i.e. the mass basis, we rotate
 the NSI operator to the mass basis by calling {\ttf RotateToB1}.

\begin{lstlisting}[frame=leftline, numbers =
  left,breaklines=true,label = ex:sin1,firstnumber=last]
  void Set_mutau(double eps){
    gsl_matrix_complex * M = gsl_matrix_complex_calloc(3,3);
    gsl_complex c {{ epsilon_mutau , 0.0 }};
    gsl_matrix_complex_set(M,2,1,c);
    gsl_matrix_complex_set(M,1,2,gsl_complex_conjugate(c));
    NSI = squids::SU_vector(M);    
    NSI.RotateToB1(params);
    gsl_matrix_complex_free(M);
  }
\end{lstlisting}

In {\ttf main.cpp} we use the new NSI class in a
multiple energy mode. In order to compare the oscillation
probabilities with and without NSI we construct two instances of {\ttf
  nuSQUIDSNSI} with {\ttf  epsilon\_mutau=0} and {\ttf
  epsilon\_mutau=1e-2}. 

\begin{lstlisting}[frame=leftline, numbers =
  left,breaklines=true,label = ex:sin1,firstnumber=last]
  double eps_mutau=1.0e-2;
  nuSQUIDSNSI nus(eps_mutau,logspace(Emin,Emax,200),numneu,antineutrino,false);
  nuSQUIDSNSI nus_zero(0.0,logspace(Emin,Emax,200),numneu,antineutrino,false);
\end{lstlisting}

In the next lines we propagate both objects and print both
fluxes to a file.

\begin{lstlisting}[frame=leftline, numbers =
  left,breaklines=true,label = ex:sin1,firstnumber=last]
  nus.EvolveState();
  nus_zero.EvolveState();

  int Nen =1000;
  double lEmin=log10(Emin);
  double lEmax=log10(Emax);
  
  std::ofstream file("fluxes_flavor.txt");

  file << "# log10(E) E flux_NSI_i flux_noNSI_i . . . ." << std::endl;
  for(double lE=lEmin; lE<lEmax; lE+=(lEmax-lEmin)/(double)Nen){
    double E=pow(10.0,lE);
    file << lE << " " << E << " ";
    for(int fl=0; fl<numneu; fl++){
      file << " " <<  nus.EvalFlavor(fl, E) << " " <<  nus_zero.EvalFlavor(fl, E);
    }
    file << std::endl;
  }
\end{lstlisting}

In the folder there is a script that allows plotting the output text
file.

\subsection{BSM extension: Lorentz Violation \textnormal{({\ttf
      examples/LV})}}
\label{sec:LV}
The Lorentz symmetry is a well established property of space-time.
As a fundamental symmetry, it should be tested and neutrino
oscillations probe part of the parameter space.

The example is technically the same as the non-standard interactions,
but we add a couple of features to illustrate good practices.

As before, the new term is added in to {\ttf HI}. For this particular
physics case the effect is positive for neutrinos and negative for
antineutrinos. This can be implemented using the integer {\ttf irho}
which labels neutrinos with $0$ and antineutrinos with $1$ when both particle types are being calculated.
Which particle types are under consideration is given by the {\ttf NT} member variable. 

\begin{lstlisting}[frame=leftline, numbers =
  left,breaklines=true,label = ex:sin1,firstnumber=last]
  squids:: SU_vector HI(unsigned int ie,unsigned int irho) const {
    squids::SU_vector potential = nuSQUIDS::HI(ie, irho);
    double sign = 1;
    if ((irho == 1 and NT==both) or NT==antineutrino){
      // antineutrino matter potential flips sign
      sign*=(-1);
    }
    // ===== HERE WE ADD THE NEW PHYSICS =====
    potential += sign*pow(E_range[ie],n_)*LVP_evol[ie]; 
    // ===== HERE WE ADD THE NEW PHYSICS =====
    return potential;
  }
\end{lstlisting}

The parameters of the Lorentz violating term are set by the function
{\ttf Set\_LV\_OpMatrix}. In doing that, the oscillation parameters are
used to rotate to the mass basis. Therefore, any change to the
oscillation parameters after setting the LV term will be inconsistent.
In order to prevent inconsistencies, we overload the function that sets
the mixing parameters {\ttf Set\_MixingAngle} and the phases {\ttf
  Set\_CPPhase} turning the label {\ttf lv\_parameters\_set} to {\ttf
  false} and forcing the need to call {\ttf Set\_LV\_OpMatrix} again.
This enforces the order of the `Set' function calls.

\begin{lstlisting}[frame=leftline, numbers =
  left,breaklines=true,label = ex:sin1,firstnumber=last]
    void Set_MixingAngle(unsigned int i, unsigned int j,double angle){
      nuSQUIDS::Set_MixingAngle(i,j,angle);
      lv_parameters_set = false;
    }

    void Set_CPPhase(unsigned int i, unsigned int j,double angle){
      nuSQUIDS::Set_CPPhase(i,j,angle);
      lv_parameters_set = false;
    }
\end{lstlisting}

\subsection{Decoherence or averaged oscillations\\ \textnormal{({\ttf
      examples/Astrophysical\_neutrino\_flavor\_ratio})}}

This file demonstrates how to calculate the astrophysical flavor ratio by means
of using the `averaged out' approximation. We do this in two ways in this
example: first by means of nuSQuIDS fast averaging functionality, 
and second explicitly using the PMNS matrix and the formula given in the literature.
For simplicity, we will do this example in the single energy mode, but it
can be performed in the multiple energy mode, as well.

In this example we set {\ttf N\_neutrino = 3} and {\ttf Type = "neutrino"}.

\begin{lstlisting}[frame=leftline, numbers =
  left,breaklines=true,label = ex:sin1]
  nuSQUIDS nus(3,neutrino);
\end{lstlisting}

We use the standard parametrization.

\begin{lstlisting}[frame=leftline, numbers =
  left,breaklines=true,label = ex:sin1]
  nus.Set_MixingAngle(0,1,0.563942);
  nus.Set_MixingAngle(0,2,0.154085);
  nus.Set_MixingAngle(1,2,0.785398);

  nus.Set_SquareMassDifference(1,7.65e-05);
  nus.Set_SquareMassDifference(2,0.00247);
  
  nus.Set_CPPhase(0,2,0.0);
\end{lstlisting}

We construct and set the body and the track; in this case we use vacuum
oscillations.

\begin{lstlisting}[frame=leftline, numbers =
  left,breaklines=true,label = ex:sin1]
  std::shared_ptr<Vacuum> vacuum = std::make_shared<Vacuum>();
  std::shared_ptr<Vacuum::Track> vacuum_track = std::make_shared<Vacuum::Track>(1.e3*units.kparsec);
  nus.Set_Body(vacuum);
  nus.Set_Track(vacuum_track);
\end{lstlisting}

Here we set the initial state for the flavor, a pion-produced flavor composition:

\begin{lstlisting}[frame=leftline, numbers =
  left,breaklines=true,label = ex:sin1]
  marray<double,1> ini_state({3},{1,2,0});
  nus.Set_initial_state(ini_state,flavor);
\end{lstlisting}

We evolve the neutrinos:

\begin{lstlisting}[frame=leftline, numbers =
  left,breaklines=true,label = ex:sin1]
  nus.EvolveState();
\end{lstlisting}

{\ttf nuSQuIDS} can calculate the average oscillation probability
where the oscillation frequencies are larger than some value;
we call this value {\ttf scale}. When this mode is used it is often
valuable to know which frequencies have been averaged out.
NuSQuIDS does this by sting true a boolean vector {\ttf is\_avg} provided to the
evaluation function.

\begin{lstlisting}[frame=leftline, numbers =
  left,breaklines=true,label = ex:sin1]
  nus.EvolveState();
  double scale = 0.;
  std::vector<bool> is_avg(3);
\end{lstlisting}

To always get an averaged result, we can set the scale to the maximum
double numerical limit.

\begin{lstlisting}[frame=leftline, numbers =
  left,breaklines=true,label = ex:sin1]
  scale = std::numeric_limits<double>::max();
\end{lstlisting}

We output the result for the averaged and non-averaged computations.

\begin{lstlisting}[frame=leftline, numbers =
  left,breaklines=true,label = ex:sin1]
    std::cout << "Out state" << std::endl;
  for (double EE : nus.GetERange()){
    std::cout << EE/units.GeV << " ";
    for(int i = 0; i < 3; i++){
      std::cout << nus.EvalFlavor(i, scale, is_avg);
      std::cout << " (" << (is_avg[i] ? "avg." : "no avg.")  << ") ";
    }
    std::cout << std::endl;
  }

  scale = 0.0;

  std::cout << "Out state" << std::endl;
  for (double EE : nus.GetERange()){
    std::cout << EE/units.GeV << " ";
    for(int i = 0; i < 3; i++){
      std::cout << nus.EvalFlavor(i, scale, is_avg);
      std::cout << " (" << (is_avg[i] ? "avg." : "no avg.")  << ") ";
    }
    std::cout << std::endl;
  }
\end{lstlisting}

Notice that the average calculation is done only on the
oscillations given by $H_0$ when any oscillation is evaluated. Fast
oscillations from other terms, such as $H_I$, won't be
averaged out.

\subsection{Atmospheric mode: Standard \textnormal{({\ttf
      examples/Atm\_default})}}
\label{sec:atmexample}
In this example we show how to use the atmospheric mode. This mode is
a compact way to treat a set of {\ttf nuSQUIDS} objects distributed in
zenith angle.
This allows us to propagate the energy-zenith dependent atmospheric
neutrino flux through the Earth.

First, we construct the {\ttf nuSQUIDSAtm} object. The parameters of the
constructors are the list of cosine of the zenith angle values ({\ttf
  linspace(czmin,czmax,40)}) and the following arguments as for the {\ttf nuSQUIDS}
multiple energy constructor.

\begin{lstlisting}[frame=leftline, numbers =
  left,breaklines=true,label = ex:sin1]
  double Emin=1.e1*units.GeV;
  double Emax=1.e6*units.GeV;
  double czmin=-1;
  double czmax=0;

  nuSQUIDSAtm<> nus_atm(linspace(czmin,czmax,40),logspace(Emin,Emax,100),numneu,both,interactions);
\end{lstlisting}

In this case the initial state is an {\ttf marray} of rank four with double values.  
The first index is for the zenith angle, the second for the energy, the
third for particle type (neutrino or antineutrino), and the last one
for the neutrino flavor.

In this example we fill the multi-dimensional array with the initial state of the
system where the function {\ttf flux\_function} would be the corresponding
atmospheric flux in terms of energy and zenith angle.

\begin{lstlisting}[frame=leftline, numbers =
  left,breaklines=true,label = ex:sin1,firstnumber=last]
  marray<double,4> inistate{nus_atm.GetNumCos(),nus_atm.GetNumE(),2,numneu};
  std::fill(inistate.begin(),inistate.end(),0);
  for ( int ci = 0 ; ci < nus_atm.GetNumCos(); ci++){
    for ( int ei = 0 ; ei < nus_atm.GetNumE(); ei++){
      for ( int rho = 0; rho < 2; rho ++ ){
        for (int flv = 0; flv < numneu; flv++){
          inistate[ci][ei][rho][flv] = (flv == 1) ? flux_function(e_range[ei], cz_range[ci]) : 0.0;//set 1 only to the muon flavor
        }
      }
    }
  }
  nus_atm.Set_initial_state(inistate,flavor);
\end{lstlisting}

To evolve the full state we call {\ttf EvolveState}, as usual. 

\begin{lstlisting}[frame=leftline, numbers =
  left,breaklines=true,label = ex:sin1,firstnumber=last]
nus_atm.EvolveState();
\end{lstlisting}

Finally, we evaluate and print the flux in a file. The
atmospheric mode has an interpolation implemented that allows
evaluating the flux at any flavor, energy, and cosine-zenith. See~\ref{sec:nusquidsatm}
for more details.

\begin{lstlisting}[frame=leftline, numbers =
  left,breaklines=true,label = ex:sin1,firstnumber=last]
  int Nen=700;
  int Ncz=100;
  double lEmin=log10(Emin);
  double lEmax=log10(Emax);;

  file << "# log10(E) cos(zenith) E flux_i . . . ." << std::endl;
  for(double cz=czmin;cz<czmax;cz+=(czmax-czmin)/(double)Ncz){
    for(double lE=lEmin; lE<lEmax; lE+=(lEmax-lEmin)/(double)Nen){
      double E=pow(10.0,lE);
      file << lE << " " << cz << " " << E;
      for(int fl=0; fl<numneu; fl++){
	file << " " <<  nus_atm.EvalFlavor(fl,cz, E);
      }
      file << std::endl;
    }
    file << std::endl;
  }
\end{lstlisting}

\subsection{Atmospheric mode: BSM \textnormal{({\ttf examples/Atm\_BSM})}}
\label{sec:atmBSM}
The atmospheric mode is implemented such  that it can be used with any
{\ttf nuSQUIDS} derived class. In this example, we use the atmospheric mode
with the NSI {\ttf nuSQUIDS} extension shown in the NSI example~\ref{sec:NSI}.

In the folder we include a copy of the NSI header file {\ttf NSI.h}.
The main file is essentially the same as the atmospheric default mode
example with the following changes.
The class type template argument is the {\ttf nuSQUIDS} derived class. The
constructor takes a list of cosine zenith angle values and then the
derived class constructor arguments. 

\begin{lstlisting}[frame=leftline, numbers =
  left,breaklines=true,label = ex:sin1,firstnumber=last]
  double epsilon_mutau=1e-2;
  nuSQUIDSAtm<nuSQUIDSNSI> nus_atm(linspace(czmin,czmax,40),epsilon_mutau,logspace(Emin,Emax,100),numneu,both,true);
\end{lstlisting}

\subsection{Extended neutrino sources \textnormal{({\ttf
      examples/Extended\_Source})}}
\label{sec:extended_source}

Some neutrino evolution problems cannot be formulated as an starting flux
of neutrinos, as shown in the preceding examples.
For example, in the case of solar neutrinos the emission of neutrinos happens
over an extended region of the Sun; the same is true for atmospheric neutrinos
which are produced predominantly 20~km above the Earth surface, but which have varying production
height of tens of kilometers.
{\ttf nuSQuIDS} provides a way to handle these scenarios, which we will illustrate in this example.

A source term is added to the {\ttf nuSQuIDS} kinetic equations Eq.~\eqref{eq:kinetic_equations},
which allows adding (or subtracting) neutrinos from the ensemble.
This term depends on the location of the ensemble in the traversed body.
For this reason it is in the body class where the user needs to implement this term, as it is
a property of the object being traversed rather than the ensemble.

In this example, we will start with an empty neutrino ensemble that traverses a neutrino-emitting
vacuum. We define the following \texttt{EmittingVacuum} class that inherits from \texttt{Vacuum}:

\begin{lstlisting}[frame=leftline, numbers =
  left,breaklines=true,label = ex:sin1,firstnumber=last]
  double epsilon_mutau=1e-2;
class EmittingVacuum: public Vacuum {
  private:
    const unsigned int flavor;
    const double decay_length;
  public :
    EmittingVacuum(unsigned int flavor, double decay_length):
      flavor(flavor),decay_length(decay_length){}
    void injected_neutrino_flux(marray<double,3>& flux, const GenericTrack& track, const nuSQUIDS& nusquids) override {
      double x_cur = track.GetX();
      for(unsigned int ei=0; ei<nusquids.GetNumE(); ei++){
        for(unsigned int rhoi=0; rhoi<nusquids.GetNumRho(); rhoi++){
          for(unsigned int flv=0; flv<nusquids.GetNumNeu(); flv++){
            flux[ei][rhoi][flv] = (flv == flavor) ? exp(-x_cur/decay_length) : 0.0;
          }
        }
      }
    }
  };
\end{lstlisting}

The constructor of this class has the following signature:

\begin{lstlisting}[frame=leftline, numbers =
  left,breaklines=true,label = ex:sin1,firstnumber=last]
    EmittingVacuum(unsigned int flavor, double decay_length)
\end{lstlisting}
where we ask the user to provide two parameters: a flavor (\texttt{decay\_length}) to inject and an injection length (\texttt{decay\_length}).
The flavor is specified in the standard {\ttf nuSQuIDS} flavor notation: 0 for electrons, 1 for muons, and 2 for taus, etc.
The injection of neutrinos in our example will follow an exponential emission energy profile, namely
\begin{equation}
  S(t) = \Pi_\alpha \exp(-x/L),
\end{equation}
where $\Pi_\alpha$ represents the projector of the neutrino state of flavor $\alpha$, $x$ is the distance along the trajectory,  and $L$ is our exponential decay length.
For a given body to emit neutrinos the following member function with the following signature must be overridden:
\begin{lstlisting}[frame=leftline, numbers =
  left,breaklines=true,label = ex:sin1,firstnumber=last]
    void injected_neutrino_flux(marray<double,3>& flux, const GenericTrack& track, const nuSQUIDS& nusquids) override
\end{lstlisting}
This function is given a container, \texttt{flux}, which has three indexes that correspond to the energy index, neutrino or antineutrino index, and flavor index.
Additionally, it is given constant references to the track being traversed and the nuSQUIDS object that is making the query of the body.
The implementation can use the track information to find the current position in the ensemble evolution and can use the nuSQUIDS object
to obtain additional information that may be needed to fill in the container.
A common need is to check the physical energies corresponding to the flavor indices, which can be obtained by calling the {\ttf GetERange} member function of the {\ttf nuSQUIDS} object.
It is important to note that the type of track that is passed by nuSQUIDS to the body object member function is a \texttt{GenericTrack}, which is necessaryto give this function a common signature for all the body types. If needed, the implementation of the member function can cast this to the specific track type for the body class.
Finally, we have added at the end of the signature above the optional keyword \texttt{override}, which ensures that this member function is overriding the correct signature used
internally by {\ttf nuSQuIDS}. This keyword will trigger a compile-time error, if the user accidentally declares a new member function and thus guarantees that the correct member function
is been overridden.

Once your body has been constructed and passed to the {\ttf nuSQUIDS} object by means of the usual \texttt{Set\_Body} and \texttt{Set\_Track} member functions, it is also necessary to also indicate that continuous neutrino injection into the ensemble should be considered in the calculation.
This is not enabled by default, in order to avoid additional calls to virtual member functions that just add zero to the differential equation.
In order to enable this one needs to set
\begin{lstlisting}[frame=leftline, numbers =
  left,breaklines=true,label = ex:sin1,firstnumber=last]
    nus.Set_NeutrinoSources(true);
\end{lstlisting}
where \texttt{nus} is the {\ttf nuSQUIDS} object that we have created.
The result of running this code can be seen in Fig.~\ref{fig:extended_sources}, where the starting conditions are an empty 
ensemble, and the injection is a muon-neutrino only source. The tau and electron contributions that appear are from oscillations.

\begin{figure}[h!]
  \centering
  \includegraphics[width=0.6\textwidth]{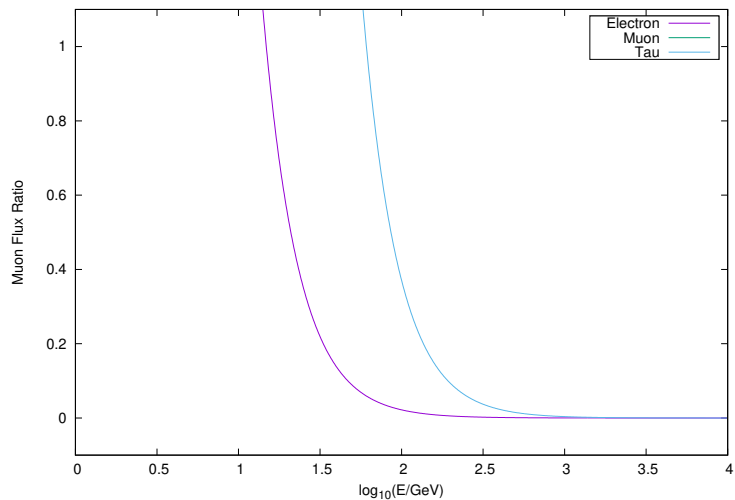}
  \caption{Final flux obtained from the extended source emission as a function of the neutrino energy.}
  \label{fig:extended_sources}
\end{figure}

\section{Description of the code} 
\label{sec:code} 

$\nu$-SQuIDS is a {\ttf C++} code built using the SQuIDS
framework~\citep{SQUIDS}. It is designed to propagate neutrinos
through media while taking into account flavor oscillations and
non-coherent interactions. 

In order to allow the user to compute simple oscillation probabilities
for a single neutrino energy the code has a simplified mode.
In this mode the neutrino energy is fixed and only coherent interactions
are treated.
In this case, only Eq.~\eqref{eq:schrodinger} is relevant for the
neutrino propagation, and the {\ttf nuSQuIDS} class implements {\ttf
  SQuIDS::H0} as in Eq.~\eqref{eq:h0} and {\ttf SQuIDS::HI} as given
in Eq.~\eqref{eq:hi}. 

In the default mode a statistical ensemble of neutrinos is
considered. The ensemble is described by means of a set of {\ttf
  SU\_vector} objects located at fixed  energy nodes spaced over the
energy region under consideration. Besides defining {\ttf SQuIDS::H0}
and {\ttf SQuIDS::HI}, as in the simplified single energy mode, the
following functions are also defined: {\ttf SQuIDS::GammaRho} by
Eqs.~\eqref{eq:gammarhoa} and \eqref{eq:gammarhob}, and {\ttf
  SQuIDS::InteractionsRho} in Eqs.~\eqref{eq:Fterm} and
\eqref{eq:antiFterm}.
Furthermore, in the latter
Eq.~\eqref{eq:antiFterm} $\tau$-regeneration is implemented assuming
instantaneous $\tau$ decay.

While the {\ttf nuSQuIDS} class implements all the necessary
differential equations, one must also specify the neutrino propagation
environment, propagation trajectory, and the relevant cross-sections.  
When interactions are considered the {\ttf nuSQuIDS} instance
will automatically construct appropriate {\ttf NeutrinoCrossSections}
and {\ttf TauDecaySpectra} objects to evaluate cross sections and
$\tau$ physics respectively. However, the user can also replace these
default versions if desired. The user must
explicitly specify the neutrino propagation medium and trajectory
through relevant specialization of {\ttf Body} and {\ttf Body::Track}.
Several implementations of {\ttf Body} and {\ttf Track} covering
common physics cases are supplied with the library. 

Finally, {\ttf nuSQuIDS} provides a set of functions to evaluate the
neutrino ensemble flavor and mass composition. The code also has the
capability to store the system state in an HDF5~\citep{folk1999hdf5}
file for later use.

\subsection{Body \& Track \label{sec:body_track}}

{\ttf Body} and {\ttf Body::Track} are abstract {\ttf C++} classes
which are used to represent the environment in which neutrinos
are propagated ({\ttf Body}) and the propagation path inside it ({\ttf
  Body::Track}). We will frequently use the shorthand {\ttf Track} for {\ttf Body::Track},
 when it should be clear from context that this is the type of {\ttf Track} 
 corresponding to a particular implementation of {\ttf Body}.

The evolution of the system depends on a single parameter which is a
{\ttf double} member of {\ttf Track}. The
interplay between the evolution of this parameter and the trajectory
inside a given body is what is encoded in the {\ttf Body} and 
{\ttf Body::Track} classes. 
Namely, a {\ttf Body} is defined as a matter density and electron
fraction depending on the position in a 3 dimensional space, $\rho_m(\vec{r})$ and
$Y_e(\vec{r})$, the role of the {\ttf Track} would be
equivalent to the trajectory ($\vec{r}(x)$) and the current position ($x$).

In the context of the code, the two main virtual functions that the
user should provide in order to define a non-trivial {\ttf Body} are:
\begin{itemize}
\item 
  \begin{lstlisting}
    virtual double density(std::shared_ptr<Track>);
  \end{lstlisting}
This function returns the density in ${\rm g}/{\rm cm}^3$ for a
given {\ttf Track} position.
\item 
  \begin{lstlisting}
    virtual double ye(std::shared_ptr<Track>);
  \end{lstlisting}
Returns the electron fraction at a given {\ttf Track} position.
\end{itemize}

In order to allow the user to store the information of the
object the following members are important,

\begin{itemize}
\item  
  \begin{lstlisting}
    std::vector<double> BodyParams;
  \end{lstlisting}
  Double vector that contains all the parameters that are
  needed to compute the density and electron fraction from the parameter
  {\ttf x}.
  
\item  
  \begin{lstlisting}
    bool is_constant_density = false;
  \end{lstlisting}  
  This variable is {\ttf true} if the density of the object is constant. This is
  used to set the fast computations internally.
\end{itemize}

Other public functions of the {\ttf Body} object are,

\begin{itemize}
\item  
  \begin{lstlisting}
    virtual void Serialize(hid_t group) const=0;
  \end{lstlisting}
  This is an abstract function whose argument is an HDF5 location
  where the user should store the body properties.

  \item  
  \begin{lstlisting}
    static std::shared_ptr<Body> Deserialize(hid_t group);
  \end{lstlisting}
  This is an abstract function whose argument is an HDF5 location
  with the body information to be used for the user to recover the body.
  
\item  
  \begin{lstlisting}
    unsigned int GetId() const {return 0;}
  \end{lstlisting}
  It returns the Id of the body which is hard-coded by the user.

\item  
  \begin{lstlisting}
    std::string GetName() const {return name;}
  \end{lstlisting}
  It returns the name of the object which is hard-coded by the user.

\item  
  \begin{lstlisting}
    const std::vector<double>& GetBodyParams() const
    { return BodyParams;}
  \end{lstlisting}
  It returns a constant reference to the vector of parameters that define body. 
  
  \item  
  \begin{lstlisting}
    virtual bool IsConstantDensity() const
    {return is_constant_density;}
  \end{lstlisting}
  It returns true if the body has constant density, false otherwise.
  
\item  
  \begin{lstlisting}
    virtual void SetIsConstantDensity(bool icd)
    {is_constant_density = icd;}
  \end{lstlisting}
  Set true or false the constant density. 
\end{itemize}

Furthermore, the {\ttf Track} object has the following protected
variables, which are the initial, final and current
evolution parameter values, by default in units of eV$^{-1}$
\begin{itemize}
\item  
  \begin{lstlisting}
  double x;
  \end{lstlisting}
  Current position.
\item  
  \begin{lstlisting}
    double xini;
  \end{lstlisting}
  Initial position.
\item  
  \begin{lstlisting}
    double xend;
  \end{lstlisting}
  Final position.
\end{itemize}
And the following public functions are provided,
\begin{itemize}
\item  
  \begin{lstlisting}
    Track(double x,double xini, double xend):
    x(x), xini(xini),xend(xend) {}
  \end{lstlisting}
  Constructor where we specify current, initial, and final positions. 
\item  
  \begin{lstlisting}
    Track(double xini, double xend):
    Track(xini,xini,xend) {}
  \end{lstlisting}
  Constructor where we specify initial and final positions. The
  current position is set to the initial. 
  \item  
  \begin{lstlisting}
    virtual void Serialize(hid_t group) const=0;
  \end{lstlisting}
  This is an abstract function whose argument is an HDF5 location
  where the user should store the track properties.
  \item  
  \begin{lstlisting}
    static std::shared_ptr<Body::Track> Deserialize(hid_t group);
  \end{lstlisting}
  This is an abstract function whose argument is an HDF5 location
  with the track information to be used for the user to recover the track.
\item  
  \begin{lstlisting}
    void SetX(double y);
  \end{lstlisting}
  Sets the current position along the trajectory.  
\item  
  \begin{lstlisting}
    double GetX() const;
  \end{lstlisting}
  Returns the current value of the evolution parameter {\ttf x}.
\item  
  \begin{lstlisting}
    double GetInitialX() const;
  \end{lstlisting}    
  Returns the initial value of the evolution parameter {\ttf x}.
\item  
  \begin{lstlisting}
    double GetFinalX() const;
  \end{lstlisting}          
  Returns the final value of the evolution parameter {\ttf x}
\item  
  \begin{lstlisting}
    static std::string GetName() {return "BodyTrack";};
  \end{lstlisting}
  Returns the name of the track object hard-coded by the user. 
\item  
  \begin{lstlisting}
    std::vector<double> GetTrackParams() const 
  \end{lstlisting}           
  Returns a vector of doubles that define the trajectory.
\item 
   \begin{lstlisting}
     virtual void FillDerivedParams(std::vector<double>& TrackParams)
     const{};
  \end{lstlisting}           
  Should be implemented by derived classes to append their
  additional parameters to TrackParams
  
\item 
  \begin{lstlisting}
    void ReverseTrack() 
  \end{lstlisting}
  Interchanges initial and final positions.
\end{itemize}

Since {\ttf Body} and {\ttf Track} are abstract classes they themselves do not perform any task, but rather their specializations specify the real neutrino propagation environment and how it relates to its trajectory. $\nu$-SQuIDS implements the most commonly used environments and trajectory configurations. The user can create new classes in order to extend $\nu$-SQuIDS functionality.

The {\ttf Body} classes specializations implemented in $\nu$-SQuIDS
are the following: {\ttf Vacuum}, {\ttf ConstantDensity}, {\ttf
  VariableDensity}, {\ttf Earth}, {\ttf EarthAtm}, {\ttf Sun}, and {\ttf SunASnu}.

In the following sub-sections we will describe the specific
constructors and the functions that are not members of the parent class.

\subsubsection{Vacuum \label{sec:vacuum}}

\begin{itemize}
\item {\ttf Vacuum}
  \begin{lstlisting}
    Vacuum():Body(){}
  \end{lstlisting}
  Initializes a {\ttf Vacuum} environment. 
\item {\ttf Vacuum::Track}
  \begin{lstlisting}
    Track(double x,double xini,double xend):
    Body::Track(x,xini,xend){};
    Track(double xini,double xend):
    Track(xini,xini,xend){};
    Track(double xend):Track(0.0,xend){}
  \end{lstlisting}
  Initialize the corresponding {\ttf Track}, setting the current ({\ttf
    x}), initial ({\ttf xini}), and final ({\ttf xend}) neutrino position in ${\rm eV}^{-1}$.
\end{itemize}

\subsubsection{ConstantDensity \label{sec:constant_density}}

\begin{itemize}
\item {\ttf ConstantDensity}
  \begin{lstlisting}
    ConstantDensity(double density,double ye);
  \end{lstlisting}
  Initializes a {\ttf ConstantDensity} environment with constant
  density ({\ttf density} in ${\rm g}/{\rm cm}^3$) and electron
  fraction ({\ttf ye}).
\item {\ttf ConstantDensity::Track}
  \begin{lstlisting}
    Track(double x,double xini,double xend):
    Body::Track(x,xini,xend){};
    Track(double xini,double xend):
    Track(xini,xini,xend){};
    Track(double xend):Track(0.0,xend){}
  \end{lstlisting}
  Initialize the corresponding {\ttf Track}, setting the current ({\ttf
  x}), initial ({\ttf xini}), and final ({\ttf xend}) neutrino position in ${\rm eV}^{-1}$.
\end{itemize}

\subsubsection{VariableDensity \label{sec:variable_density}}

\begin{itemize}
\item {\ttf VariableDensity}
  \begin{lstlisting}
    VariableDensity(std::vector<double> x,
        std::vector<double> density,
        std::vector<double> ye);
  \end{lstlisting}
  Initializes a {\ttf VariableDensity} environment given three equal size arrays specifying the density and electron fraction at given positions. An object will be created that interpolates using {\ttfamily gsl\_spline}~\citep{gough2009gnu} along the {\ttf x} array to get the density and electron fraction as continuous functions.
  \item {\ttf VariableDensity::Track}
  \begin{lstlisting}
    Track(double x,double xini,double xend):
    Body::Track(x,xini,xend){};
    Track(double xini,double xend):
    Track(xini,xini,xend){};
    Track(double xend):Track(0.0,xend){}
  \end{lstlisting}
  Initialize the corresponding {\ttf Track}, setting the current ({\ttf
  x}), initial ({\ttf xini}), and final ({\ttf xend}) neutrino position in ${\rm eV}^{-1}$.
\end{itemize}

\subsubsection{Earth \label{sec:earth}}
The {\ttf Earth} body specification is designed to propagate neutrinos
in the Earth from two points on the surface. Since the Earth in the
PREM model is assumed to be spherically symmetric the length of the
path is enough to determine the trajectory. {\ttfamily AkimaSpline} is used to interpolate $\rho$ and $y_e$ as a function of radius to the earth center.
\begin{itemize}
\item {\ttf Earth}
  \begin{lstlisting}
    Earth();
  \end{lstlisting}
  Initializes an {\ttf Earth} environment as defined by the PREM~\citep{dziewonski1981preliminary}.
  \begin{lstlisting}
    Earth(std::string earthmodel);
  \end{lstlisting}
  Initializes an {\ttf Earth} environment as defined by a table given in the file specified by {\ttf filepath}. The table should have three columns: radius (where 0 is center and 1 is surface), density (${\rm g}/{\rm cm}^3$), and $y_e$ (dimensionless). 
  \begin{lstlisting}
    Earth(std::vector<double> x,std::vector<double> rho,
    std::vector<double> ye);
  \end{lstlisting}
  Initialize an {\ttf earth} whose radial density is specified by the
  values in the vector {\ttf rho} in ${\rm g}/{\rm cm}^3$, the
  electron fraction in the vector {\ttf ye}, and the radial positions
  in the vector {\ttf x} in centimeters. 

  \begin{lstlisting}
    double GetRadius() const;
  \end{lstlisting}
  Returns the radius of the Earth in eV$^{-1}$.

\item {\ttf Earth::Track}
  \begin{lstlisting}
    Track(double x,double xini,double xend,double baseline):
    Body::Track(x,xini,xend),baseline(baseline){};
    Track(double xini,double xend,double baseline):
    Track(xini,xini,xend,baseline){};
  \end{lstlisting}
  Initialize the corresponding {\ttf Track} setting the current ({\ttf
    x}), initial ({\ttf xini}), and final ({\ttf xend}) neutrino position in ${\rm eV}^{-1}$.
  
  \begin{lstlisting}
    Track(double baseline):Track(0.,baseline,baseline){}
  \end{lstlisting}
  Constructor that sets the path in the earth for a given baseline
  {\ttf baseline}.

  \begin{lstlisting}
    double GetBaseline() const;
  \end{lstlisting}
  Returns the baseline in eV$^{-1}$.
    
\end{itemize}

\subsubsection{EarthAtm\label{sec:earthatm}}
This body includes both the density profile of the Earth and a simple model of its atmosphere. 
Paths end at the surface of the Earth, and are expressed in terms of their local zenith angle at the endpoint. 
The deafult atmosphere height is 22 km, and a scale height of 7.594 km is used for its density. 
\begin{itemize}
\item {\ttf EarthAtm}
  \begin{lstlisting}
    EarthAtm();
  \end{lstlisting}
  Initializes an {\ttf Earth} environment as defined by the PREM~\citep{dziewonski1981preliminary}.
  \begin{lstlisting}
    EarthAtm(std::string earthmodel);
  \end{lstlisting}
  Initializes an {\ttf EarthAtm} environment as defined by a table given in the file specified by {\ttf filepath}. The table should have three columns: relative radius (dimensionless, where 0 is the center and 1 the is surface), density (${\rm g}/{\rm cm}^3$), and electron fraction $y_e$ (dimensionless). 
  \begin{lstlisting}
    EarthAtm(std::vector<double> x,std::vector<double> rho,
    std::vector<double> ye);
  \end{lstlisting}
  Initialize an {\ttf EarthAtm} whose radial density is specified by the
  values in the vector {\ttf rho} in ${\rm g}/{\rm cm}^3$, the
  electron fraction in the vector {\ttf ye}, and the relative radial positions
  in the vector {\ttf x}. 
  \begin{lstlisting}
    void SetAtmosphereHeight(double height);
  \end{lstlisting}
  Change the height of the top of the atmosphere. 
  \begin{lstlisting}
    EarthAtm::Track MakeTrack(double phi);
  \end{lstlisting}
  Create a track which arrives at the surface with the given zenith angle and an initial position at the top of the atmosphere.
  \begin{lstlisting}
    EarthAtm::Track MakeTrackWithCosine(double cosphi);
  \end{lstlisting}
  The same as {\ttf MakeTrack}, but takes the cosine of the zenith angle, which is a more convenient variable in many contexts.

\item {\ttf EarthAtm::Track}
  \begin{lstlisting}
    Track(double x, double phi, 
          double earth_radius, double atmheight);
    Track(double phi, double earth_radius, double atmheight);
  \end{lstlisting}
  Initialize the corresponding {\ttf Track} by specifying the zenith
  angle in radians ({\ttf phi}), radius of the Earth ({\ttf earth\_radius}) in eV$^{-1}$, height of the atmosphere ({\ttf atmheight}) in eV$^{-1}$, and the current position along the
  track ({\ttf x}) in eV$^{-1}$.
  
  \begin{lstlisting}
    double GetBaseline() const;
  \end{lstlisting}
  Returns the baseline in eV$^{-1}$.

  \begin{lstlisting}
    static Track makeWithCosine(double cosphi, 
                                double earth_radius, 
                                double atmheight);
  \end{lstlisting}
  Create a track with the cosine of the zenith angle ({\ttf cosphi}) instead of the angle. 
  
\end{itemize}

\subsubsection{{Sun}\label{sec:sun}}
This specification of the object allows to define the Sun and radial
trajectories that start from the center of the sun.
\begin{itemize}
\item {\ttf Sun}
  \begin{lstlisting}
    Sun();
  \end{lstlisting}
  Initializes a {\ttf Sun} environment as defined by the {\it
    Standard Solar Model}~\citep{bahcall2005new}.
  
  \begin{lstlisting}
    Sun(std::vector<double> x,std::vector<double> rho,
        std::vector<double> xh);
  \end{lstlisting}
  Initialize a {\ttf Sun} whose radial density is specified by the
  values in the vector {\ttf rho} in ${\rm g}/{\rm cm}^3$, the
  electron fraction in the vector {\ttf ye}, and the radial positions
  in the vector {\ttf x} in centimeters. 

  \begin{lstlisting}
    Sun(std::string sunmodel);
  \end{lstlisting}
  Initialize a {\ttf Sun} object from a text file that should have the same format as the {\it
  Standard Solar Model}~\citep{bahcall2005new}. Namely, the second column one must run from zero to one representing
  the center and surface of the Sun respectively. The
  fourth column must contain the Sun density in ${\rm g}/{\rm cm}^3$ at
  a given position, while the seven column must contain
  the hydrogen fraction, $r_h$, which is related to the electron fraction by $ye = 0.5*(1.0+r_h(r))$.
  Internally we assume that the solar radius is 695980.0~kilometers.

\item {\ttf Sun::Track}
  \begin{lstlisting}
    Sun::Track(double xend);
  \end{lstlisting}
  This constructor sets the trajectory starting at the Sun center and ending at a distance 
  {\ttf xend} from it.
  \item {\ttf Sun::Track}
  \begin{lstlisting}
    Sun::Track(double xini, double xend);
  \end{lstlisting}
  Initialize the corresponding {\ttf Track} by the initial position in the sun {\ttf xini} and {\ttf xend} along the solar radius.
\end{itemize}

\subsubsection{{SunASnu}\label{sec:sunasnu}}
This specification of the object allows to define the Sun and the
trajectories with different impact parameters.
\begin{itemize}
\item {\ttf Sun}
  \begin{lstlisting}
    SunASun();
  \end{lstlisting}
  Initializes an {\ttf Sun} environment as defined by the {\it
    Standard Solar Model}~\citep{bahcall2005new}.
  \begin{lstlisting}
    SunASnu(std::vector<double> x,std::vector<double> rho,
            std::vector<double> xh);
  \end{lstlisting}
  Initialize an {\ttf Sun} whose radial density is specified by the
  values in the vector {\ttf rho} in ${\rm g}/{\rm cm}^3$, the
  electron fraction in the vector {\ttf ye}, and the radial positions
  in the vector {\ttf x} in centimeters. 

  \begin{lstlisting}
    SunASnu(std::string sunmodel);
  \end{lstlisting}
  Initialize a {\ttf SunASnu} object from a text file that should have the same format as the {\it
  Standard Solar Model}~\citep{bahcall2005new}. Namely, the second column one must run from zero to one representing
  the center and surface of the Sun respectively. The
  fourth column must contain the Sun density in ${\rm g}/{\rm cm}^3$ at
  a given position, while the seven column must contain
  the hydrogen fraction, $r_h$, which is related to the electron fraction by $ye = 0.5*(1.0+r_h(r))$.
  Internally we assume that the solar radius is 695980.0~kilometers.

\item {\ttf SunASun::Track}
  \begin{lstlisting}
    Track(double x,double xini,double b_impact);
    Track(double xini,double b_impact):Track(xini,xini,b_impact){};
    Track(double b_impact_):Track(0.0,b_impact_){}
  \end{lstlisting}
  This constructor sets the trajectory starting at a distance 
  {\ttf xini}, with current position {\ttf x}, and impact factor {\ttf
    b\_impact} in eV$^{-1}$.
\end{itemize}

\subsection{NeutrinoCrossSections\label{sec:neutrino_cross_section}}
\label{sec:xs}
This object can be queried to obtain neutrino cross section information used when considering neutrino non-coherent interactions. The {\ttf NeutrinoCrossSections} is a base abstract class, which the user has to subclass and implement the relevant neutrino cross section for the problem at hand. The user must specify the total cross section per flavor and per interaction type (charge and neutral current), as well as the single differential cross sections with respect to the outgoing neutrino energy.

\subsubsection{NeutrinoCrossSections\label{sec:neutrino_cross_section_class}}

First, we define enumerations to label flavor, neutrino, and interaction type.
\begin{itemize}
  \item {\ttf NeutrinoFlavor}
  \begin{lstlisting}
    enum NeutrinoFlavor
    {electron = 0, muon = 1, tau = 2, sterile = 3};
 \end{lstlisting}
  Enumeration that is used to specify the neutrino flavor.
  \item {\ttf NeutrinoType}
  \begin{lstlisting}
    enum NeutrinoType {neutrino = 0, antineutrino = 1};
  \end{lstlisting}
  Enumeration used to specify {\ttf neutrino} and {\ttf antineutrino} particle type.
  \item {\ttf Current}
  \begin{lstlisting}
    enum Current {CC, NC, GR};
  \end{lstlisting}
  Enumeration used to specify charged ({\ttf CC}), neutral ({\ttf NC}) current interactions, 
  and Glashow resonant interactions ({\ttf GR}).
\end{itemize}

Second we list the public abstract virtual functions.

\begin{itemize}
  \item Total cross section
  \begin{lstlisting}
    virtual double TotalCrossSection(double Enu,
    	NeutrinoFlavor flavor, NeutrinoType neutype,
    	Current current) const;
  \end{lstlisting}
  Abstract virtual function that given a neutrino energy ({\ttf Enu})
    in eV, neutrino flavor ({\ttf flavor}), neutrino type ({\ttf
    neutype}), and interaction type ({\ttf current}) returns the 
  total cross section in ${\rm cm}^2$.
  \item Single differential cross section
  \begin{lstlisting}
    virtual double SingleDifferentialCrossSection(double E1,
        double E2, NeutrinoFlavor flavor, NeutrinoType neutype,
    	Current current) const;
  \end{lstlisting}
  Abstract virtual function that given an incident neutrino energy
  ({\ttf E1}) in eV, outgoing neutrino energy ({\ttf E2}) in eV,
    neutrino flavor ({\ttf flavor}), neutrino type ({\ttf neutype}), and
  interaction type ({\ttf current}) returns the differential cross
  section with respect to the outgoing neutrino energy in ${\rm
    cm}^2{\rm GeV}^{-1}$.  
  \item Double differential cross section
  \begin{lstlisting}
    virtual double DoubleDifferentialCrossSection(double E, 
        double x, double y,NeutrinoFlavor flavor,
        NeutrinoType neutype, Current current) const;
  \end{lstlisting}
  Virtual function such that given the neutrino energy ({\ttf E}),
  Bjorken-x ({\ttf x}), and y ({\ttf y}) should return the double
   differential cross section. Its implementation is not required to
   run {\ttf nuSQUIDS} and, by default, when evaluated, unless
   overwritten, it will throw an error.
 \item Averaged cross sections
   \begin{lstlisting}
     virtual double AverageTotalCrossSection(double EMin,
     double EMax, NeutrinoFlavor flavor, NeutrinoType neutype,
     Current current) const;

     virtual double AverageSingleDifferentialCrossSection(double E1,
     double E2Min, double E2Max, NeutrinoFlavor flavor,
     NeutrinoType neutype, Current current) const;
   \end{lstlisting}
   Averaged total and single differential cross section in energy
   range {\ttf Emin}-{\ttf Emax} and {\ttf E2min}-{\ttf E2max} respectively. 
\end{itemize}

\subsubsection{NullCrossSections\label{sec:null_cross_section}}
This class is a simple implementation of the a null cross section.
\begin{lstlisting}
  NullCrossSections(){}

  double TotalCrossSection(double Enu, NeutrinoFlavor flavor,
  NeutrinoType neutype, Current current) const override { return 0;}

  double SingleDifferentialCrossSection(double E1, double E2,
  NeutrinoFlavor flavor, NeutrinoType neutype, Current current)
  const override { return 0;}
  
  double DoubleDifferentialCrossSection(double E, double x, double y,
  NeutrinoFlavor flavor, NeutrinoType neutype, Current current)
  const override { return 0;}
\end{lstlisting}

\subsubsection{NeutrinoDISCrossSectionsFromTables\label{sec:neutrino_cross_section_from_tables}}

This class uses pre-calculated deep inelastic scattering (DIS) cross section tables
which are provided by {\ttf nuSQuIDS} or by the user. 
In the code two different cross sections are available by default:
{\ttf csms.h5} is a perturbative QCD next-to-leading order calculation~\cite{CooperSarkar:2011pa} using the HERA
parton distribution functions~\cite{Chekanov:2002pv}, recalculated for proton and neutron targets, 
and updated to account for the non-zero mass of the tau, reducing the $\nu_\tau$ 
cross sections at low energies compared to the originally published version. 
{\ttf nusigma}~\cite{nusigma} is a first order QCD
calculation using the {\ttf CTEQ6} parton distribution functions on an
iso-scalar target. In the {\ttf csms} calculation the mass of the tau
is neglected, thus the neutrinos cross section is the same for all
flavors; this is not the case for the {\ttf nusigma} calculation.
Both correspond to deep inelastic scattering which is the dominant
neutrino interaction with nucleons above $O({\rm 10~GeV})$. 

The cross sections are loaded from tables included in
{\ttfamily nuSQUIDS/data/xsections/}, or provided by the user. 
The cross section object can be constructed from a single {\ttf HDF5}
file that contains both single and total cross sections or by a set of
twelve text files, containing all cross sections related to a single target species 
(e.g. all cross sections for DIS neutrino interactions with protons). 

When an {\ttf HDF5} file is used, it must have an internal structure as shown 
in Table~\ref{tab:hdf_cross_layout}. All cross sections must be tabulated with the same set of incident neutrino energies, and each singly-differential cross section must be tabulated for the same set of out-going lepton energies, expressed in the $z$ variable described in Sec.~\ref{sec:differential_tabulation}. 
These common energy and $z$ values are stored in the one-dimensional {\ttf energies} and {\ttf zs} datasets. 
For the energies, more specifically, the base-10 logarithms of the energies in GeV are stored. 
All total cross section tables are one-dimensional datasets containing the base-10 logarithm of the cross section values in cm$^2$, and all singly-differential cross section tables are two-dimensional datasets with dimensions incident and out-going energy containing the base-10 logarithm of the cross section values in cm$^2$. 

When text files are used, they must be organized into three groups, 
one for each active neutrino flavor, with a corresponding 
filename component ({\ttf electron\_}, {\ttf muon\_}, or {\ttf tau\_}). 
The four text files within each flavor set must end with the suffixes
{\ttf nu\_dsde\_CC.dat}, {\ttf nubar\_dsde\_NC.dat}, {\ttf nu\_sigma\_CC.dat}, and
{\ttf nubar\_sigma\_NC.dat} for the singly differential neutrino, singly differential 
antineutrino, total neutrino, and total antineutrino cross sections, respectively. 
Each total cross section file must contain two columns: 
The incident neutrino energy in GeV and the total cross section in cm$^2$. 
Each singly-differential cross section file must have three columns:
The incident neutrino energy in GeV, 
outgoing neutrino energy encoded via the unitless $z$ variable described in the next section, 
and the differential cross section in cm$^2$. 

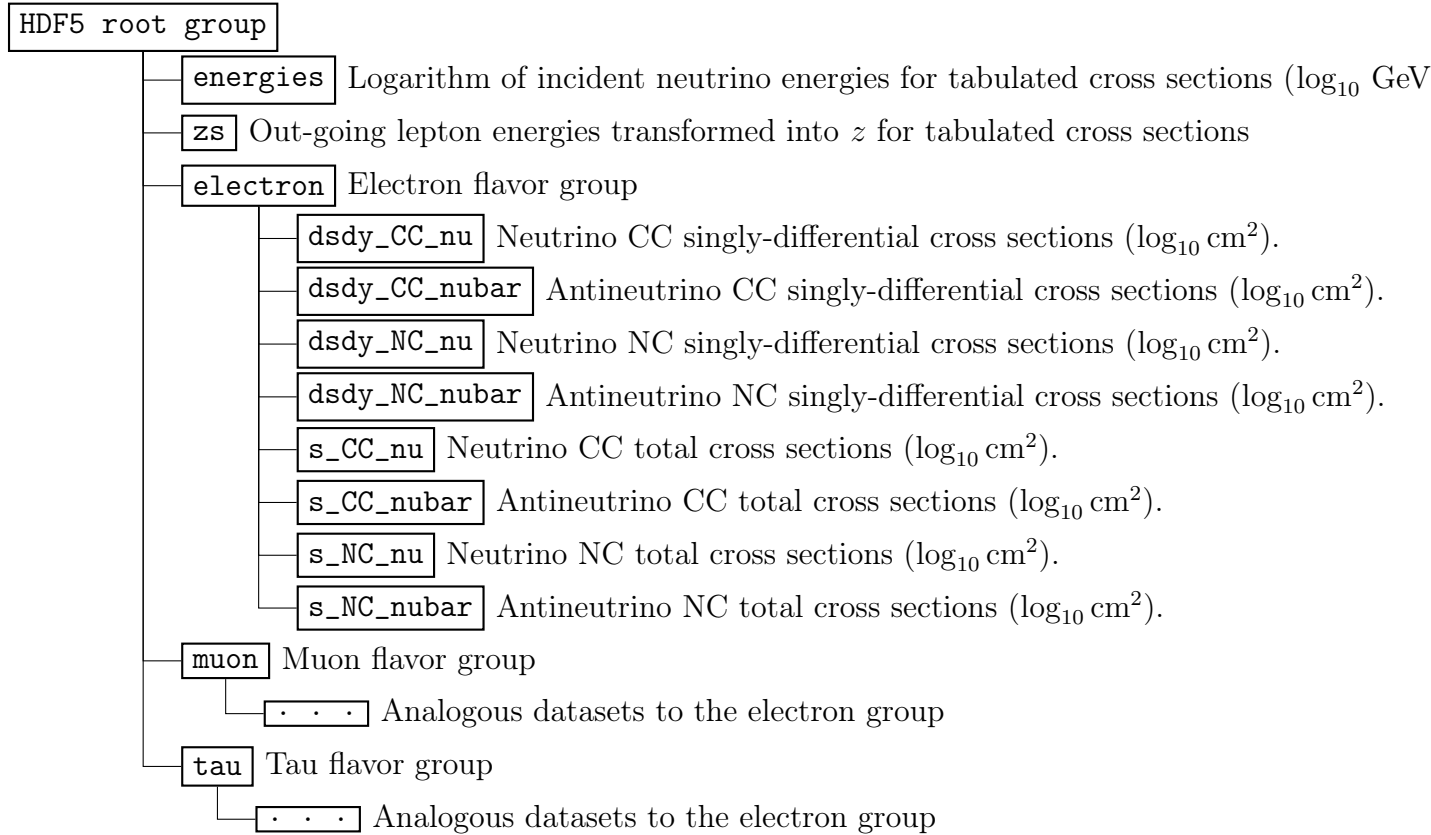
\begin{figure}[]
  \centering
  \begin{tikzpicture}[
    grow via three points={one child at (0.5,-0.7) and
      two children at (0.5,-0.7) and (0.5,-1.4)},
    edge from parent path={(\tikzparentnode.south) |- (\tikzchildnode.west)}]
    \node {{\ttf HDF5 root group}}
    child { node [label=right:{Logarithm of incident neutrino energies for tabulated cross sections ($\log_{10}$ GeV)}]  {\ttf energies}}
    child { node [label=right:{Out-going lepton energies transformed into $z$ for tabulated cross sections}]  {\ttf zs}}
    child { node [label=right:{Electron flavor group}] {\ttf electron}
      child { node [label=right:{Neutrino CC singly-differential cross sections (${\log_{10}\rm cm^2}$).}]  {\ttf dsdy\_CC\_nu}}
      child { node [label=right:{Antineutrino CC singly-differential cross sections (${\log_{10}\rm cm^2}$).}]  {\ttf dsdy\_CC\_nubar}}
      child { node [label=right:{Neutrino NC singly-differential cross sections (${\log_{10}\rm cm^2}$).}]  {\ttf dsdy\_NC\_nu}}
      child { node [label=right:{Antineutrino NC singly-differential cross sections (${\log_{10}\rm cm^2}$).}]  {\ttf dsdy\_NC\_nubar}}
      child { node [label=right:{Neutrino CC total cross sections (${\log_{10}\rm cm^2}$).}]  {\ttf s\_CC\_nu}}
      child { node [label=right:{Antineutrino CC total cross sections (${\log_{10}\rm cm^2}$).}]  {\ttf s\_CC\_nubar}}
      child { node [label=right:{Neutrino NC total cross sections (${\log_{10}\rm cm^2}$).}]  {\ttf s\_NC\_nu}}
      child { node [label=right:{Antineutrino NC total cross sections (${\log_{10}\rm cm^2}$).}]  {\ttf s\_NC\_nubar}}
    }
    child [missing] {}				
    child [missing] {}				
    child [missing] {}
    child [missing] {}
    child [missing] {}			
    child [missing] {}
    child [missing] {}
    child [missing] {}
    child { node [label=right:{Muon flavor group}] {\ttf muon}
      child { node [label=right:{Analogous datasets to the electron group}]  {\ttf .\ .\ .}}
    }
    child [missing] {}
    child { node [label=right:{Tau flavor group}] {\ttf tau}
      child { node [label=right:{Analogous datasets to the electron group}]  {\ttf .\ .\ .}}
    }
    ;
  \end{tikzpicture}
  \caption{HDF5 cross section format. `CC' and `NC' are abbreviations for charged-current and neutral current, respectively. For the common case that two flavors (e.g. electron and muon) have indistinguishable cross section values, one group may be a link to the other within the file.}
  \label{tab:hdf_cross_layout}
\end{figure}

\subsubsection{Constructors}

\begin{itemize}
\item Default constructor.
  \begin{lstlisting}
    NeutrinoDISCrossSectionsFromTables();
  \end{lstlisting}
  This constructor load the defaults {\ttf csms.h5} file
\end{itemize}

The files to load the different cross-sections can be specified
explicitly as follows. 
\begin{lstlisting}
  NeutrinoDISCrossSectionsFromTables csh5('./csms.h5');
  NeutrinoDISCrossSectionsFromTables cstxt('./nusigma_');
\end{lstlisting}

\subsubsection{Functions}

\begin{itemize}
\item Total cross sections
  \begin{lstlisting}
    double TotalCrossSection(double Enu,
    	NeutrinoFlavor flavor, NeutrinoType neutype,
    	Current current) const;
      \end{lstlisting}
      
  Returns the total cross section at an energy {\ttf Enu} in eV, neutrino
  flavor {\ttf flavor}, neutrino type {\ttf neutype}, and {\ttf
    current} can be either {\ttf NC} or {\ttf CC} for neutral or
  charge current DIS cross sections. This function performs a linear interpolation in
  the logarithm of the neutrino energy to estimate the cross section from a reference table.
     
\item Single differential cross sections              
  \begin{lstlisting}
    double SingleDifferentialCrossSection(double E1, double E2,
    	NeutrinoFlavor flavor, NeutrinoType neutype,
    	Current current) const;
  \end{lstlisting}
      Function that given an incident neutrino energy ({\ttf E1}) in eV an outgoing lepton energy ({\ttf E2}), as well as neutrino flavor,
       type, and process, returns the differential cross section with
       respect to the outgoing lepton energy in ${\rm cm}^2/{\rm
         GeV}$. This function interpolates the cross section from a reference table by means of a bi-linear interpolation.
\end{itemize}

\subsubsection{GlashowResonanceCrossSection\label{sec:neutrino_cross_section_glashow}}

This class implements the formulas in~\citep{Gandhi:1998ri} in order to calculate the electron antineutrino Glashow resonance cross section contribution.

\subsubsection{Constructors}

\begin{itemize}
\item Default constructor.
  \begin{lstlisting}
    GlashowResonanceCrossSection();
  \end{lstlisting}
\end{itemize}

\subsubsection{Functions}

\begin{itemize}
\item Total cross sections
  \begin{lstlisting}
    double TotalCrossSection(double Enu,
    	NeutrinoFlavor flavor, NeutrinoType neutype,
    	Current current) const;
  \end{lstlisting}
     Returns the total cross section in cm$^2$ at an energy {\ttf Enu} in eV, neutrino flavor {\ttf flavor}, and neutrino type {\ttf neutype}.
     If the flavor is not electron and neutrino type is not antineutrino it returns zero.
\item Single differential cross section
  \begin{lstlisting}
    double SingleDifferentialCrossSection(double E1, double E2,
    	NeutrinoFlavor flavor, NeutrinoType neutype,
    	Current current) const;
  \end{lstlisting}
  Returns the single differential cross section in cm$^2$/GeV for an incident neutrino energy ({\ttf
    E1}) in eV an outgoing neutrino energy ({\ttf E2}) in eV, neutrino flavor ({\ttf flavor}),
  and neutrino type ({\ttf neutype}).
     If the flavor is not electron and neutrino type is not antineutrino it returns zero.
\end{itemize}

\subsection{TauDecaySpectra\label{sec:tau_decay_spectra}}

This object can be queried to obtain $\tau$ decay physics into leptons
and hadrons. The formulas implemented in this class were taken
from~\citep{Dutta:2000jv}. It is only used when $\tau$-regeneration is
activated and it returns the following quantities on the energy
nodes, 
\begin{equation}
\frac{dN^{lep/had}_{dec} (E_\tau, E_\nu)}{dE_\nu} {\rm ~~and~~ }
\frac{d\bar{N}^{lep/had}_{dec} (E_\tau, E_\nu)}{dE_\nu},
\label{eqn:tau-dist}
\end{equation}
{\it i.e.} the neutrino and antineutrino spectral distributions from $\tau$
leptonic and hadronic decay modes.

\subsubsection{Constructors and Initializing Functions}

\begin{itemize}
\item Default constructor.
  \begin{lstlisting}
    TauDecaySpectra();
  \end{lstlisting}
\item Constructor and initializing function with memory reservation.
  \begin{lstlisting}
    TauDecaySpectra(marray<double,1> E_range);
    void Init(marray<double,1> E_range);
  \end{lstlisting}
This constructor and initialization function calculates and stores the
$\tau$ decay spectra on nodes specified by the one dimensional array
{\ttf E\_range} in eV.
\end{itemize}

\subsubsection{Functions}

The following functions assume that the $\tau$ and $\bar{\tau}$ have
the same decay distribution. 

\begin{itemize}
\item (Anti)Neutrino spectra with respect to neutrino energy.
  \begin{lstlisting}
    double dNdEnu_All(int e1,int e2) const;
  \end{lstlisting}
  Returns neutrino decay spectra evaluated from energy node
  {\ttfamily e1} to energy node {\ttfamily e2} when $\tau$ decays into
  leptons or hadrons.  
  \begin{lstlisting}
    double dNdEnu_Lep(int e1,int e2) const;
  \end{lstlisting}
  Returns neutrino decay spectra evaluated between energy nodes
  {\ttfamily e1} and {\ttfamily e2} when $\tau$ decays into leptons. 
\item (Anti)Neutrino spectra with respect to $\tau$ energy.
  \begin{lstlisting}
    double dNdEle_All(int e1,int e2) const;
  \end{lstlisting}
  Returns neutrino decay spectra evaluated between energy nodes
  {\ttfamily e1} and {\ttfamily e2} when $\tau$ decays into leptons
  or hadrons with respect to the initial $\tau$ energy.
  \begin{lstlisting}
    double dNdEle_Lep(int e1,int e2) const;
  \end{lstlisting}
  Returns neutrino decay spectra evaluated between energy nodes
  {\ttfamily e1} and {\ttfamily e2} when $\tau$ decays into
  leptons with respect to the initial $\tau$ energy.
\item Get the $\tau$ branching ratio to leptons.
  \begin{lstlisting}
    double GetTauToLeptonBranchingRatio() const;
  \end{lstlisting}
  Returns the $\tau$ branching ratio to leptons.
\item Get the $\tau$ branching ratio to hadrons.
  \begin{lstlisting}
    double GetTauToHadronBranchingRatio() const;
  \end{lstlisting}
  Returns the $\tau$ branching ratio to hadrons.
\end{itemize}

\subsection{nuSQUIDS\label{sec:nusquids}}

This object is a specialization of the {\ttf SQUIDS} class~\citep{SQUIDS} that implements the
differential equations as described in Sec.~\ref{sec:theory}. In
particular, it is used to specify the propagation {\ttf Body} and its
associated {\ttf Track}. Moreover, it uses the {\ttf
  NeutrinoCrossSections} and {\ttf TauDecaySpectra} in order to
evaluate the neutrino cross sections and $\tau$ decay spectra; the
latter is only used when {\it $\tau$} regeneration is
enabled. Furthermore, it enables the user to modify the neutrino
oscillation parameters as well as the differential equation numerical
precision. Finally, it also has the capability to create and read HDF5
files that store the program results and configuration. 

\subsubsection{Constructors}

\begin{itemize}
\item Default constructor.
  \begin{lstlisting}
    nuSQUIDS();
  \end{lstlisting}
\item Move constructor.
  \begin{lstlisting}
    nuSQUIDS(nuSQUIDS&&);
  \end{lstlisting}
Constructs a {\ttf nuSQUIDS} object from an r-value reference.  
\item Single energy mode constructor.
  \begin{lstlisting}
    nuSQUIDS(unsigned int numneu, NeutrinoType NT);
  \end{lstlisting}
This constructor and initialization function initializes {\ttfamily nuSQUIDS} in the
single energy mode. {\ttfamily numneu} specifies the number of
neutrino flavors which can go from two to six, while {\ttfamily NT} can be
set to  {\ttfamily neutrino} or {\ttfamily antineutrino}. 
\item Multiple energy mode constructor.
  \begin{lstlisting}
    nuSQUIDS(marray<double,1> E_vector,
    unsigned int numneu,NeutrinoType NT = both,
    bool iinteraction = false,
    std::shared_ptr<NeutrinoCrossSections> ncs = nullptr)
  \end{lstlisting}
This constructor and initialization function initializes {\ttfamily nuSQUIDS} in the
multiple energy mode. We need to provide the following arguments:
list of neutrino energy nodes
(\lstinline[columns=fixed,breaklines=true]{E_vector}),
number of neutrino flavors ({\ttf numneu}), neutrino or
anti-neutrino type ({\ttf NT}),  non-coherent scattering
interactions ({\ttf iinteraction}), and neutrino cross section object
pointer ({\ttf ncs}).  

\item Constructing from a $\nu$SQuIDS-HDF5 file
  \begin{lstlisting}
    nuSQUIDS(std::string hdf5_filename, std::string grp = "/",
    std::shared_ptr<InteractionStructure> int_struct = nullptr)
  \end{lstlisting}
This constructor initializes {\ttfamily nuSQUIDS} from a 
previously generated $\nu$SQuIDS HDF5 file. {\ttfamily filepath} must
specify the full path of the HDF5 file, {\ttfamily grp} specifies the
location on the HDF5 file structure where the object will be saved (by default
it will be saved on the {\ttfamily root} of the HDF5 file), and {\ttf
  int\_struct} can specify the cross-section object instead of loading it from
the file.
\end{itemize}

\subsubsection{Functions}

\textbf{Functions to evaluate flavor and mass composition}

\begin{itemize}
\item Flavor composition evaluator (single energy mode)
  \begin{lstlisting}
    double EvalFlavor(unsigned int flv) const;
  \end{lstlisting}
Returns the content of a given neutrino flavor specified by {\ttfamily
  flv} ({\ttfamily 0 = $e$}, {\ttfamily 1 = $\mu$}, {\ttfamily 2 =
  $\tau$}, ...). This function can only be used in the single energy
mode. 
\item Flavor composition evaluator (multiple energy mode)
  \begin{lstlisting}
    double EvalFlavorAtNode(unsigned int flv, unsigned int ie, 
                            unsigned int rho=0) const;
    double EvalFlavor(unsigned int flv, double enu,
                      unsigned int rho=0) const;
    double EvalFlavor(unsigned int flv,double enu,
                      unsigned int rho,double scale,
                      std::vector<bool>& avr) const;
  \end{lstlisting}
{\ttfamily EvalFlavorAtNode} returns the content of a given neutrino
flavor specified by {\ttfamily flv} ({\ttfamily 0 = $e$}, {\ttfamily 1
  = $\mu$}, {\ttfamily 2 = $\tau$, ...}) at an energy node {\ttfamily
  ie}. Furthermore, {\ttfamily EvalFlavor} returns the approximate
content of a given flavor for a specific neutrino energy  {\ttf enu}
by interpolating in the interaction basis. In each function, when
considering  {\ttf  NT = both}, the parameter {\ttf rho} toggles
between {\ttf neutrino (0)} and {\ttf antineutrino (1)}.
The last overload takes two additional arguments: a {\ttf scale}, such
that all $H_0$ induced oscillation frequencies larger than this scale
will be averaged, and a vector of booleans, which entries will be set to
true if the corresponding oscillations frequencies have been averaged
out.

\item Mass composition evaluator (single energy mode)
  \begin{lstlisting}
    double EvalMass(unsigned int eig) const;
  \end{lstlisting}
Returns the content of a given neutrino mass eigenstate specified by {\ttfamily eig} ({\ttfamily 0 = $\nu_1$}, {\ttfamily 1 = $\nu_2$}, {\ttfamily 2 = $\nu_3$, ...}). This function can only be used in the single energy mode.
\item Mass composition evaluator (multiple energy mode)
  \begin{lstlisting}
    double EvalMassAtNode(unsigned int eig, unsigned int ie,
                           unsigned int rho=0) const;
    double EvalMass(unsigned int eig, double enu,
                     unsigned int rho=0) const;
    double EvalMass(unsigned int flv,double enu,
                    unsigned int rho,double scale,
                    std::vector<bool>& avr) const;
  \end{lstlisting}
{\ttfamily EvalMassAtNode} returns the content of a given neutrino mass
eigenstate specified by {\ttfamily eig} ({\ttfamily 0 = $\nu_1$},
{\ttfamily 1 = $\nu_2$}, {\ttfamily 2 = $\nu_3$, ...}) at an energy
node {\ttfamily ie}. Furthermore, {\ttfamily EvalMass} returns the
approximate content of a given mass eigenstate for a specific neutrino
energy  {\ttf enu} by interpolating in the interaction basis. In each
function, when considering  {\ttf  NT = both}, the parameter {\ttf
  rho} toggles between {\ttf neutrino (0)} and {\ttf antineutrino
  (1)}. The last function gets two additional arguments: a {\ttf scale}, such
that all $H_0$ induced oscillation frequencies larger than this scale
will be averaged, and a vector of booleans, which entries will be set to
true if the corresponding oscillations frequencies have been averaged
out.
\end{itemize}

\textbf{Functions to evolve the neutrino ensemble}

\begin{itemize}
\item Evolve state
  \begin{lstlisting}
    void EvolveState();
  \end{lstlisting}
Once the neutrino propagation problem has been set up, this function
evolves the neutrino state from its initial position to its final
position specified by the {\ttfamily track}. 
\end{itemize}

\textbf{Functions obtain properties of the nuSQUIDS object as well as the state}

\begin{itemize}
\item Get energy nodes values
  \begin{lstlisting}
    marray<double,1> GetERange() const;
  \end{lstlisting}
  Returns a one-dimensional array containing the energy nodes
  positions given in eV. 
  \item Get number of energy nodes
  \begin{lstlisting}
    unsigned int GetNumE() const;
  \end{lstlisting}
  Returns the number of energy nodes.
  \item Get number of neutrino flavors
  \begin{lstlisting}
    unsigned int GetNumNeu() const;
  \end{lstlisting}
  Returns the number of neutrino flavors.
  \item Get Hamiltonian at current position
  \begin{lstlisting}
    SU_vector GetHamiltonian(unsigned int ie, 
                             unsigned int rho = 0);
  \end{lstlisting}
  Returns the {\ttf SU\_vector} that represents the (anti)neutrino
  Hamiltonian at the current position, at a given energy node {\ttf
    ie}, and  {\ttf rho} specifies whether the neutrino or antineutrino hamiltonian is returned.
  \item Get the state of the system 
  \begin{lstlisting}
    const squids::SU_vector& GetState(unsigned int ei,
        unsigned int rho = 0) const;
  \end{lstlisting}
  Returns the {\ttf SU\_vector} that represents the (anti)neutrino state at given energy node {\ttf ie}, 
  and {\tt rho} specifies whether the neutrino or antineutrino state is returned.
  \item Get the flavor projector
  \begin{lstlisting}
    SU_vector GetFlavorProj(unsigned int ie,
         unsigned int rho = 0) const;
  \end{lstlisting}
  Returns an {\ttf SU\_vector} that represents the flavor projector for the energy node {\ttf ie}, and 
  {\ttf rho} specifies if neutrinos or antineutrinos are requested.
  \item Get the mass projector
  \begin{lstlisting}
    SU_vector GetMassProj(unsigned int ie,
         unsigned int rho = 0) const;
  \end{lstlisting}
  Returns an {\ttf SU\_vector} that represents the mass projector for the energy node {\ttf ie}, and 
  {\ttf rho} specifies if neutrinos or antineutrinos are requested.
  \item Get {\ttfamily Body}
  \begin{lstlisting}
    std::shared_ptr<Body> GetBody() const;
  \end{lstlisting}
  Returns the {\ttf Body} instance currently stored in the {\ttf nuSQUIDS} object.
  \item Get {\ttfamily Track}
  \begin{lstlisting}
    std::shared_ptr<Track> GetTrack() const;
  \end{lstlisting}
  Returns the {\ttf Track} instance currently stored in the {\ttf nuSQUIDS} object.
  \item Get mixing angle
  \begin{lstlisting}
    double Get_MixingAngle(unsigned int i, unsigned int j) const;
  \end{lstlisting}
  Returns the $\theta_{i,j}$ mixing angle where {\ttf i} and {\ttf j} are zero based indices.
  \item Get CP phase
  \begin{lstlisting}
    double Get_CPPhase(unsigned int i, unsigned int j) const;
  \end{lstlisting}
  Returns the $\delta_{i,j}$ $CP$ phase where {\ttf i} and {\ttf j} are zero based indices.
  \item Get square mass difference
  \begin{lstlisting}
    double Get_SquareMassDifference(unsigned int i) const;
  \end{lstlisting}
  Returns the $\Delta m^2_{i0}$ in ${\rm eV}^2$ where $1\le i < {\rm \tt numneu}$.
  \end{itemize}
  
\textbf{Functions to set properties of the nuSQUIDS object as well as the state}

\begin{itemize}
  \item Set {\ttfamily Body}
  \begin{lstlisting}
    void Set_Body(std::shared_ptr<Body>);
  \end{lstlisting}
  Sets the {\ttf Body} instance in which the neutrino propagation will take place.
  \item Set {\ttfamily Track}
  \begin{lstlisting}
    void Set_Track(std::shared_ptr<Track>);
  \end{lstlisting}
  Sets the {\ttf Track} instance which describes the neutrino propagation inside a
  given {\ttf Body}.
  \item Set the initial state
  \begin{lstlisting}
    void Set_initial_state(marray<double,1> state, Basis basis);
    void Set_initial_state(marray<double,2> state, Basis basis);
    void Set_initial_state(marray<double,3> state, Basis basis);
  \end{lstlisting}
  {\ttf Set\_initial\_state} sets the initial neutrino (and
  antineutrino) state. The states  can be specified for the single and
  multiple energy modes using the appropiate {\ttf
    C++} signature. In each case {\ttf basis} can be either {\ttf
    mass} or {\ttf flavor}. 
  \begin{itemize}
  	\item {\ttf marray<double,1> state}: 
	Can only be used in single energy mode and is defined by 
	{\ttf state[$\alpha$]  = $\phi_\alpha$ } where $\alpha$ is a
        flavor or mass eigenstate index. 
	\item {\ttf marray<double,2> state}: Can only be used in
          multiple energy mode and is defined by {\ttf
            state[ei][$\alpha$]  = $\phi_\alpha (E$[ei]$),$ } i.e. the
          flavor (mass) eigenstate composition at a given energy node
          {\ttf ei}. 
	\item {\ttf marray<double,3> state}: Can only be used in
          multiple energy mode and is defined by {\ttf
            state[ei][$\rho$][$\alpha$]  = $\phi^{\rho}_\alpha
            (E$[ei]$),$ } i.e. the flavor (mass) eigenstate
          composition at a given energy node {\ttf ei}, and where
          $\rho = 0 \equiv {\rm neutrino}$ and $\rho = 1 \equiv {\rm
            antineutrino}$. 
  \end{itemize}
  \item Set energy
  \begin{lstlisting}
    void Set_E(double enu);
  \end{lstlisting}
  Set the neutrino energy. This function can only be used in the
  single energy mode and {\ttf enu} has to be in natural units.
  \item Enable progress bar
  \begin{lstlisting}
    void Set_ProgressBar(bool opt);
  \end{lstlisting}
  If {\ttf opt} is {\ttf true} a progress bar will be printed to
  indicate the calculation progress. 
  \item Enable tau regeneration
  \begin{lstlisting}
    void Set_TauRegeneration(bool opt);
  \end{lstlisting}
  If {\ttf opt} is {\ttf true}, {\ttf NT} is {\ttf both}, and non-coherent 
    interactions are enabled, tau regeneration effects will be included.
  \item Set mixing angle
  \begin{lstlisting}
    void Set_MixingAngle(unsigned int i, unsigned int j,
        double angle);
  \end{lstlisting}
  Sets the $\theta_{i,j}$ mixing angle to {\ttf angle}, where {\ttf i}
  and {\ttf j} are zero-based indexes, and {\ttf angle} must be given in radians.
  \item Set CP phase
  \begin{lstlisting}
    void Set_CPPhase(unsigned int i, unsigned int j,
        double phase);
  \end{lstlisting}
  Sets the $\delta_{i,j}$ $CP$ phase to {\ttf phase}, where {\ttf i} and
  {\ttf j} are zero-based indexes, and {\ttf phase} must be given in radians.
  \item Set square mass difference
  \begin{lstlisting}
    void Set_SquareMassDifference(unsigned int i, doble dmsq);
  \end{lstlisting}
  Sets the $\Delta m^2_{i0}$ to {\ttf dams}, given in ${\rm eV}^2$,
  where $1\le i < {\rm \tt numneu}$. 
  \item Set parameters to default
  \begin{lstlisting}
    void Set_MixingParametersToDefault();
  \end{lstlisting}
  Sets the mixing angles and square mass mixings to the best fit point
  given in~\citep{Gonzalez-Garcia:2014bfa}. 
  \item Set the basis of solution
  \begin{lstlisting}
    void Set_Basis(Basis basis);
  \end{lstlisting}
  Sets the basis in which the evolution will be performed. Two options
  are available: {\ttf mass} and {\ttf interaction}, the latter being
  the default. 
\item Set neutrino cross-section
  \begin{lstlisting}
    void SetNeutrinoCrossSections(
        std::shared_ptr<NeutrinoCrossSections> xs)
  \end{lstlisting}
  Sets the neutrino cross-section~\ref{sec:xs} given by the shared
  pointer {\ttf xs} to the nuSQuIDS object.
\end{itemize}
  
\textbf{HDF5 interface functions}
 
\begin{itemize}
  \item Write nuSQUIDS object into an HDF5 file.
  \begin{lstlisting}
    void WriteStateHDF5(std::string hdf5_filename,
                        std::string group = "/",
                        bool save_cross_sections = true, 
                        std::string cross_section_grp_loc = "")
                        const;
  \end{lstlisting}
  Writes the current {\ttf nuSQUIDS} configuration and state into an HDF5 file
  for later use. {\ttf hdf5\_filename} specifies the output filename, {\ttf group} is the
  location in the HDF5 structure where the {\ttf nuSQUIDS} object will be saved; by default
  the root of the HDF5 will be used. Furthermore, {\ttf
  save\_cross\_sections} and {\ttf cross\_section\_grp\_loc}
  specify if and where in the HDF5 file structure will the cross sections be saved, respectively
   See Figure~\ref{fig:nusquids_hdf5} for details of the nuSQUIDS HDF5 structure.
  \item Add to the HDF5 write funtion.
  \begin{lstlisting}
    virtual void AddToWriteStateHDF5(hid_t hdf5_loc_id) const;
  \end{lstlisting}
  Enables the user to add content to the HDF5 file. An HDF5
    location identifier,{\ttf hdf5\_loc\_id}, will be provided by {\ttf nuSQUIDS}. Using this
    identifier the user can store relevant
  information about the derived nuSQuIDS class.
  \item Read nuSQUIDS object from an HDF5 file.
    \begin{lstlisting}
    void ReadStateHDF5(std::string hdf5_filename,
                       std::string group = "/",
                       std::string cross_section_grp_loc = "");
  \end{lstlisting}
  Reads a previously generated HDF5 file and sets the {\ttf nuSQUIDS} object
  accordingly, i.e. it configures it and loads the saved state. {\ttf
    hdf5\_filename} specifies the input filename, {\ttf group} is the
  location in the HDF5 structure where the {\ttf nuSQUIDS} object is;
  by default the root of the HDF5 is assumed. Furthermore, {\ttf
    cross\_section\_grp\_loc} specify where the cross sections are in
  the HDF5 file structure. See Figure~\ref{fig:nusquids_hdf5}. 
  \item Add to the HDF5 read file.
  \begin{lstlisting}
    virtual void AddToReadStateHDF5(hid_t hdf5_loc_id);
  \end{lstlisting}
  Enables the user to read user defined content from the HDF5 file. An
  HDF5 location,{\ttf hdf5\_loc\_id}, is provided so the user can
  interface with the HDF5 library. For a correct implementation this
  functions has to be written consistently with the {\ttf
    AddToWriteStateHDF5} function.
\end{itemize}

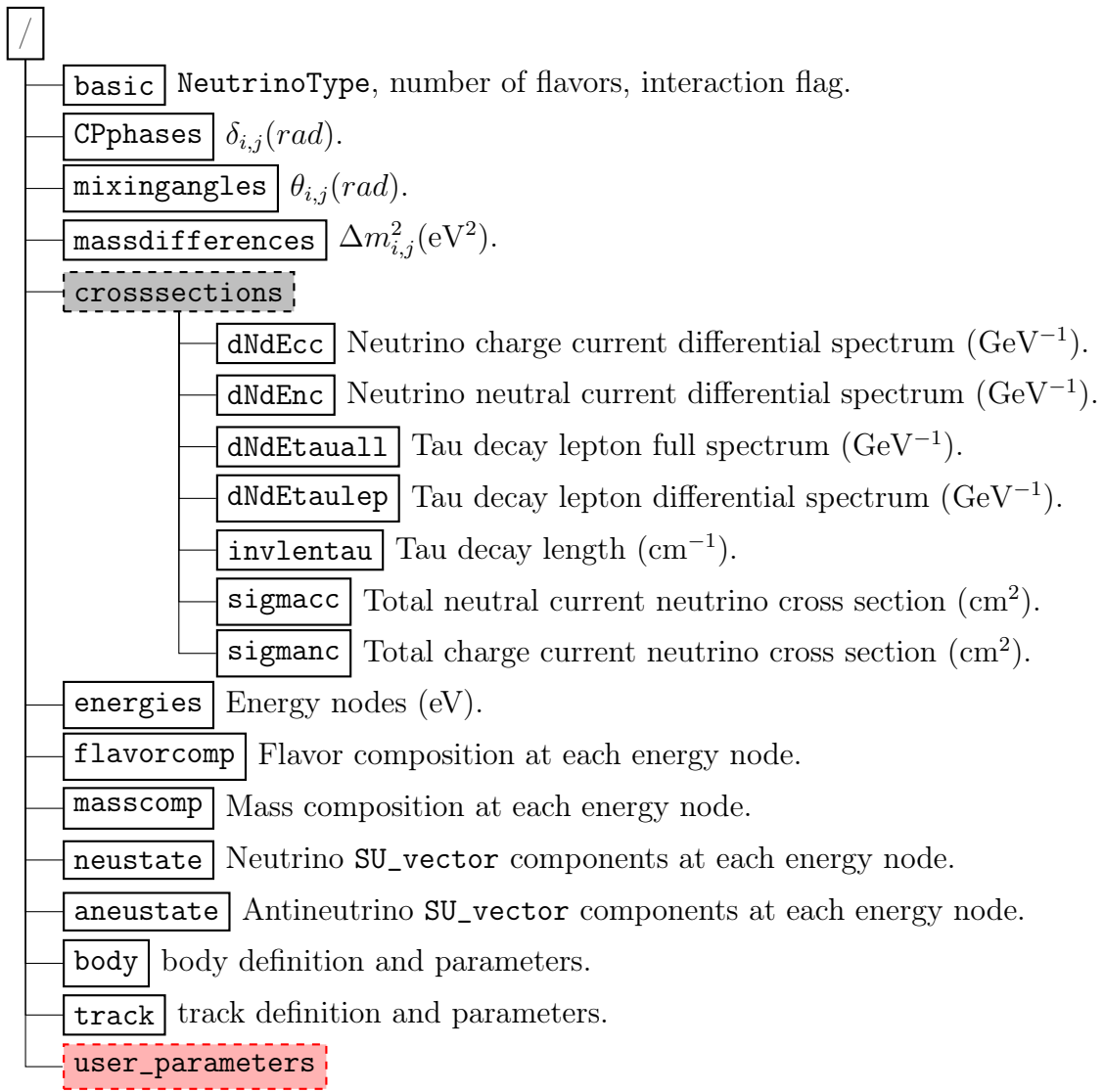
\begin{figure}[htb!]
\centering
\begin{tikzpicture}[%
  grow via three points={one child at (0.5,-0.7) and
  two children at (0.5,-0.7) and (0.5,-1.4)},
  edge from parent path={(\tikzparentnode.south) |- (\tikzchildnode.west)}]
  \node {/}
    child { node [label=right:{{\ttf NeutrinoType}, number of flavors, interaction flag.}] {\ttf basic}}	
    child { node [label=right:{$\delta_{i,j} (rad).$}] {\ttf CPphases}}		
    child { node [label=right:{$\theta_{i,j} (rad).$}] {\ttf mixingangles}}
    child { node [label=right:{$\Delta m^2_{i,j} (\rm eV^2).$}] {\ttf massdifferences}}
    child { node [optional] {\ttf crosssections}
      child { node [label=right:{Neutrino charge current differential spectrum (${\rm GeV^{-1}}$).}]  {\ttf dNdEcc}}
      child { node [label=right:{Neutrino neutral current differential spectrum (${\rm GeV^{-1}}$).}] {\ttf dNdEnc}}
      child { node [label=right:{Tau decay lepton full spectrum (${\rm GeV^{-1}}$).}] {\ttf dNdEtauall}}
      child { node [label=right:{Tau decay lepton differential spectrum (${\rm GeV^{-1}}$).}] {\ttf dNdEtaulep}}
      child { node [label=right:{Tau decay length (${\rm cm^{-1}}$).}] {\ttf invlentau}}
      child { node [label=right:{Total neutral current neutrino cross section (${\rm cm^2}$).}]{\ttf sigmacc}}
      child { node [label=right:{Total charge current neutrino cross section (${\rm cm^2}$).}] {\ttf sigmanc}}
    }
    child [missing] {}				
    child [missing] {}				
    child [missing] {}
    child [missing] {}
    child [missing] {}			
    child [missing] {}
    child [missing] {}			
    child { node [label=right:{Energy nodes ({\rm eV}).}] {\ttf energies}}	
    child { node [label=right:{Flavor composition at each energy node.}] {\ttf flavorcomp}}
    child { node [label=right:{Mass composition at each energy node.}] {\ttf masscomp}}
    child { node [label=right:{Neutrino {\ttf SU\_vector} components at each energy node.}] {\ttf neustate}}
    child { node [label=right:{Antineutrino {\ttf SU\_vector} components at each energy node.}] {\ttf aneustate}}
    child { node [label=right:{body definition and parameters.}] {\ttf body}}
    child { node [label=right:{track definition and parameters.}] {\ttf track}}
    child { node [selected] {\ttf user\_parameters}}
    ;
\end{tikzpicture}
\caption{Structure of {\ttf nuSQUIDS} HDF5 file. The {\ttf
    crosssections} group will only be written when interactions are
  enabled. Furthermore, {\ttf user\_parameters} is by default empty
  and can be set/accessed by {\ttf AddToWriteStateHDF5/AddToReadStateHDF5}.} 
\label{fig:nusquids_hdf5}
\end{figure}

\subsection{nuSQUIDSAtm\label{sec:nusquidsatm}}

Atmospheric neutrino oscillations are a major and instrumental part of
research in contemporary neutrino physics. Experiments like
SuperKamiokande, IceCube, and Antares have used atmospheric neutrinos
to measure neutrino mass splittings and mixing angles. Furthermore,
proposed extensions like HyperKamiokande, PINGU, and ORCA ought to
improve the current measurements and have sensitivity to 
neutrino mass ordering. This class allows to propagate a set of full
energy spectrum of neutrinos for different zenith angles.
It implements functions to easily set the initial conditions and also
interpolations of the fluxes.
\subsubsection{Constructors}

\begin{itemize}
\item Constructor with {\ttf costh} range.
  \begin{lstlisting}
    template<typename... ArgTypes> nuSQUIDSAtm(
        marray<double,1> costh_array, ArgTypes&&... args)
  \end{lstlisting}
This constructor creates a set of {\ttf nuSQUIDS} or derived {\ttf
  nuSQUIDS} objects with a set of zenith angles given by the {\ttf
  marray} {\ttf costh\_array}. The arguments given in {\ttf arg} are the
corresponding arguments of the nuSQUIDS or derived nuSQUIDS class constructor.

\item Constructing from a nuSQuIDSAtm-HDF5 file
  \begin{lstlisting}
    nuSQUIDSAtm(std::string hdf5_filename)
  \end{lstlisting}
This constructor initializes {\ttfamily nuSQUIDS} from a 
previously generated $\nu$SQuIDS-HDF5 file. The resulting {\ttfamily nuSQUIDS} 
object will be given in {\it single} or {\it multiple} energy mode
depending on the HDF5 file configuration. {\ttfamily filepath} must specify the full
path of the HDF5 file. Furthermore,
{\ttfamily grp} specifies the location in the HDF5 file structure
where the object will be saved; by default
it will be saved on the {\ttfamily root} of the HDF5 file.

\item Move constructor.
  \begin{lstlisting}
    nuSQUIDSAtm(nuSQUIDSAtm&&);
  \end{lstlisting}

  \subsubsection{Functions}
\item Set initial state.
  \begin{lstlisting}
    void Set_initial_state(const marray<double,3>& ini_flux,
        Basis basis=flavor);
    void Set_initial_state(const marray<double,4>& ini_flux,
        Basis basis=flavor);
  \end{lstlisting}
  Sets the initial state of the system given in the marray {\ttf
    ini\_flux}, {\ttf basis} is the basis in which the state is
  defined. Different {\ttf marray} can be used for different cases:

  \begin{itemize}
  \item {\ttf marray<double,3> state}: Can only be used in
    multiple energy mode and is defined by {\ttf
      state[czi][ei][$\alpha$]  = $\phi_\alpha (E$[ei]$,costh$[czi]$),$ } i.e. the
    flavor (mass) eigenstate composition at a given energy and cosine
    zenith node {\ttf ei} and {\ttf czi}. 
  \item {\ttf marray<double,4> state}: Can only be used in
    multiple energy mode and is defined by {\ttf
      state[czi][ei][$\rho$][$\alpha$]  = $\phi^{\rho}_\alpha
      (E$[ei]$,costh$[czi]$),$ } i.e. the flavor (mass) eigenstate
    composition at a given energy and zenith node {\ttf ei} and {\ttf czi}, and where
    $\rho = 0 \equiv {\rm neutrino}$ and $\rho = 1 \equiv {\rm
      antineutrino}$. 
  \end{itemize}
  
\item Evolve function
  \begin{lstlisting}
    void EvolveState();
  \end{lstlisting}
  Function that evolves the system.

\item Evaluate the flux for a given flavor.
  \begin{lstlisting}
    double EvalFlavor(unsigned int flv,double costh,
        double enu, unsigned int rho,
        double scale,std::vector<bool> avr) const;

    double EvalFlavor(unsigned int flv,double costh,
    double enu,unsigned int rho = 0,
        bool randomize_production_height = false) const;
  \end{lstlisting}
  It returns the flux for the flavor {\ttf flv} at the value of cosine
  zenith given by {\ttf costh} and energy given by {\ttf enu}.
  {\ttf rho} is {\ttf neutrino} or {\ttf antineutrino}.
  Some arguments can also be set: If a frequency is higher than {\ttf
    scale} it will be averaged out, and the corresponding entry in the
  Boolean vector {\ttf avr} will be set to {\ttf true}.
  In the second case the randomization of the production height of the
  neutrino can be set to $true$, by default is $false$.
  
\item Read and write function.
  \begin{lstlisting}
    void ReadStateHDF5(std::string hdf5_filename);
    void WriteStateHDF5(std::string hdf5_filename,
        bool overwrite = true) const;
  \end{lstlisting}
  
  These functions read and write the state of the system in the file
  {\ttf hdf5\_filename}, in {\ttf WriteStateHDF5} {\ttf overwrite} may be set to
  {\ttf true} or {\ttf false}.

\item Set functions as in nuSQUIDS.
  \begin{lstlisting}
    void Set_MixingParametersToDefault();
    void Set_MixingAngle(unsigned int i,
                         unsigned int j,double angle);
    void Set_CPPhase(unsigned int i,
                     unsigned int j,double angle);
    void Set_SquareMassDifference(unsigned int i,double sq);
    void Set_h(double h);
    void Set_h_max(double h);
    void Set_h_min(double h);
    void Set_abs_error(double eps);
    void Set_rel_error(double eps);
    void Set_GSL_step(gsl_odeiv2_step_type const * opt);
    void Set_ProgressBar(bool opt);
    void Set_TauRegeneration(bool opt);
    void Set_GlashowResonance(bool opt);
    void Set_IncludeOscillations(bool opt);
    void Set_AllowConstantDensityOscillationOnlyEvolution(bool opt);
    void Set_PositivityConstrain(bool opt);
    void Set_PositivityConstrainStep(double step);
    void SetNeutrinoCrossSections(
                   std::shared_ptr<NeutrinoCrossSections> xs);
  \end{lstlisting}
  All these functions do a recursive call to the function with
  the same name in all the nuSQUIDS objects.

\item Set earth model.
  \begin{lstlisting}
    void Set_EarthModel(std::shared_ptr<EarthAtm> earth);
  \end{lstlisting}
  Sets the body given by {\ttf earth} to all the nuSQUIDS objects in
  every node.

\item Sets the number of threads
  \begin{lstlisting}
    void Set_EvalThreads(unsigned int nThreads);
  \end{lstlisting}
  The evolution can be done in a multi-thread with the number of
  threads specified by {\ttf nThreads}.

\item Set absolute error in a given node.
  \begin{lstlisting}
    void Set_abs_error(double eps, unsigned int idx);
  \end{lstlisting}
  Sets the GSL absolute error {\ttf} in the cosine zenith node
  given by $idx$.

\item Number of threads.
  \begin{lstlisting}
    unsigned int Get_EvalThreads() const{
  \end{lstlisting}
  Returns the number of threads used in the evaluation.
\item Get mixing angles.
  \begin{lstlisting}
    double Get_MixingAngle(unsigned int i, unsigned int j) const;
  \end{lstlisting}
  Returns the mixing angle of the first cosine zenith node in the
  rotation plane given by ({\ttf i}, {\ttf j}).

\item Get square mass difference.
  \begin{lstlisting}
    double Get_SquareMassDifference(unsigned int i) const;
  \end{lstlisting}
  Returns the square mass difference value given by {\ttf i}.
 \item Number of energy nodes.
  \begin{lstlisting}
    size_t GetNumE() const;
  \end{lstlisting}
  Gives the number of energy nodes.

 \item Number of cosine zenith nodes.
  \begin{lstlisting}
    size_t GetNumCos() const;
  \end{lstlisting}
  Gives the number of cosine zenith nodes.

\item Number of rho.
  \begin{lstlisting}
    unsigned int GetNumRho() const;
  \end{lstlisting}
  Gives the number of {\ttf rho}, for neutrino-antineutrino case it
  will be two. 

\item  Energy array. 
  \begin{lstlisting}
    marray<double,1> GetERange() const;
  \end{lstlisting}
  Returns an {\ttf marray} with the value of the energies in the
  energy nodes.

\item Cosine zenith array.
  \begin{lstlisting}
    marray<double,1> GetCosthRange() const;
  \end{lstlisting}
  Returns an {\ttf marray} with the value of the cosine zenith in the
  cosine zenith nodes.

\item Get the nuSQUIDS object.
  \begin{lstlisting}
    BaseSQUIDS& GetnuSQuIDS(unsigned int ci);
  \end{lstlisting}
  Returns the nuSQUIDS object in the cosine zenith node {\ttf ci}.

\item Get the array of nuSQUIDS.
  \begin{lstlisting}
      std::vector<BaseSQUIDS>& GetnuSQuIDS();
    \end{lstlisting}
    Returns a {\ttf std::vector} with the nuSQUIDS objects in all the
    cosine zenith nodes.
\end{itemize}

\section{Python Interface}
\label{sec:python}

As the particle physics community has transitioned from {\ttfamily FORTRAN} to {\ttfamily C++} based frameworks, 
it is also a current trend to be able to interface analysis software with high-level interpreted 
languages such as {\ttfamily Mathematica}, {\ttfamily R}, and {\ttfamily Python}. 
Of these languages we have decided to implement bindings with Python due to the well-developed 
{\ttfamily C++}-{\ttfamily Python} bindings given by the {\ttfamily Boost} library.

\subsection{Installation}

In order to install nuSQuIDS python bindings, additional libraries are required, namely, {\ttfamily Boost.Python} ($\ge1.54$) and {\ttfamily Python.numpy} ($\ge1.7$). 
Upon installing these new prerequisites you can run the following command to generate a Makefile to compile the bindings

\begin{lstlisting}[language=Bash]
./configure --with-python-bindings
\end{lstlisting}

This command will try to autodetect the location of the aforementioned mentioned libraries.
The following configuration script flags are influencial and can be used to set pre-requisites locations:
\begin{itemize}
  \item Specify the Python executable
  \begin{lstlisting}
    --python-bin=PYTHON_EXECUTABLE
  \end{lstlisting}
  The provided {\ttf Python} executable will be used to find the python version, libraries, and includes.
  \item Set Boost library and include paths
  \begin{lstlisting}
    --with-boost=DIR
    --with-boost-incdir=DIR
    --with-boost-libdir=DIR
  \end{lstlisting}
  The {\ttf incdir} and {\ttf libdir} options are used to set the boost include and library directories. 
    The first option is a convenient flag that sets both directories to: {\ttf DIR/lib} and {\ttf DIR/include} for the library and include directory.

\end{itemize}
This will produce a {\ttf Makefile} in the {\ttf resources/python/src} directory of nuSQuIDS. This Makefile will be compiled when you run {\ttf make} on the main nuSQuIDS folder. This will produce a shared library in {\ttf resources/python/pybindings}; in order to be able to use the python bindings you need to add this directory to the system variable {\ttf PYTHONPATH}. For example, in the Bash shell
\begin{lstlisting}[language=Python]
export PYTHONPATH=NUSQUIDS_DIR/resources/python/bindings:$PYTHONPATH
\end{lstlisting}

After successful installation you can import the python bindings in the following manner
\begin{lstlisting}[language=Python]
import nuSQUIDSpy as nsq
\end{lstlisting}
where here we have introduced the alias {\ttfamily nsq} for the nuSQuIDS python module. We can further extend the
capabilities of nuSQuIDS in {\ttfamily Python} by means of the the {\ttfamily nuSQUIDSTools} module.
When it is loaded the nuSQuIDS functions and objects get overloaded with additional functionalities. In order to enable them, after loading the {\ttfamily nuSQUIDSpy} module, do
\begin{lstlisting}[language=Python]
import nuSQUIDSTools
\end{lstlisting}

\subsection{Description of the interface}

The interface is implemented in two files {\ttf nuSQUIDSpy.h} and {\ttf nuSQUIDSpy.cpp}. 
Additionally in {\ttf resources/python/inc} the file {\ttf container\_conversions.h} 
is provided to enable translation between {\ttf std::vector<type>} to  {\ttf List} among {\ttf C++} and {\ttf Python}. 
The python bindings header contains {\ttf C++} templates that facilitate registration of {\ttf nuSQuIDS} python bindings.

In the bindings source file the following classes have been registered:
\begin{itemize}
  \item \nuSQUIDS,
  \item {\ttf nuSQUIDSAtm},
  \item {\ttf NeutrinoCrossSections},
  \item {\ttf NullCrossSections},
  \item {\ttf NeutrinoDISCrossSectionsFromTables},
  \item {\ttf TauDecaySpectra},
  \item {\ttf Body}, {\ttf Body.Track}
  \item {\ttf ConstantDensity}, {\ttf ConstantDensity.Track}
  \item {\ttf VariableDensity}, {\ttf VariableDensity.Track}
  \item {\ttf Earth}, {\ttf Earth.Track}
  \item {\ttf Sun}, {\ttf Sun.Track}
  \item {\ttf SunASnu}, {\ttf SunASnu.Track}
  \item {\ttf EarthAtm}, {\ttf EarhAtm.Track}
\end{itemize}
for all these classes we have implemented the public members described in this document with 
the same names as in the {\ttf C++} interface. Additionally, two classes from SQuIDS have been exposed to {\ttf Python}:
\begin{itemize}
  \item {\ttf SU\_vector} and
  \item {\ttf Const}.
\end{itemize}
See~\cite{SQUIDS} for details on these classes. Also, the following enumerations have been made available in {\ttf Python}
\begin{itemize}
  \item \lstinline[columns=fixed,breaklines=true]{GSL_STEP_FUNCTIONS} enumeration contains the following members: 
    \begin{itemize}
      \item \lstinline[columns=fixed,breaklines=true]{GSL_STEP_RK2},
      \item \lstinline[columns=fixed,breaklines=true]{GSL_STEP_RK4},
      \item \lstinline[columns=fixed,breaklines=true]{GSL_STEP_RKF45},
      \item \lstinline[columns=fixed,breaklines=true]{GSL_STEP_RKCK},
      \item \lstinline[columns=fixed,breaklines=true]{GSL_STEP_RK8PD},
      \item and \lstinline[columns=fixed,breaklines=true]{GSL_STEP_MSADAMS}.
    \end{itemize}
    These can be used in the {\ttf Set\_GSL\_step} {\ttf nuSQUIDS}  member function.
  \item {\ttf Basis}, which contains the members:
    \begin{itemize}
      \item {\ttf mass} ({\ttf Basis::mass}),
      \item {\ttf flavor} ({\ttf Basis::flavor}),
      \item and {\ttf interaction} ({\ttf Basis::interaction}).
    \end{itemize}
  \item {\ttf NeutrinoCrossSections\_NeutrinoFlavor}, which contains the members:
    \begin{itemize} 
      \item {\ttf electron} ({\ttf NeutrinoCrossSections::NeutrinoFlavor::electron}),
      \item {\ttf muon} ({\ttf NeutrinoCrossSections::NeutrinoFlavor::muon}),
      \item {\ttf tau} ({\ttf NeutrinoCrossSections::NeutrinoFlavor::tau}),
      \item and {\ttf sterile} ({\ttf NeutrinoCrossSections::NeutrinoFlavor::sterile}).
    \end{itemize}
  \item {\ttf NeutrinoCrossSections\_NeutrinoType}, which contains the members:
    \begin{itemize} 
      \item {\ttf neutrino} ({\ttf NeutrinoCrossSections::NeutrinoType::neutrino})
      \item and {\ttf antineutrino} ({\ttf NeutrinoCrossSections::NeutrinoType::antineutrino}).
    \end{itemize}
  \item {\ttf NeutrinoCrossSections\_Current}, which contains the members:
    \begin{itemize}
      \item {\ttf CC} ({\ttf NeutrinoCrossSections::Current::CC}),
      \item {\ttf NC} ({\ttf NeutrinoCrossSections::Current::NC}),
      \item and {\ttf GR} ({\ttf NeutrinoCrossSections::Current::GR}).
    \end{itemize}
  \item {\ttf Neutrinotype}, which contains the members:
    \begin{itemize}
      \item {\ttf neutrino} ({\ttf NeutrinoType::neutrino}),
      \item {\ttf antineutrino} ({\ttf NeutrinoType::antineutrino}),
      \item and {\ttf both} ({\ttf NeutrinoType::both}).
    \end{itemize}
\end{itemize}

Finally, conversion between {\ttf Numpy.Array} and {\ttf marray<double,DIM>} has been made available up to dimension four.
Higher dimensionality multidimensional array conversion is possible by adding the following lines to the boost python module
\begin{lstlisting}[language=C++]
to_python_converted< marray<double, DIM>, marray_to_numpyarray<DIM> >();
marray_from_python<DIM>();
\end{lstlisting}
where {\ttf DIM} is the array dimensionality to register.

\subsection{Python walk-through}

In this section we will illustrate the usage of the {\ttf Python} interface. 
These examples can be found in the IPython notebook located in {\ttf resources/python/example}.
We start by importing the modules
\begin{lstlisting}[language=Python, frame=leftline, numbers=left, breaklines=true]
import numpy as np
import matplotlib.pyplot as plt
import nuSQUIDSpy as nsq
import nuSQUIDSTools
\end{lstlisting}
Throughout the example we will use the shorthand {\ttf nsq} for the {\ttf nuSQUIDS} python bindings.

\subsubsection{Single-energy mode}

To start, like in the {\ttf C++} case, we need to create a $\nu$-SQuIDS object. 
To begin this demonstration we will create a simple single-energy mode three-flavor neutrino oscillation calculator.
Thus we just need to specify the number of neutrinos (3) and if we are dealing with neutrinos or antineutrinos.
\begin{lstlisting}[language=Python, breaklines=true]
nuSQ = nsq.nuSQUIDS(3,nsq.NeutrinoType.neutrino)
\end{lstlisting}
<<<<<<< HEAD

nuSQuIDS inputs should be given in natural units.
In order to make this convenient, as in the {\ttf C++} code, we make use of the {\ttf SQuIDS} class {\ttf Const}.
We can instantiate it as follows
\begin{lstlisting}[language=Python, breaklines=true]
units = nsq.Const()
\end{lstlisting}

As in the {\ttf C++} $\nu$-SQuIDS interface one can propagate the neutrinos in various environments
(see~\ref{sec:body} for further details). We can start by considering oscillactions in vacuum by setting:
\begin{lstlisting}[language=Python, breaklines=true]
nuSQ.Set_Body(nsq.Vacuum())
\end{lstlisting}

Since we have to specify that we are considering vacuum propagation, 
we must construct a trajectory inside that object. This can be done using the {\ttf Track} class-property of a given {\ttf Body}.
Each {\ttf Body} will have its own {\ttf Track} subclass and its constructors. We can set and construct a vacuum trajectory in the following way:
\begin{lstlisting}[language=Python, breaklines=true]
nuSQ.Set_Track(nsq.Vacuum.Track(100.0*units.km))
\end{lstlisting}

Next we have to set the neutrino energy, which can be done as follows
\begin{lstlisting}[language=Python, breaklines=true]
nuSQ.Set_E(1.0*units.GeV)
\end{lstlisting}

Now we have to set the initial neutrino state. We can provide this state in either the flavor or mass basis.
We can do this using the {\ttf Set\_initial\_state} function, providing it with a numpy array and a {\ttf nsq.Basis}.
If the basis is {\ttf nsq.Basis.flavor} then the list must contain {\ttf np.array([}$\phi_e$,$\phi_\mu$,$\phi_\tau${\ttf ])}, 
similarly if enumeration is {\ttf nsq.Basis.mass} the list must specify $\phi_i$.
In this example we set the initial state to $\nu_\mu$,
\begin{lstlisting}[language=Python, breaklines=true]
nuSQ.Set_initial_state(np.array([0.,1.,0.]),nsq.Basis.flavor)
\end{lstlisting}

Finally, we calculate the system evolution by
\begin{lstlisting}[language=Python, breaklines=true]
nuSQ.EvolveState()
\end{lstlisting}

After this runs, {\ttf nuSQuIDS} has evolved the state and holds it in memory.
We can inquire for the flavor composition at this stage by
\begin{lstlisting}[language=Python, breaklines=true]yy
print([nuSQ.EvalFlavor(i) for i in range(3)])
\end{lstlisting}

which results in
\begin{lstlisting}[language=Python, breaklines=true]yy
[0.004675502573140013, 0.9022064087866011, 0.09311808864025879]
\end{lstlisting}

$\nu$-SQuIDS knows everything about the neutrino state at the current moment, it also knows what we did with it so far, where it went, what mixing parameters were used, etc. It would be convenient to store this information. One way of doing this is to serialize the $\nu$-SQuIDS status. We can do this in the following way
\begin{lstlisting}[language=Python, breaklines=true]
nuSQ.WriteStateHDF5("current_state.hdf5")
\end{lstlisting}

Everything that is in the $\nu$-SQuIDS object is now in that file. We can use that file to create a new $\nu$-SQuIDS object and do another calculation, we can stop the calculation midway and use it to restart, we can explore that file with other analysis tools, etc. In particular, the `ReadStateHDF5` will return us to the given configuration.
\begin{lstlisting}[language=Python, breaklines=true]
nuSQ.ReadStateHDF5("current_state.hdf5")
\end{lstlisting}

We can use the current tool to try to calculate $P(\nu_\mu \to \nu_e)$ as a function of energy. We can do the following
\begin{lstlisting}[language=Python, frame=leftline, numbers=left, breaklines=true]
energy_values = np.linspace(1,10,40)
nu_mu_to_nu_e = []
for Enu in energy_values:
    nuSQ.Set_E(Enu*units.GeV)
    nuSQ.Set_initial_state(np.array([0.,1.,0.]),nsq.Basis.flavor)
    nuSQ.EvolveState()
    nu_mu_to_nu_e.append(nuSQ.EvalFlavor(0))

plt.figure(figsize = (8,6))
plt.xlabel(r"$E_\nu [{\rm GeV}]$")
plt.ylabel(r"$P(\nu_\mu \to \nu_e)$")
plt.plot(energy_values,nu_mu_to_nu_e, lw = 2, color = 'blue')
\end{lstlisting}
This results in the plot shown in Figure~\ref{fig:nusquids_python_osc} left panel.
\begin{figure}[h!]
  \label{fig:nusquids_python_osc}
  \centering
  \includegraphics[width=0.45\textwidth]{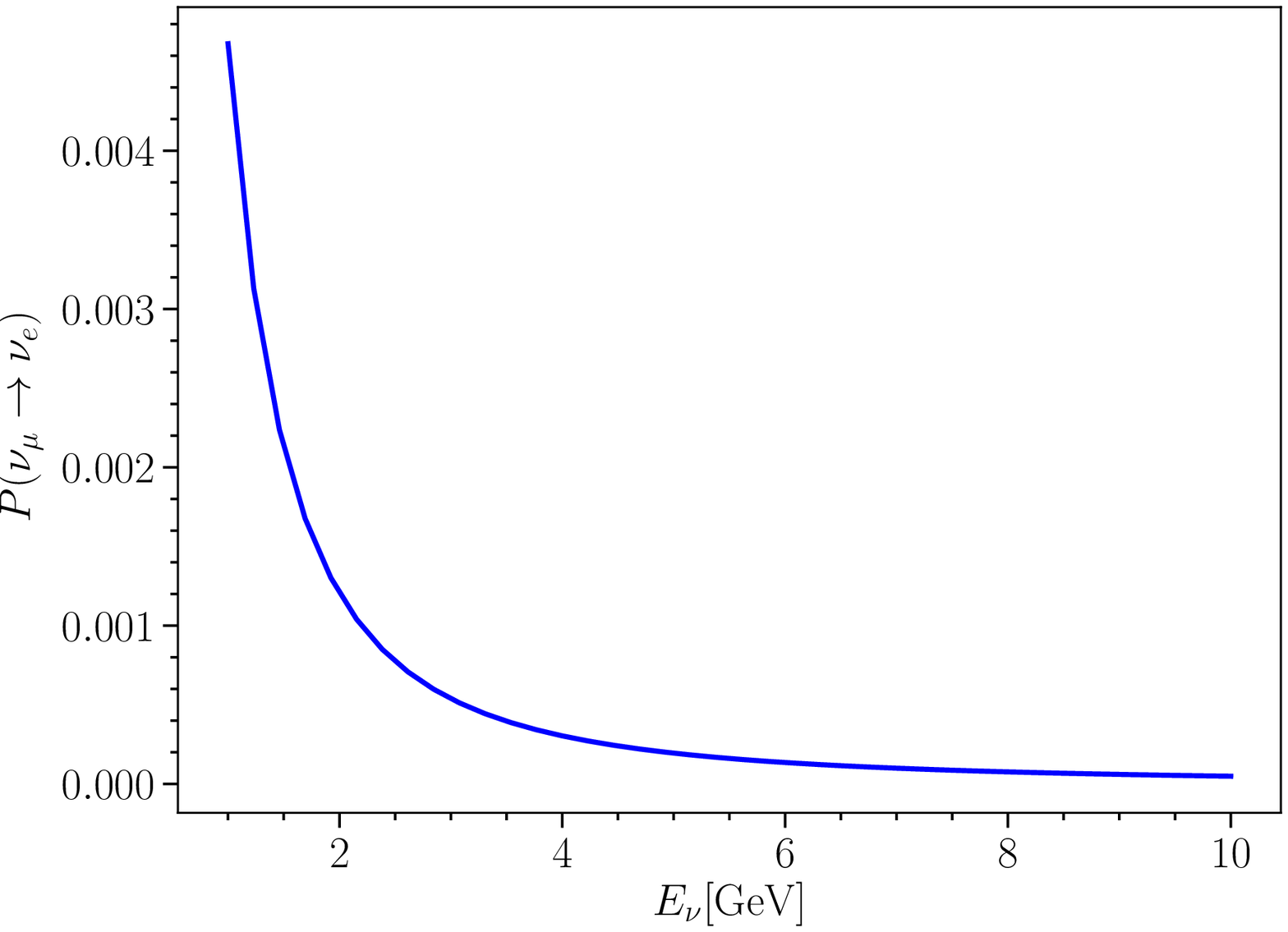}
  \includegraphics[width=0.45\textwidth]{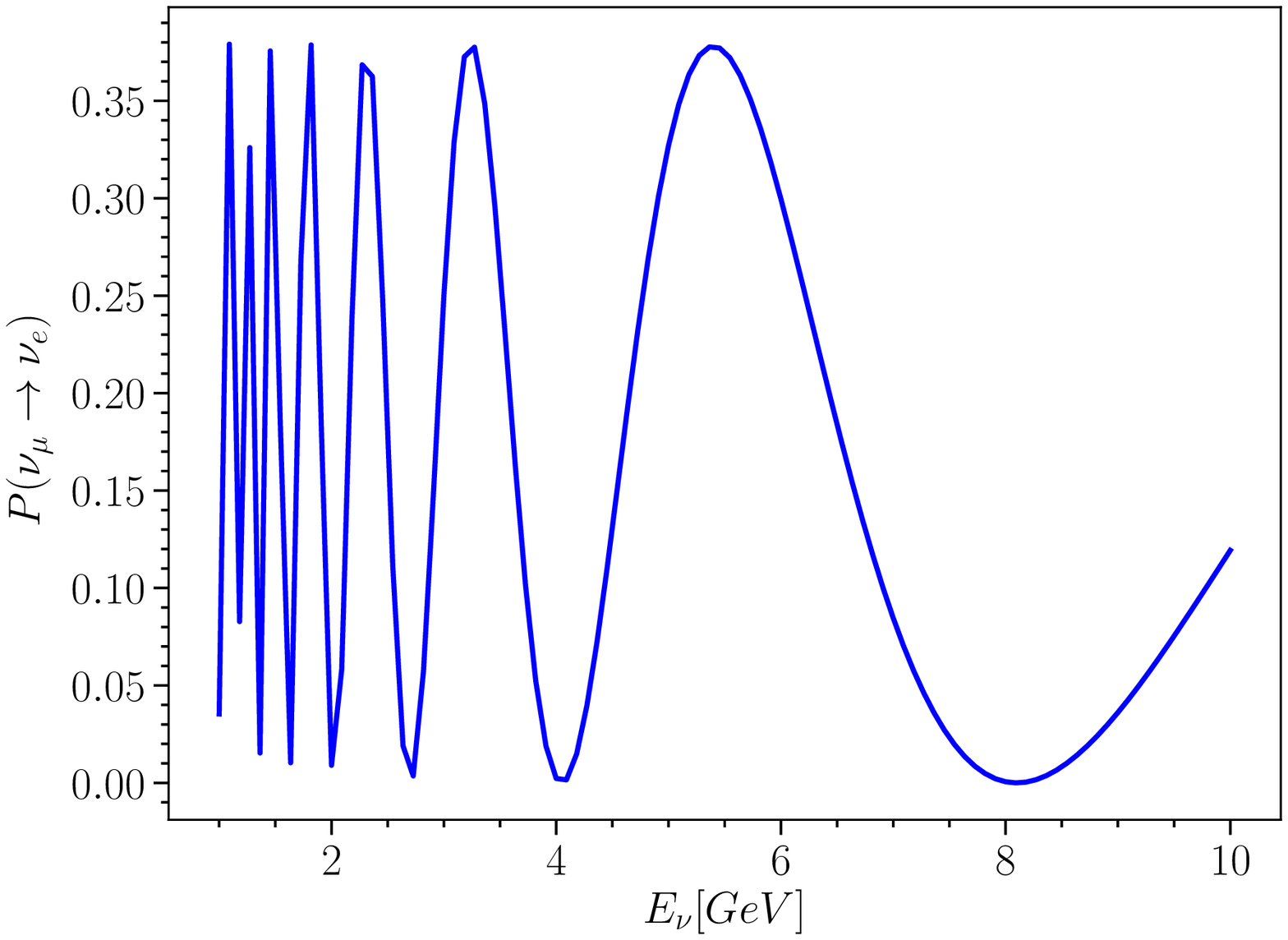}
  \caption{Muon-neutrino to electron neutrino oscillation probability in vacuum as a function of the neutrino energy for a baseline of 100 km. 
  The left panel is for the default standard oscillation parameters and the right panel has $\Delta m^2_{21}$ set to $1~{\rm eV}^2$}
\end{figure}

nuSQuIDs has some predefined oscillation mixing angles ($\theta_{ij}$) and mass splittings ($\Delta m^2_{ij}$),
which we have taken from~\cite{Esteban:2020cvm}.
We can change the square mass differences by using the following function
\begin{lstlisting}[language=Python, breaklines=true]
nuSQ.Set_SquareMassDifference(1,2.0e-1) # sets dm^2_{21} in eV^2.
\end{lstlisting}

Running the same script again results in the plot shown in Fig.~\ref{fig:nusquids_python_osc} right panel.

We can also try to modify the mixing angles, for example
\begin{lstlisting}[language=Python, breaklines=true]
nuSQ.Set_MixingAngle(0,1,1.2) # sets \theta_{12} in radians.
nuSQ.Set_MixingAngle(0,2,0.3) # sets \theta_{23}} in radians.
nuSQ.Set_MixingAngle(1,2,0.4) # sets \theta_{23}} in radians.
\end{lstlisting}
We can also restore the default mixing angles and mass differences to default by doing
\begin{lstlisting}[language=Python, breaklines=true]
nuSQ.Set_MixingParametersToDefault()
\end{lstlisting}

\subsubsection{Multiple-energy mode}

In this section we will demonstrate the use of {\ttf nuSQuIDS} in the multiple energy mode.
To begin we have to specify the $\{E_i\}$ grid where the equation will be solved.
The energy nodes can be an arbitrary ordered list of energy in eV.
The most common choice is to have linearly-spaced nodes or logarithmically-spaced nodes.
For constructing numpy array of linear and logarithmic scales nuSQuIDS provides two
 convenient functions called {\ttf linspace} and {\ttf logspace }respectively. The 
following lines define the energy grid and construct the nuSQUIDS object.

\begin{lstlisting}[language=Python, frame=leftline, numbers=left, breaklines=true]
interactions = False

E_min = 1.0*units.GeV
E_max = 10.0*units.GeV
E_nodes = 101

energy_nodes = nsq.logspace(E_min,E_max,E_nodes)

neutrino_flavors = 3

nuSQ = nsq.nuSQUIDS(energy_nodes,neutrino_flavors,nsq.NeutrinoType.neutrino,interactions)
\end{lstlisting}

We can propagate a neutrino ensemble through the Earth in an atmospheric neutrino telescope setting.
We can do that by setting the following Body and Track:

\begin{lstlisting}[language=Python, frame=leftline, numbers=left, breaklines=true]
earth = nsq.EarthAtm()
nuSQ.Set_Body(earth)
nuSQ.Set_Track(earth.MakeTrackWithCosine(-1))
\end{lstlisting}

For the initial flux we will assume that it's given by $\phi_\nu = N_0 E^{-2}$,
with flavor composition $\phi_e:\phi_\mu:\phi_\tau$ = $0:1:0$.
$\nu$-SQuIDS input flux is a {\ttf numpy.ndarray} formatted in the following way:
\begin{equation}
InputState \doteq [[\phi^1_e,\phi^1_\mu,\phi^1_\tau],...,[\phi^i_e,\phi^i_\mu,\phi^i_\tau],...,[\phi^n_e,\phi^n_\mu,\phi^n_\tau]]
\end{equation}
where $i = 1$ to $n$ and $n$ is the number of energy nodes. We can implement this in the following way
\begin{lstlisting}[language=Python, frame=leftline, numbers=left, breaklines=true]
N0 = 1.0e18
Eflux = lambda E: N0*E**-2
Einitial = (Eflux(nuSQ.GetERange()).reshape((101,1)))*(np.array([0.,1.,0.]).reshape(1,3))
\end{lstlisting}
where {\ttf N0} is an arbitrary normalization constant, but we have chosen it such that the input flux is $O(1)$.

\begin{lstlisting}[language=Python, frame=leftline, numbers=left, breaklines=true]
nuSQ.Set_initial_state(Einitial,nsq.Basis.flavor)

nuSQ.Set_rel_error(1.0e-17)
nuSQ.Set_abs_error(1.0e-17)

nuSQ.EvolveState()
\end{lstlisting}

Then we can evaluate the resulting flux and plot the resulting oscillation 
probabilities by using the folowing script

\begin{lstlisting}[language=Python, frame=leftline, numbers=left, breaklines=true]
e_range = np.linspace(1.0,10.0,200)

nu_e = np.array([nuSQ.EvalFlavor(0,EE*units.GeV,0) 
                           for EE in e_range])
nu_mu = np.array([nuSQ.EvalFlavor(1,EE*units.GeV,0) 
                           for EE in e_range])
nu_tau = np.array([nuSQ.EvalFlavor(2,EE*units.GeV,0) 
                           for EE in e_range])

total = nu_e + nu_mu + nu_tau

plt.figure(figsize = (8,6))

plt.xlabel(r"$E_\nu [{\rm GeV}]$")
plt.ylabel(r"$\phi_\alpha$")

plt.plot(e_range,nu_e, lw = 2, color = 'red', label = r"$\nu_e$")
plt.plot(e_range,nu_mu, lw = 2, color = 'blue', label = r"$\nu_\mu$")
plt.plot(e_range,nu_tau, lw = 2, color = 'green', label = r"$\nu_\tau$")
plt.plot(e_range,total, lw = 2, color = 'black', label = r"Total")
plt.legend(fancybox = True, fontsize = 10)
plt.semilogy()
\end{lstlisting}
The result of this is shown in the left panel of Fig.~\ref{fig:nusquids_atm_python_no_interaction}. Our final example is 
performing the full nuSQuIDS calculation with neutrinos and antineutrinos, but this time including interactions

\begin{figure}[h!]
  \label{fig:nusquids_atm_python_no_interaction}
  \centering
  \includegraphics[width=\textwidth]{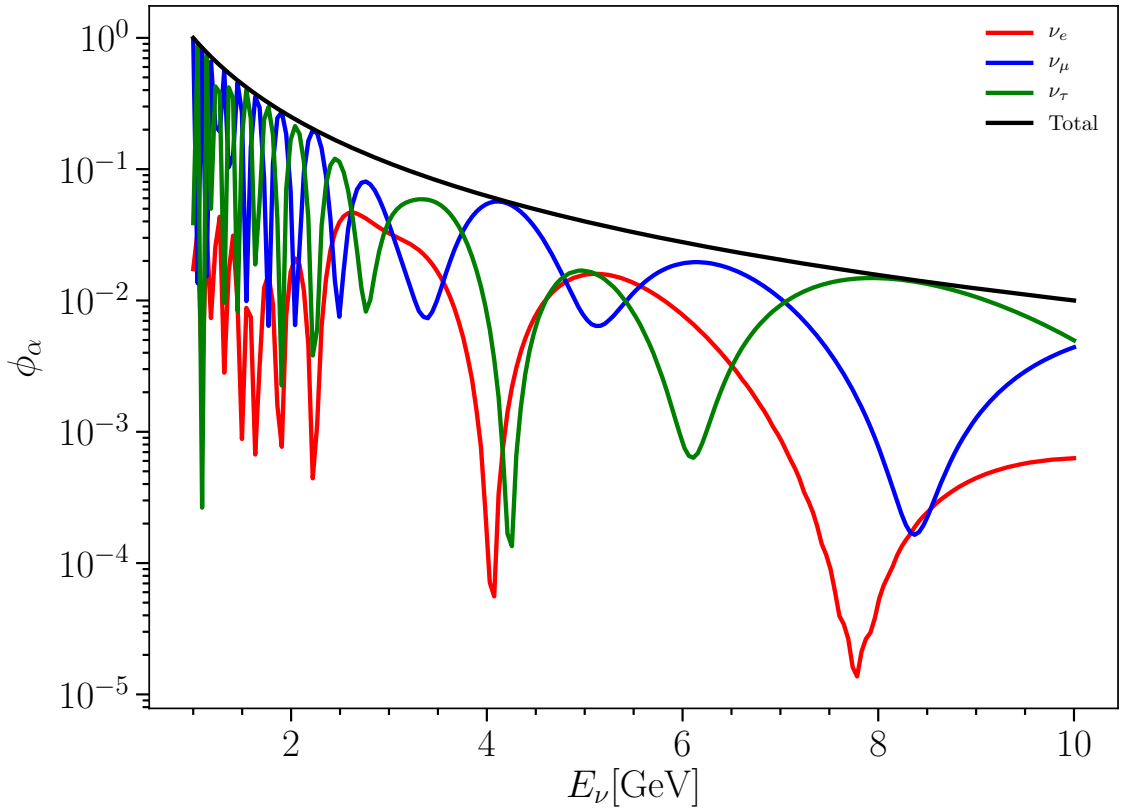}
  \caption{Result of propagating a power-law flux with spectral index of -1 through the Earth considering only oscillations and no interactions.}
\end{figure}

\begin{lstlisting}[language=Python, frame=leftline, numbers=left, breaklines=true]
interactions = True

E_min = 1.0e2*units.GeV
E_max = 1.0e8*units.GeV
E_nodes = 200
energy_nodes = nsq.logspace(E_min,E_max,E_nodes)

neutrino_flavors = 3

=======

nuSQuIDS inputs should be given in natural units.
In order to make this convenient, as in the {\ttf C++} code, we make use of the {\ttf SQuIDS} class {\ttf Const}.
We can instantiate it as follows
\begin{lstlisting}[language=Python, breaklines=true]
units = nsq.Const()
\end{lstlisting}

As in the {\ttf C++} $\nu$-SQuIDS interface one can propagate the neutrinos in various environments
(see~\ref{sec:body} for further details). We can start by considering oscillactions in vacuum by setting:
\begin{lstlisting}[language=Python, breaklines=true]
nuSQ.Set_Body(nsq.Vacuum())
\end{lstlisting}

Since we have to specify that we are considering vacuum propagation, 
we must construct a trajectory inside that object. This can be done using the {\ttf Track} class-property of a given {\ttf Body}.
Each {\ttf Body} will have its own {\ttf Track} subclass and its constructors. We can set and construct a vacuum trajectory in the following way:
\begin{lstlisting}[language=Python, breaklines=true]
nuSQ.Set_Track(nsq.Vacuum.Track(100.0*units.km))
\end{lstlisting}

Next we have to set the neutrino energy, which can be done as follows
\begin{lstlisting}[language=Python, breaklines=true]
nuSQ.Set_E(1.0*units.GeV)
\end{lstlisting}

Now we have to set the initial neutrino state. We can provide this state in either the flavor or mass basis.
We can do this using the {\ttf Set\_initial\_state} function, providing it with a numpy array and a {\ttf nsq.Basis}.
If the basis is {\ttf nsq.Basis.flavor} then the list must contain {\ttf np.array([}$\phi_e$,$\phi_\mu$,$\phi_\tau${\ttf ])}, 
similarly if enumeration is {\ttf nsq.Basis.mass} the list must specify $\phi_i$.
In this example we set the initial state to $\nu_\mu$,
\begin{lstlisting}[language=Python, breaklines=true]
nuSQ.Set_initial_state(np.array([0.,1.,0.]),nsq.Basis.flavor)
\end{lstlisting}

Finally, we calculate the system evolution by
\begin{lstlisting}[language=Python, breaklines=true]
nuSQ.EvolveState()
\end{lstlisting}

After this runs, {\ttf nuSQuIDS} has evolved the state and holds it in memory.
We can inquire for the flavor composition at this stage by
\begin{lstlisting}[language=Python, breaklines=true]yy
print([nuSQ.EvalFlavor(i) for i in range(3)])
\end{lstlisting}

which results in
\begin{lstlisting}[language=Python, breaklines=true]yy
[0.004675502573140013, 0.9022064087866011, 0.09311808864025879]
\end{lstlisting}

$\nu$-SQuIDS knows everything about the neutrino state at the current moment, it also knows what we did with it so far, where it went, what mixing parameters were used, etc. It would be convenient to store this information. One way of doing this is to serialize the $\nu$-SQuIDS status. We can do this in the following way
\begin{lstlisting}[language=Python, breaklines=true]
nuSQ.WriteStateHDF5("current_state.hdf5")
\end{lstlisting}

Everything that is in the $\nu$-SQuIDS object is now in that file. We can use that file to create a new $\nu$-SQuIDS object and do another calculation, we can stop the calculation midway and use it to restart, we can explore that file with other analysis tools, etc. In particular, the `ReadStateHDF5` will return us to the given configuration.
\begin{lstlisting}[language=Python, breaklines=true]
nuSQ.ReadStateHDF5("current_state.hdf5")
\end{lstlisting}

We can use the current tool to try to calculate $P(\nu_\mu \to \nu_e)$ as a function of energy. We can do the following
\begin{lstlisting}[language=Python, frame=leftline, numbers=left, breaklines=true]
energy_values = np.linspace(1,10,40)
nu_mu_to_nu_e = []
for Enu in energy_values:
    nuSQ.Set_E(Enu*units.GeV)
    nuSQ.Set_initial_state(np.array([0.,1.,0.]),nsq.Basis.flavor)
    nuSQ.EvolveState()
    nu_mu_to_nu_e.append(nuSQ.EvalFlavor(0))

plt.figure(figsize = (8,6))
plt.xlabel(r"$E_\nu [{\rm GeV}]$")
plt.ylabel(r"$P(\nu_\mu \to \nu_e)$")
plt.plot(energy_values,nu_mu_to_nu_e, lw = 2, color = 'blue')
\end{lstlisting}
This results in the plot shown in Figure~\ref{fig:nusquids_python_osc} left panel.
\begin{figure}[h!]
  \label{fig:nusquids_python_osc}
  \centering
  \includegraphics[width=0.45\textwidth]{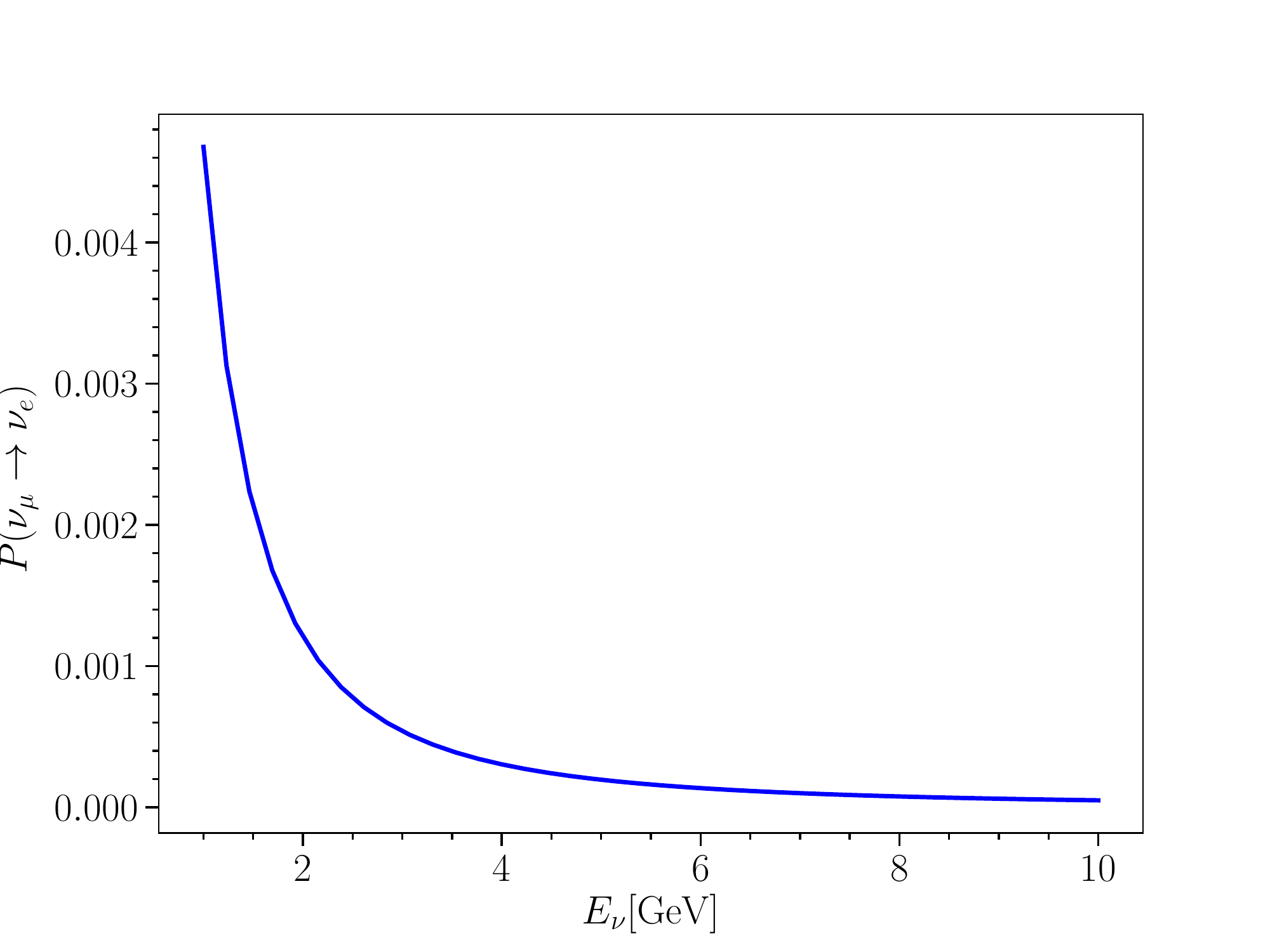}
  \includegraphics[width=0.45\textwidth]{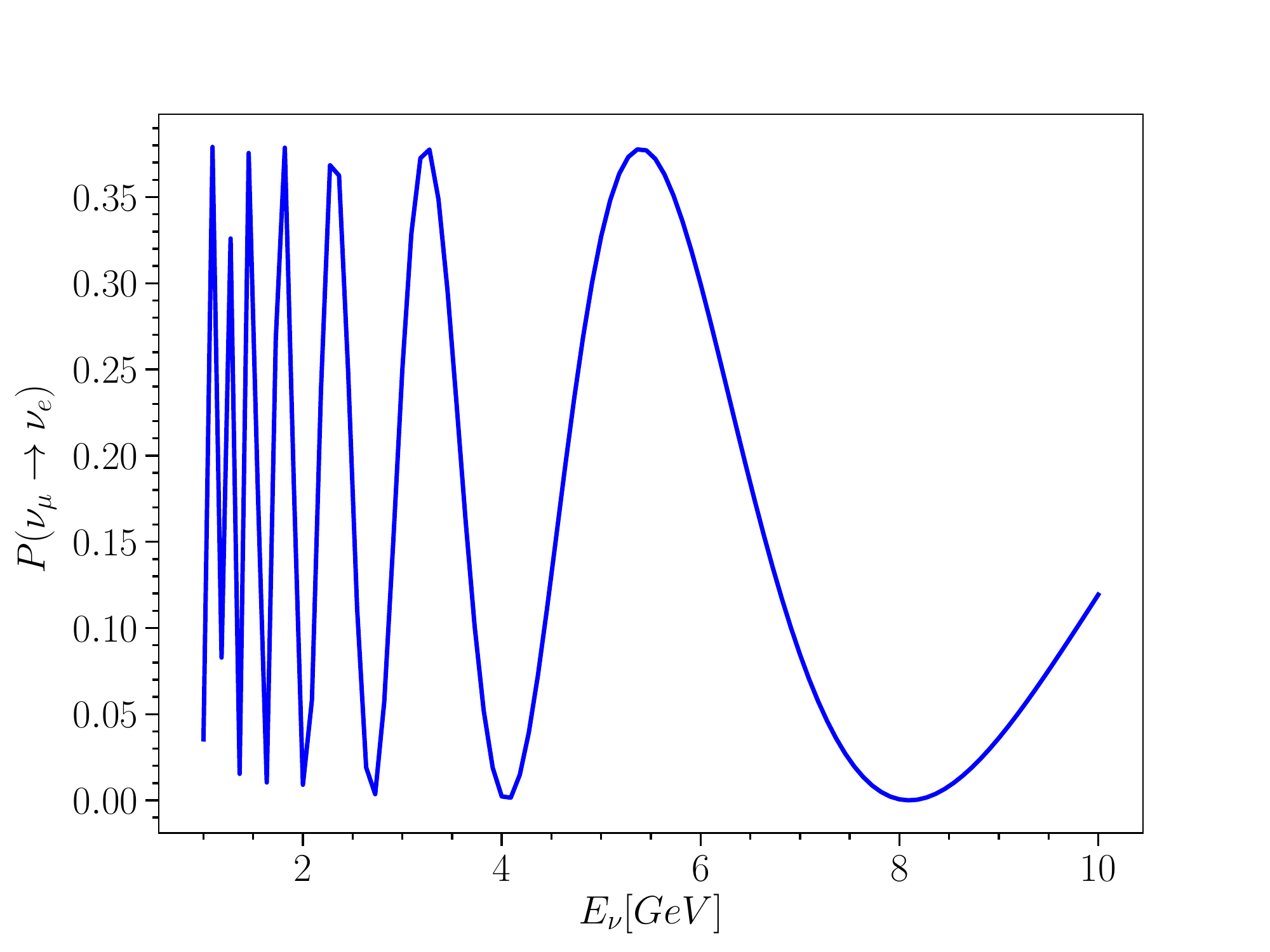}
  \caption{Muon-neutrino to electron neutrino oscillation probability in vacuum as a function of the neutrino energy for a baseline of 100 km. 
  The left panel is for the default standard oscillation parameters and the right panel has $\Delta m^2_{21}$ set to $1~{\rm eV}^2$}
\end{figure}

nuSQuIDs has some predefined oscillation mixing angles ($\theta_{ij}$) and mass splittings ($\Delta m^2_{ij}$),
which we have taken from~\cite{Esteban:2020cvm,Gonzalez-Garcia:2021dve}.
We can change the square mass differences by using the following function
\begin{lstlisting}[language=Python, breaklines=true]
nuSQ.Set_SquareMassDifference(1,2.0e-1) # sets dm^2_{21} in eV^2.
\end{lstlisting}

Running the same script again results in the plot shown in Fig.~\ref{fig:nusquids_python_osc} right panel.

We can also try to modify the mixing angles, for example
\begin{lstlisting}[language=Python, breaklines=true]
nuSQ.Set_MixingAngle(0,1,1.2) # sets \theta_{12} in radians.
nuSQ.Set_MixingAngle(0,2,0.3) # sets \theta_{23}} in radians.
nuSQ.Set_MixingAngle(1,2,0.4) # sets \theta_{23}} in radians.
\end{lstlisting}
We can also restore the default mixing angles and mass differences to default by doing
\begin{lstlisting}[language=Python, breaklines=true]
nuSQ.Set_MixingParametersToDefault()
\end{lstlisting}

\subsubsection{Multiple-energy mode}

In this section we will demonstrate the use of {\ttf nuSQuIDS} in the multiple energy mode.
To begin we have to specify the $\{E_i\}$ grid where the equation will be solved.
The energy nodes can be an arbitrary ordered list of energy in eV.
The most common choice is to have linearly-spaced nodes or logarithmically-spaced nodes.
For constructing numpy array of linear and logarithmic scales nuSQuIDS provides two
 convenient functions called {\ttf linspace} and {\ttf logspace }respectively. The 
following lines define the energy grid and construct the nuSQUIDS object.

\begin{lstlisting}[language=Python, frame=leftline, numbers=left, breaklines=true]
interactions = False

E_min = 1.0*units.GeV
E_max = 10.0*units.GeV
E_nodes = 101

energy_nodes = nsq.logspace(E_min,E_max,E_nodes)

neutrino_flavors = 3

nuSQ = nsq.nuSQUIDS(energy_nodes,neutrino_flavors,nsq.NeutrinoType.neutrino,interactions)
\end{lstlisting}

We can propagate a neutrino ensemble through the Earth in an atmospheric neutrino telescope setting.
We can do that by setting the following Body and Track:

\begin{lstlisting}[language=Python, frame=leftline, numbers=left, breaklines=true]
earth = nsq.EarthAtm()
nuSQ.Set_Body(earth)
nuSQ.Set_Track(earth.MakeTrackWithCosine(-1))
\end{lstlisting}

For the initial flux we will assume that it's given by $\phi_\nu = N_0 E^{-2}$,
with flavor composition $\phi_e:\phi_\mu:\phi_\tau$ = $0:1:0$.
$\nu$-SQuIDS input flux is a {\ttf numpy.ndarray} formatted in the following way:
\begin{equation}
InputState \doteq [[\phi^1_e,\phi^1_\mu,\phi^1_\tau],...,[\phi^i_e,\phi^i_\mu,\phi^i_\tau],...,[\phi^n_e,\phi^n_\mu,\phi^n_\tau]]
\end{equation}
where $i = 1$ to $n$ and $n$ is the number of energy nodes. We can implement this in the following way
\begin{lstlisting}[language=Python, frame=leftline, numbers=left, breaklines=true]
N0 = 1.0e18
Eflux = lambda E: N0*E**-2
Einitial = (Eflux(nuSQ.GetERange()).reshape((101,1)))*(np.array([0.,1.,0.]).reshape(1,3))
\end{lstlisting}
where {\ttf N0} is an arbitrary normalization constant, but we have chosen it such that the input flux is $O(1)$.

\begin{lstlisting}[language=Python, frame=leftline, numbers=left, breaklines=true]
nuSQ.Set_initial_state(Einitial,nsq.Basis.flavor)

nuSQ.Set_rel_error(1.0e-17)
nuSQ.Set_abs_error(1.0e-17)

nuSQ.EvolveState()
\end{lstlisting}

Then we can evaluate the resulting flux and plot the resulting oscillation 
probabilities by using the folowing script

\begin{lstlisting}[language=Python, frame=leftline, numbers=left, breaklines=true]
e_range = np.linspace(1.0,10.0,200)

nu_e = np.array([nuSQ.EvalFlavor(0,EE*units.GeV,0) 
                           for EE in e_range])
nu_mu = np.array([nuSQ.EvalFlavor(1,EE*units.GeV,0) 
                           for EE in e_range])
nu_tau = np.array([nuSQ.EvalFlavor(2,EE*units.GeV,0) 
                           for EE in e_range])

total = nu_e + nu_mu + nu_tau

plt.figure(figsize = (8,6))

plt.xlabel(r"$E_\nu [{\rm GeV}]$")
plt.ylabel(r"$\phi_\alpha$")

plt.plot(e_range,nu_e, lw = 2, color = 'red', label = r"$\nu_e$")
plt.plot(e_range,nu_mu, lw = 2, color = 'blue', label = r"$\nu_\mu$")
plt.plot(e_range,nu_tau, lw = 2, color = 'green', label = r"$\nu_\tau$")
plt.plot(e_range,total, lw = 2, color = 'black', label = r"Total")
plt.legend(fancybox = True, fontsize = 10)
plt.semilogy()
\end{lstlisting}
The result of this is shown in the left panel of Fig.~\ref{fig:nusquids_atm_python_no_interaction}. Our final example is 
performing the full nuSQuIDS calculation with neutrinos and antineutrinos, but this time including interactions

\begin{figure}[h!]
  \label{fig:nusquids_atm_python_no_interaction}
  \centering
  \includegraphics[width=\textwidth]{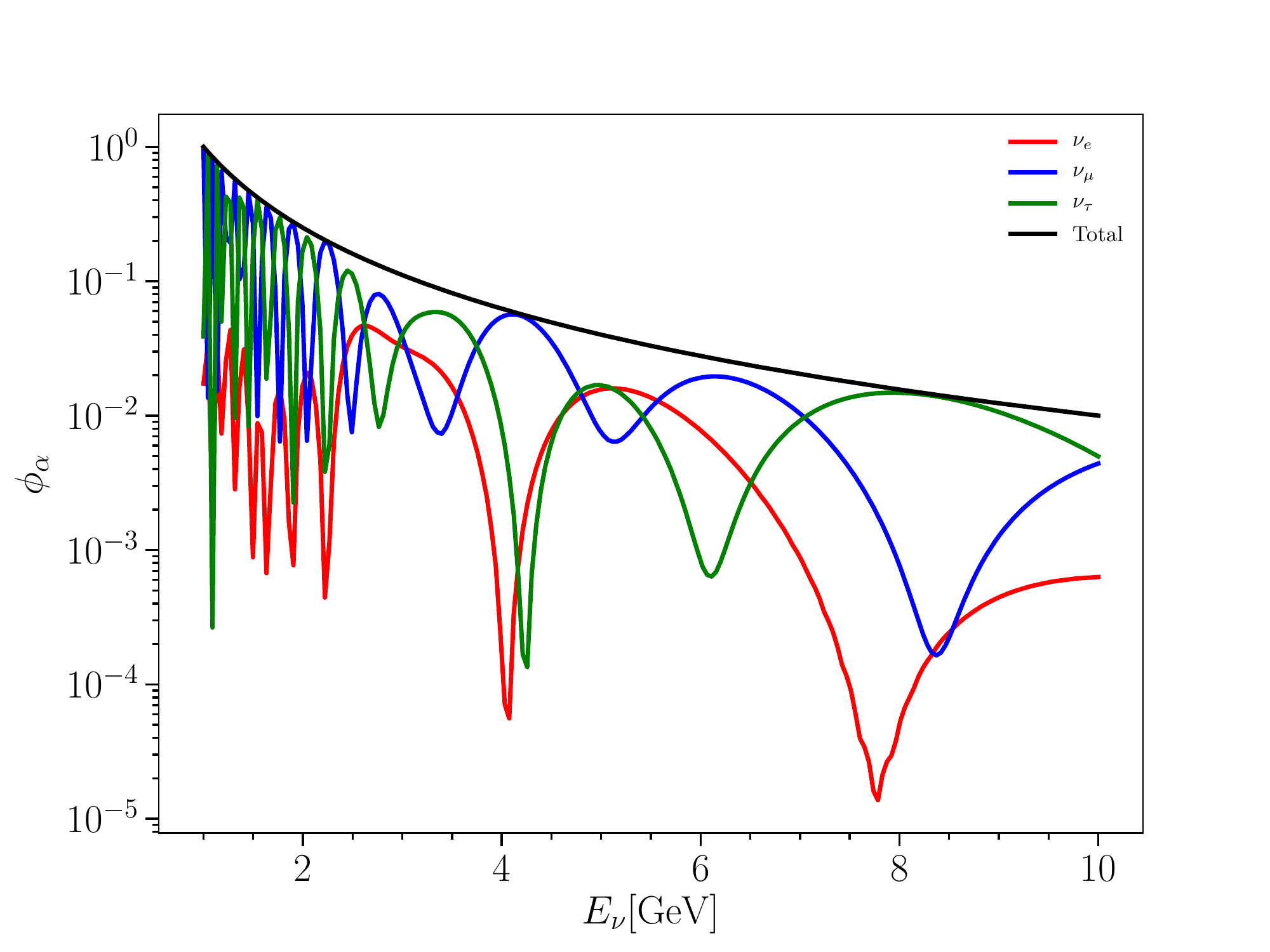}
  \caption{Result of propagating a power-law flux with spectral index of -1 through the Earth considering only oscillations and no interactions.}
\end{figure}

\begin{lstlisting}[language=Python, frame=leftline, numbers=left, breaklines=true]
interactions = True

E_min = 1.0e2*units.GeV
E_max = 1.0e8*units.GeV
E_nodes = 200
energy_nodes = nsq.logspace(E_min,E_max,E_nodes)

neutrino_flavors = 3

>>>>>>> c6a8be92b35067939fa5267703c6d8ea03949c30
nuSQ = nsq.nuSQUIDS(energy_nodes,neutrino_flavors,nsq.NeutrinoType.both,interactions)
\end{lstlisting}
In this case we need to specify the neutrino and antineutrino initial fluxes. Those have the following format:
\begin{equation}
InputState \doteq [[[\phi^1_e,\phi^1_\mu,\phi^1_\tau],[\bar{\phi}^1_e,\bar{\phi}^1_\mu,\bar{\phi}^1_\tau]],...,[[\phi^i_e,\phi^i_\mu,\phi^i_\tau],[\bar{\phi}^i_e,\bar{\phi}^i_\mu,\bar{\phi}^i_\tau]],...,[[\phi^n_e,\phi^n_\mu,\phi^n_\tau],[\bar{\phi}^n_e,\bar{\phi}^n_\mu,\bar{\phi}^n_\tau]]] \nonumber \nonumber
\end{equation}
where $i = 1$ to $n$ and $n$ is the number of energy nodes and $\phi$ ($\bar{\phi}$) is the neutrino (antineutrino) flux.
We can implement this as follows
\begin{lstlisting}[language=Python, frame=leftline, numbers=left, breaklines=true]
N0 = 1.0e18; Power = -1.0
Eflux = lambda E: N0*E**Power

InitialFlux = np.zeros((200,2,3))
for i,E in enumerate(nuSQ.GetERange()):
    InitialFlux[i][0][0] = 0.0
    InitialFlux[i][1][0] = 0.0
    InitialFlux[i][0][1] = Eflux(E)
    InitialFlux[i][1][1] = Eflux(E)
    InitialFlux[i][0][2] = Eflux(E)
    InitialFlux[i][1][2] = Eflux(E)
\end{lstlisting}

\begin{lstlisting}[language=Python, frame=leftline, numbers=left, breaklines=true]
nuSQ.Set_MixingParametersToDefault()
earth = nsq.EarthAtm()
nuSQ.Set_Body(earth)
nuSQ.Set_Track(earth.MakeTrackWithCosine(-1))
nuSQ.Set_initial_state(InitialFlux,nsq.Basis.flavor)
\end{lstlisting}
An important phenomena in this scenario is that of tau-regeneration, which we can enable by
\begin{lstlisting}[language=Python, breaklines=true]
nuSQ.Set_TauRegeneration(True)
\end{lstlisting}

After doing the evolution we can evaluate the resulting flux by doing:
\begin{lstlisting}[language=Python, frame=leftline, numbers=left, breaklines=true]
nu_tau = np.array([ nuSQ.EvalFlavorAtNode(2,ie,0) for ie,EE in enumerate(nuSQ.GetERange())])
nu_tau_bar = np.array([ nuSQ.EvalFlavorAtNode(2,ie,1) for ie,EE in enumerate(nuSQ.GetERange())])

nu_mu = np.array([ nuSQ.EvalFlavorAtNode(1,ie,0) for ie,EE in enumerate(nuSQ.GetERange())])
nu_mu_bar = np.array([ nuSQ.EvalFlavorAtNode(1,ie,1) for ie,EE in enumerate(nuSQ.GetERange())])
\end{lstlisting}

and plot it by using this script

\begin{lstlisting}[language=Python, frame=leftline, numbers=left, breaklines=true]
plt.figure(figsize = (6,6))
plt.xlabel(r"$E_\nu [{\rm GeV}]$")
plt.ylabel(r"${\rm E_\nu} \times \phi_\nu (E_\nu)$")
e_range = nuSQ.GetERange()/units.GeV
plt.plot(e_range,0.5*(nu_tau+nu_tau_bar)*(e_range*units.GeV)/N0, lw = 3, label =r"$\nu_\tau + \bar{\nu}_\tau$", color = "blue")
plt.plot(e_range,0.5*(nu_mu+nu_mu_bar)*(e_range*units.GeV)/N0, lw = 3, label =r"$\nu_\mu + \bar{\nu}_\mu$", color = "red")
plt.axhline (1.0, color = "k", lw = 3)
plt.xlim(1.0e2,1.0e6)
plt.ylim(1.0e-2,1.0e1)
plt.grid()
plt.loglog()
plt.legend(loc = "upper right", fancybox = True, fontsize = 15)
\end{lstlisting}
The result of this is shown in Fig.~\ref{fig:nusquids_atm_python_with_interaction}.

\begin{figure}[h!]
  \label{fig:nusquids_atm_python_with_interaction}
  \centering
  \includegraphics[width=0.7\textwidth]{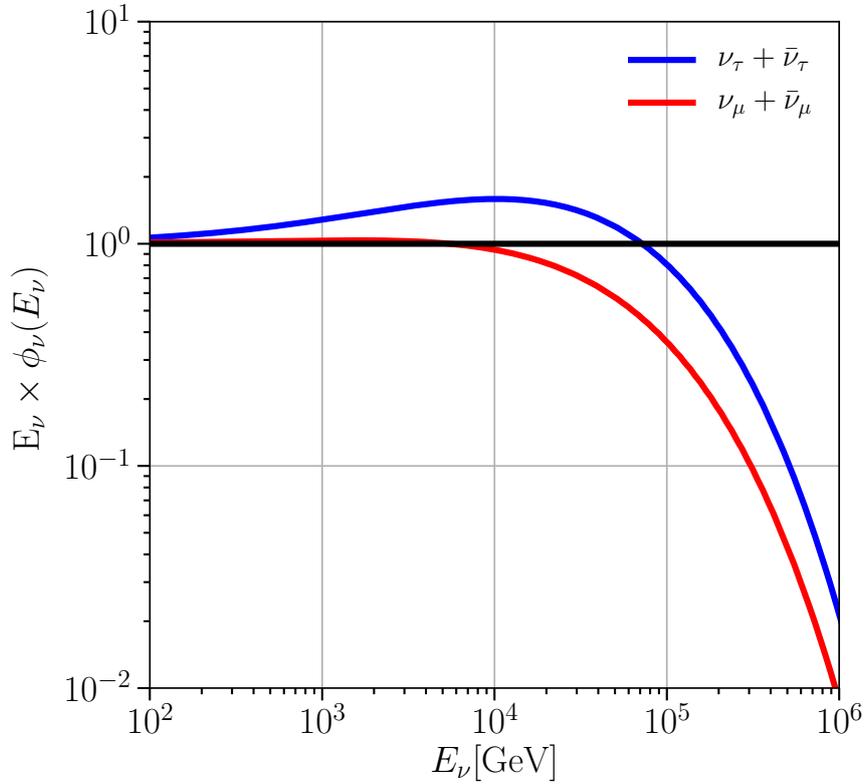} 
  \caption{Propagated flux for muon and tau neutrinos through the Earth including the effect of tau regeneration.}
\end{figure}

\subsection{{\ttf SU\_vector} operations in Python~\label{sec:python_suvector}}

SQuIDS {\ttf SU\_vector} class provides a combinent way to perform operations
that commonly arise in quantum mechanics. For this reason we have also make them
available in the {\ttf Python} interface. All of the {\ttf SU\_vector} operations have been implemented,
but due to optimizations performed in the {\ttf C++} some notes are in order.

We can compute the {\ttf SU\_vector} that represents a typical neutrino Hamiltonian 
in three flavors as
\begin{lstlisting}[language=Python, frame=leftline, numbers=left, breaklines=true]
import nuSQUIDSpy as nsq
p2=nsq.SU_vector.Projector(3,1)
dm12=1e-5
H=p2*dm12
\end{lstlisting}
In this code snippet {\ttf p2} is a {\ttf SU\_vector}, but {\ttf H} is a {\ttf SU\_vector::MultiplicationProxy}.
In the {\ttf C++} implementation these proxies appear when operations are performed on {\ttf SU\_vector} objects
to optimize memory usage, as described earlier in this document. To keep operating on these objects one needs to 
cast them onto real {\ttf SU\_vector} rather than proxies for them. This can be done as follows:
\begin{lstlisting}[language=Python, frame=leftline, numbers=left, breaklines=true]
H=nsq.SU_vector(H)
\end{lstlisting}
after which {\ttf H} is a real vector and can be operated upon. This same conversion needs to be performed explicitly
every time a proxy is encountered in {\ttf Python}.

\subsection{Extending the interface with user-defined classes~\label{sec:python_extensions}}

In this section we briefly show how to construct python bindings for user-defined classes. As an example
we have an extension class called {\ttf nuSQUIDSExt} defined as:

\begin{lstlisting}[language=C++, frame=leftline, numbers=left, breaklines=true]
class nuSQUIDSExt : public nuSQUIDS {
  ...
  public:
    nuSQUIDSExt(); // default constructor
    nuSQUIDSExt(double x); // constructor
    double FunctionExtension(double x); // function of interest
    ...
}
\end{lstlisting}
where this class has two constructors and one function of interest which we would like to expose to {\ttf Python}.
In order to construct the python module we need to make a new C++ source file with the following content.

\begin{lstlisting}[language=C++, frame=leftline, numbers=left, breaklines=true]
#include "nuSQUIDSpy.h"
#include "nuSQUIDSExt.h"

BOOST_PYTHON_MODULE(nuSQUIDSExtensionPythonModuleName)
{
  // Register all standard nuSQuIDS and nuSQuIDS atmospheric functions for the user class
  auto nusquidsext_register = RegisterBasicNuSQuIDSPythonBindings<nuSQUIDSExt>("nuSQUIDSExt");
  auto nusquidsext_atm_register = RegisterBasicAtmNuSQuIDSPythonBindings<nuSQUIDSExt>("nuSQUIDSExtAtm");
  // Register additional functions or members of the user class
  auto nusquidsext_class_object = nusquidsext_register.GetClassObject();
  nusquidsext_class_object->def("FuctionExtension",&nuSQUIDSExt::FunctionExtension);
}
\end{lstlisting}
where the header {\ttf nuSQUIDSpy.h} contains the nuSQUIDS python binding utilities and definitions, and {\ttf nuSQUIDSExt.h} is the 
header where the new class of interest is defined. In this code we have used two template structures, whose constructors are:
\begin{itemize}
  \item
    \begin{lstlisting}[language=C++, breaklines=true]
      RegisterBasicNuSQuIDSPythonBindings<nuSQUIDSDerivedType>(std::string name);
    \end{lstlisting}
    The {\ttf type} of the extension has to be a {\ttf nuSQUIDS} derived type, in this case \\
  {\ttf nuSQUIDSDerivedType} is {\ttf nuSQUIDSExt}. The string that is passed
  as an argument does not need to match the type name and is how the class will be named in {\ttf Python}. This class will register all the functions
  that it inherits as a nuSQUIDS derived class, but not any new members introduced by the user.
  \item
    \begin{lstlisting}[language=C++, breaklines=true]
      RegisterBasicAtmNuSQuIDSPythonBindings<nuSQUIDSDerivedType>(std::string name);
    \end{lstlisting}
    The {\ttf type} of the extension has to be a {\ttf nuSQUIDS} derived type, in this case \\ {\ttf nuSQUIDSDerivedType} is {\ttf nuSQUIDSExt}. The string that is passed
  as an argument does not need to match the type name and is how the class will be named in {\ttf Python}. This class will register all the functions
    that it inherits as a {\ttf nuSQUIDSAtm<Type>} derived class, but not any new members introduced by the user.
\end{itemize}

This module needs to be compiled with the same dependencies as our nominal python bindings, for which we suggest the user
to copy our Makefile and modify the name of the library to be generated to match the module name. As a final step the user
should place the resulting shared object library in the system {\ttf PYTHONPATH}.

\section{Singly-Differential Cross-Section Tabulation}
\label{sec:differential_tabulation}
Efforts to precisely and consistently capture the physics of the singly-differential DIS cross sections have lead to using a slightly unusual tabulation scheme, the details of which we describe here. 

First, in order to interpolate cross section values for arbitrary incident and out-going energies, it is convenient that values be tabulated on some rectangular grid. 
The natural choice is to use a common set of out-going energies energies for all incident energies. 
Clearly, such a set of out-going energies should extend up to the maximum tabulated incident energy. 
However, the cross section is meaningless for out-going energies greater than the incident energy, so for all table rows corresponding to incident energies less than the maximum tabulated, some of the higher out-going energy entries will be undefined, and must be filled with some dummy value (or a non-rectangular storage format must be used). 
This layout is shown in the left panel of Fig.~\ref{fig:dsdy_tab}. 
The presence of this diagonal boundary through the table between physical and unphysical regions frustrates simple interpolation schemes in the region near the boundary, as the method used must not allow the undefined or placeholder values to contribute to the calculation of the interpolated value. 
This can be addressed by switching interpolation methods depending on location in the table (such as using linear interpolation via barycentric coordinates on a triangle for points in cells with only three physically allowed corner points but typical rectangular bilinear interpolation in other cells), but this adds complexity and is hostile to using higher-order interpolation methods. 
As a more minor point, when typical rectangular storage formats are used, the unphysical region wastes storage space. 

\begin{figure}
  \centering
  \includegraphics[width=0.9\textwidth]{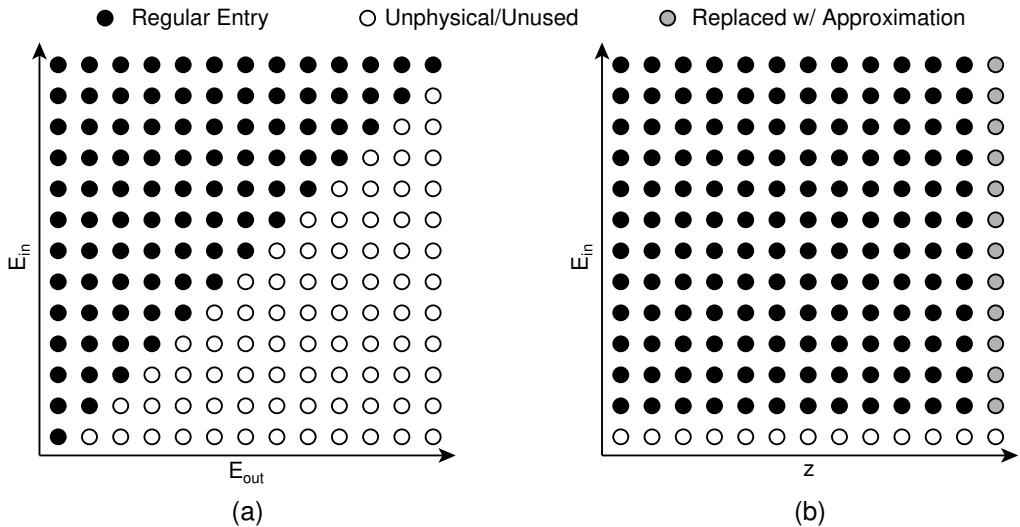}
  \caption{Schematic depiction of the two tabulation schemes discussed for singly-differential cross sections. (a) shows simple tabulation in $E_{in}$ and $E_{out}$, while (b) shows the scheme introduced here using $E_{in}$ and the transformed $z$ variable, and indicates the $z = 1$ entries which are replaced by the values chosen to approximate the peak as a function of $E_{out}$ or $z$. It is assumed that $E_{out}$ ranges over the same set of values as $E_{in}$, and that $z$ ranges over [0, 1].}
  \label{fig:dsdy_tab}
\end{figure}

Second, the cross sections have considerable curvature, which leads to systematic errors when linear interpolation is used. 
These errors are illustrated in the lower panel of Fig.~\ref{fig:cross_interp} where the relative error of the interpolated cross section grows periodically to between the tabulated cross section points. 
It should be the case that the integral of the singly-differential cross section over out-going energy for a given incident energy should equal the total cross section (up to truncation effects due to the limited domain of tabulation). 
If this is not the case, errors gradually accumulate when solving the evolution of a neutrino flux undergoing interactions with matter, producing non-physical artifacts in the final computed flux. 
This effect is particularly pernicious because it is difficult to detect in calculations using a coarse energy node spacing, with its nature only becoming clearly evident for node spacings finer than the cross section tabulation spacing. 
Importantly, neither calculation will be correct, but efforts to determine a suitably dense node spacing can be frustrated by the appearance of the unphysical artifacts when more dense spacing are tested for comparison. 

\begin{figure}
  \centering
  \includegraphics[width=0.9\textwidth]{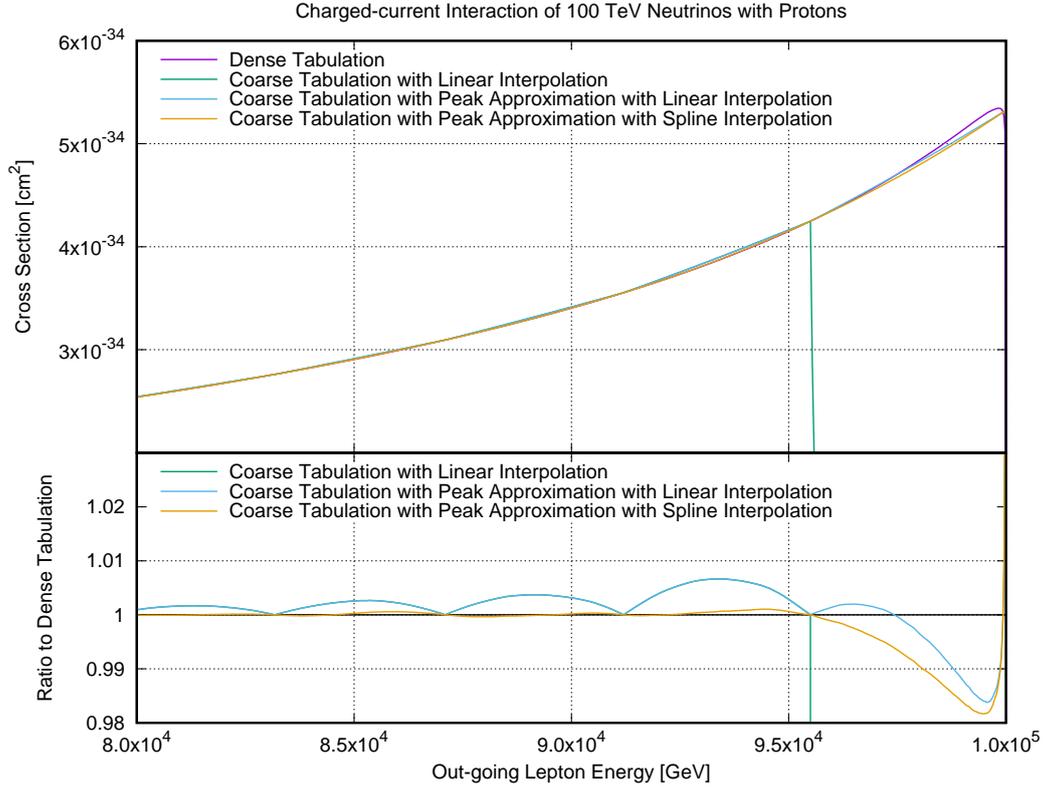}
  \caption{Illustration of problems interpolating the singly-differential DIS cross section at high energies and the effects of different mitigations discussed in the text. The `Dense Tabulation' represents an essentially ideal treatment which unfortunately uses far too many samples for general practicality (10,000 samples per decade in $E_{out}$). The `Coarse Tabulation' uses are more normal 50 samples per decade. The tabulation points for the coarse table are visible as the points in the the lower panel where the error of the interpolation disappears, since the values tabulated in both tables are identical where they both exist. The linear interpolation of the coarse table is efficient to compute, but has problematically large errors in near the centers of the table bins due to being unable to represent curvature. Approximation with cubic splines greatly reduces this type of error. Simple interpolation with the final table entry (for $E_{out} = E_{in}$) being the formally correct zero value shows a very large error due to entirely missing the peak of the function. Replacing the final sample with a carefully chosen approximation eliminates this severe effect, although the true shape of the peak still cannot be captured.}
  \label{fig:cross_interp}
\end{figure}

Third, as incident neutrino energies increase, the cross section as a function of out-going energy develops a sharp peak structure for out-going energies close to the incident energy. 
Inevitably this peak will at some energy become too narrow to be sampled accurately by the tabulation spacing, whatever it has been chosen to be. 
This problem is greatly exacerbated by the additional fact that when the out-going energy is equal to the incident energy, the cross section must be zero. 
Taken literally, this forces the final table entry in each row to be zero, which causes the interpolation between the last two table entries to fall sharply, even when one would wish for it to first rise to the peak. 
Since a large portion of the integral of the differential cross section can be contained in this peak this causes a sizable discrepancy between the integral of the (interpolated) differential cross section and the total cross section. 
This is illustrated in Fig.~\ref{fig:cross_interp} by the severe disagreement between the ideal sufficiently dense tabulation and the (practically realizable) coarse tabulation with linear interpolation over the latter's last table bin. 

To address the first and second problems, it is desirable to use a rectangular tabulation which has no unphysical entries and will permit simple use of higher order interpolation methods, such as cubic splines (or bicubic interpolation based on cubic splines, given the two-dimensional nature of the table). 
It is therefore useful to introduce a variable $z$:

\begin{equation}
  z = {E_{out} - E_{out,min} \over E_{in} - E_{out,min}}
\end{equation}

where $E_{in}$ is the incident neutrino energy, $E_{out}$ is the out-going lepton energy, and $E_{out,min}$ is the minimum tabulated out-going lepton energy. 
This variable has the property that for $E_{out}$ ranging over $[E_{out,min}, E_{in}]$, $z$ takes on values from $[0, 1]$ for all $E_{in}$. 
If one tabulates in some set of $z$ values distributed over $[0, 1]$, the range of $E_{out}$ values covered will differ for different $E_{in}$, but in a useful way, since all will be physically meaningful, and while no $E_{out}$ values except $E_{out,min}$ will generally be shared between different $E_{in}$ values, this does no particular harm, since interpolation can simply be performed in the $(E_{in}, z)$ space. 
Tabulation in $z$ eliminates the first concern and significantly mitigates the second by enabling convenient use of higher order interpolation schemes, as shown in Fig.~\ref{fig:cross_interp}, where the spline interpolation shows greatly reduced errors for all except the last bin (which merits a separate discussion) compared to the linear interpolation. 
It should be noted that $z$ is undefined for $E_{in} = E_{out,min}$, but there can be no non-zero cross section for this table row (shown in the right panel Fig.~\ref{fig:dsdy_tab} as white points) if it is included anyway, so {\ttf nuSQuIDS} will never explicitly refer to these entries and they can be safely set to any convenient values (such as values similar to the entries for the next row with greater $E_{out}$, in order to give well-behaved interpolation). 

Unfortunately, tabulation in $z$ does nothing useful to address the third concern. 
More additional, valid tabulated values are added for lower incident energies, which have less need to resolve a sharp peak. 
Instead, another approach is needed. 
As previously mentioned, the $E_{out} = E_{in}$ (or $z = 1$) column of the table (shown as gray points in the right panel of Fig.~\ref{fig:dsdy_tab}) is known to always be formally zero, and so like the $E_{in} = E_{out,min}$ row it is not explicitly needed. 
Since this column is also a major driver of the bad interpolation of the peak when it is unresolvable inside the final table bin, replacing it is doubly useful for changing the interpolation to be something more suitable. 
Since the only strictly physically correct value to place in this column is zero, any other choice is an approximation, and which approximation is `best' is a matter of interpretation. 
Since the shape of the peak itself cannot be represented faithfully, we choose to attempt to preserve its integral instead, and choose to replace the final entry in the row with one selected so that the average of the last two values is equal to the true average computed from a detailed integral of the cross section over the bin. 
That is, if $T_i$ are the tabulated cross section values, of which there are $N$ in the table row, we set:

\begin{equation}
  T_{N} = {2 \over y(E_{out,N}) - y(E_{out,N-1})}\int^{y(E_{out,N})}_{y(E_{out,N-1})}{{d\sigma \over dy} dy} - T_{N-1}
\end{equation}

where $y$ is the standard scaling variable $y = (E_{in} - E_{out}) / E_{in}$. 
This choice gives the correct integral when the table is interpolated linearly, and gives similarly close results for common choices of higher-order interpolation, and can be interpreted as `smearing out' the contribution of the peak uniformly across the table bin in a cumulative sense. 
The two curves shown in Fig.~\ref{fig:cross_interp} with `Peak Approximation' in their labels show this scheme in action. 
While some residual error in the interpolated cross section remains due to the approximation cutting through the peak rather than following it exactly, the situation is vastly improved over the interpolation without approximation which drops rapidly to zero (extremely so, due to interpolation being done in the space of the logarithm of the cross section). 
From the figure it is clear that this much more accurately captures the contribution of this bin to the integral.

\section{Test suite}
\label{sec:tests}
A test suite is provided with the library to guarantee the right
output in different systems or after any modification of the
code. All the tests are located in the folder {\ttf nuSQuIDS/test}
And they can be run using the provided script {\ttf test/run\_tests}.
A single test can be run by adding the name as an argument to the script
command.
In the following we list the tests that are provided with a brief
description of what they test.

\begin{itemize}
  
\item {\ttf test/vacuum\_osc\_prob.test}
  
  Tests that propagation with only oscillation of
  neutrinos in vacuum is equal to the analytic solution.

\item {\ttf test/constant\_density\_osc\_prob.test}
  
  Tests that propagation with only oscillation of
  neutrinos in constant density is equal to the analytic solution.

\item {\ttf test/constant\_opacity.test}
  
  Tests that propagation on a constant density with no oscillation and
  neglecting regeneration from neutral current or tau decay matches
  the expected exponential behavior.
  
\item {\ttf test/constant\_opacity\_with\_nc.test}
  
  Tests that propagation on a constant density with no oscillation 
  matches an independent numerical solution.  
  
\item {\ttf test/atmospheric\_he.test}

  Tests that the high-energy propagation without oscillation of
  neutrinos from the atmosphere does not give negative or {\ttf nan} values.

\item {\ttf test/atmospheric\_osc.test}
  
  Tests that propagation with only oscillation of
  neutrinos from the atmosphere does not give negative or {\ttf nan}
  values.
  
\item {\ttf test/body\_serialization.test}
  
  It checks that writing a body in to the hdf5 file and reading it
  back does not alter the body.  

\item {\ttf test/cross\_section\_consistency.test}
  
  Checks that the differential cross section adds, within numerical error,
  to the total cross section.
  
\item {\ttf test/earth\_osc\_prob.test}
  
  Test that three and four neutrinos oscillation probability over a
  1000km baseline in multiple and single energy modes matches with the
  numerical error to a reference independent calculation.
  
\item {\ttf test/glashow\_resonance.test}
  
  Checks that Glashow resonance integrated differential cross-section
  matches with the total cross-section and that is used while propagating. 
  
\item {\ttf test/hdf5\_atm\_in\_out.test}
  
  It checks that writing a nuSQuIDs atmospheric object into a hdf5 file and reading it
  back does not alter the object and recovers all the properties.

\item {\ttf test/hdf5\_in\_out.test}
  
  It checks that writing a nuSQuIDs object into a hdf5 file and reading it
  back does not alter the object and recovers all the properties.

\item {\ttf test/move\_assig.test}
  
  Checks that constructor of a nuSQuIDS object from an rvalue nuSQuIDS object works.
  
\item {\ttf test/mul\_energy\_constructor.test}
  
  Checks that the multiple energy constructor works.
  
\item {\ttf test/time\_reversal.test}
  
  Checks that propagating a neutrino and propagating again backwards
  in time returns to the original state.
  
\item {\ttf test/tools\_integrator.test}
  
  Checks that the provided one dimensional integrate function works
  within the numerical error.
  
\item {\ttf test/track\_concatenate\_hdf5.test}
  
  Checks that propagating by concatenating a series of tracks in
  vacuum and writing and reading that state at every step is the same
  as propagating with a single track with the total length.

\item {\ttf test/track\_concatenate.test}
  
  Checks that propagating by concatenating a series of tracks in
  vacuum is the same as propagating with a single track with the total length. 
  
\item {\ttf test/external\_flux.test}
  
  Checks that the emitting body feature works. This is done by propagating 
  an initially empty neutrino ensemble on a medium that emits neutrinos and comparing the 
  result with an analytical solution.

\end{itemize}

%\section{Additional resources}
%\label{sec:resources}

\journal{arXiv}
\fi % forjournal

\ifdefined\manualonly
\bibliographystyle{elsarticle-num}
\bibliography{manual}
\fi

\end{document}